\documentclass[usenatbib]{mnras}
 \usepackage{color}
 \usepackage{subfig}
 \usepackage{mathrsfs}
 \usepackage{amssymb}
 \usepackage{amsmath}
 \usepackage{tikz}
 \usepackage{gensymb}
 \usepackage{makecell}

\definecolor{codegreen}{rgb}{0,0.6,0}
\definecolor{codegray}{rgb}{0.5,0.5,0.5}
\definecolor{codepurple}{rgb}{0.58,0,0.82}
\definecolor{backcolour}{rgb}{0.95,0.95,0.92}
 
\newcommand{\beq}{\begin{equation}}
\newcommand{\eeq}{\end{equation}}

\newcommand{\pw}{Paczy\'nsky-Wiita}
\newcommand{\rg}{r_{\rm g}}

\newcommand{\argt}{\theta}

\newcommand{\argtp}{(\theta+\psi)}

\newcommand{\argtpa}{(\theta_{\rm A}+\psi_{\rm A})}
\newcommand{\alf}{Alfv\'en~}
\newcommand{\fmx}{(1-M^2-x^2)}
\newcommand{\fmxA}{(1-M^2-x_{\rm A}^2)}
\newcommand{\fgmx}{(G^2-M^2-x^2)}
\newcommand{\fg}{(1-G^2)}

\newcommand{\numm}{\mathcal{N}_1}
\newcommand{\nump}{\mathcal{N}_2}
\newcommand{\den}{\mathcal{D}}
\newcommand{\refp}[1]{(\ref{#1})}
\newcommand{\citeeg}[1]{\citep[e.g.][]{#1}}
\newcommand{\pder}[2]{{\partial #1\over \partial #2}}
\newcommand{\tder}[2]{{d #1\over d #2}}

\defcitealias{Polko:2010}{PMM10}
\newcommand{\capmp}{\citetalias{Polko:2010}}

\defcitealias{Polko:2013}{PMM13}
\newcommand{\capt}{\citetalias{Polko:2013}}

\defcitealias{Polko:2014}{PMM14}
\newcommand{\capp}{\citetalias{Polko:2014}}

\defcitealias{VlahakisKonigl:2003}{VK03}
\newcommand{\cavt}{\citetalias{VlahakisKonigl:2003}}
\newcommand{\cpavt}[2]{\citepalias[#1][#2]{VlahakisKonigl:2003}}

\defcitealias{VTST:2000}{VK00}
\newcommand{\cavz}{\citetalias{VTST:2000}}
\newcommand{\cpavz}[2]{\citepalias[#1][#2]{VTST:2000}}

\defcitealias{NRfortran:1993}{NR93}

\newcommand{\cpan}[2]{\citepalias[#1][#2]{NRfortran:1993}}

\newcommand{\syst}{system of equations \eqref{sys:wind}-\eqref{sys:xi}}

\newcommand{\fitn}{goodness of fit}
\newcommand{\fits}{\chi}

\newcommand{\cri}{critical}
\newcommand{\crif}{\mathcal{C}}

\newcommand{\zero}{_{\rm 0}}
\newcommand{\gr}{_{\rm g}}
\newcommand{\ata}{_{\rm A}}
\newcommand{\thea}{\theta\ata}

\newcommand{\masa}{\varpi\ata}
\newcommand{\psia}{\psi\ata}
\newcommand{\thef}{\theta_{\rm MFP}}
\newcommand{\thes}{\theta_{\rm MSP}}

\newcommand{\xa}{x\ata^2}
\newcommand{\themg}[1]{\theta_{\rm mid,\,#1}}
\newcommand{\themf}{\themg{MFP}}
\newcommand{\thems}{\themg{MSP}}

\newcommand{\mny}{\texttt{multinest}}

\usepackage{etoolbox}
\makeatletter
\newcount\c@additionalboxlevel
\newcount\c@maxboxlevel
\patchcmd\@combinedblfloats{\box\@outputbox}{%
  \stepcounter{additionalboxlevel}%
  \box\@outputbox
}{}{\errmessage{\noexpand\@combinedblfloats could not be patched}}

\AtBeginShipout{%
  \ifnum\value{additionalboxlevel}>\value{maxboxlevel}%
    \typeout{Warning: maxboxlevel might be too small, increase to %
      \the\value{additionalboxlevel}%
    }%
  \fi 
  \@whilenum\value{additionalboxlevel}<\value{maxboxlevel}\do{%
    \typeout{* Additional boxing of page `\thepage'}%
    \setbox\AtBeginShipoutBox=\hbox{\copy\AtBeginShipoutBox}%
    \stepcounter{additionalboxlevel}%
  }%
  \setcounter{additionalboxlevel}{0}%
}
\makeatother

\title[Self-similar relativistic MHD outflows]{A new method for extending solutions to the self-similar relativistic magnetohydrodynamics equations for black hole outflows}
\author[Ceccobello et al.]{Ceccobello$^{1}$\thanks{Contact e-mail: \href{mailto:c.ceccobello@uva.nl}{c.ceccobello@uva.nl}} C., Cavecchi$^{2,3}$ Y., Heemskerk$^{1}$ M.H.M., Markoff$^{1}$ S., Polko$^{4}$ P., Meier$^{5}$ D.\\
$^{1}$``Anton Pannekoek'' Instituut voor Sterrekunde, Universiteit van Amsterdam, Science Park 904, 1098 XH Amsterdam, The Netherlands\\
$^{2}$ Department of Astrophysical Sciences, Princeton University, Peyton Hall, Princeton, NJ 08544, USA\\
$^{3}$ Mathematical Sciences and STAG Research Centre, University of Southampton, SO17 1BJ, UK\\
$^{4}$ University of Maryland at College Park, Dept. of Physics, Joint Space-Science Institute,\\
3114 Physical Sciences Complex, College Park, MD 20742, USA\\
$^{5}$ Jet Propulsion Laboratory, California Institute of Technology, Pasadena, CA 91109, USA}
\begin{document}
\maketitle

\begin{abstract}
The paradigm in which magnetic fields play a crucial role in launching/collimating outflows in many astrophysical objects continues to gain support.  However, semi-analytical models including the effect of magnetic fields on the dynamics and morphology of jets are still missing due to the intrinsic difficulties in integrating the equations describing a collimated, relativistic flow in the presence of gravity.
Only few solutions have been found so far, due to the highly nonlinear character of the equations together with the need to blindly search for singularities. These numerical problems prevented a full exploration of the parameter space.
We present a new integration scheme to solve $r$-self-similar, stationary, axisymmetric magnetohydrodynamics equations describing collimated, relativistic outflows crossing smoothly all the singular points (\alf point and modified slow/fast points).  For the first time, we are able to integrate from the disk mid-plane to downstream of the modified fast point.
We discuss an ensemble of jet solutions, emphasising trends and features that can be compared to observables. 
We present, for the first time with a semi-analytical MHD model, solutions showing counter-rotation of the jet for a substantial fraction of its extent.  
We find diverse jet configurations with bulk Lorentz factors up to 10 and potential sites for recollimation between $10^3-10^7$ gravitational radii. 
Such extended coverage of the intervals of quantities, such as magnetic-to-thermal energy ratios at the base or the heights/widths of the recollimation region, makes our solutions suitable for application to many different systems where jets are launched.
\end{abstract}

\begin{keywords}
black hole physics - magnetohydrodynamics (MHD) - galaxies: jets - stars: jets - methods: numerical
\end{keywords}

\section{Introduction}
\label{intro}

Since their discovery, relativistic collimated outflows of matter have been observed in many astrophysical objects and they are known to be associated with accretion flows.   
Jets reveal themselves at different scales and redshifts, showing an extreme diversity in energetics, shapes and emission. Jets are found to be characteristic features of black hole systems, such as X-ray binaries (XRBs) and active galactic nuclei (AGN), as well as of young stellar objects (YSOs), explosive transients such as tidal disruption events (TDEs)  and gamma-ray bursts (GRBs). 
Observations suggest that jets are an energetically important component of the system that hosts them, because the jet power appears to be comparable to the accretion power \citep[see e.g.][for a more recent discussion]{RawlingsSaunders:1991,NemmenTchekhovskoy:2015}.
Significant evidence has been found of the effect of jets not only on the immediate proximity of the central object, but also on their surrounding environment, where they deposit the energy extracted from the accretion flow  \citep[e.g.][]{Gallo:2005,Fabian:2012}. 
To launch, accelerate and collimate a relativistic outflow over such large distances, magnetic fields need to be invoked. Observational evidence, such as polarization measurements both in the radio \citep{MartiVidal:2015} and in the hard X-rays \citep{Laurent:2011}, support the idea that ordered magnetic fields are a key ingredient in the jet phenomena, and have a significant effect on their emission as well. 
Understanding what causes the jet to be launched from an accretion disk, which mechanisms determine its shape and extension and where the radiation is produced in these systems is one of the most fundamental questions in astrophysics and needs to be addressed promptly. 

Multiwavelength continuum emission from jets and disks is observed for all accreting, jet-launching sources, XRBs in particular can provide essential pieces of information because extensive monitoring of these sources shows that they go through a duty cycle multiple times during their lives \citep{Fender:2004}. 
During an outburst, they spend most of their time in the \textit{low-luminosity hard state} where they exhibit a mildly relativistic steady-state jet launched from a likely recessed disk. In this state, the jet dominates the total power and shows a characteristic power-law spectrum that may extend to high energies. 
Many outbursts show a rapid increase in luminosity, bringing the system close to its Eddington limit. The standard paradigm has the disk inner radius moving closer to the black hole while the jet becomes ballistic, emitting superluminal knots. Eventually the jet switches off and the emission is dominated by the disk. The spectrum becomes softer and loses its non-thermal high-energy component (\textit{high-luminosity soft state}). Finally, the system slowly decays into a hard state and the cycle restarts  \citep[see, e.g.][for a recent review]{BelloniMotta:2016}.
When the emission is dominated by the compact steady-state jet, its characteristic synchrotron emission can span several orders of magnitude in frequency. At the wavelength range where the synchrotron transitions from the optically thin ($\tau < 1$) to the optically thick regime ($\tau > 1$), the spectrum shows a break and it becomes a flat/inverted power-law, characteristic of the self-absorbed synchrotron \citep[see ][for a recent review]{Romero:2017}. 
The region in the jet corresponding to the break frequency is believed to be the site where particles are first accelerated \citep{Markoff:2010}, potentially by internal shocks \citep{Malzac:2014}.
The jet break can be inferred from observations to occur over a fairly large range of distances from the black holes $\sim10-10^4$ gravitational radii (hereafter $\rg$), and it has been seen to span 4 orders of magnitude in frequency during a state transition in a single object \citep{Russell:2014}.
Similar spectral features and a duty cycle $10^7-10^8$ times longer are seen in AGN as well, and a power-unification scenario has been proposed independently by  \citet{Merloni:2003} and \citet{Falcke:2004}. They showed that AGN and XRBs of the same relative luminosity, rescaled by the mass of the black hole, can be explained with the same physics framework.
While XRBs provide unique constraints on the emission of jets and accretion disks thanks to the multiwavelength monitoring of the activity of the source during state-transition episodes, AGN are ideal for studying the structure of jets and the dynamical processes that shape them.  \\
Indeed, in the case of nearby AGN, high spatial resolution observations are now possible with very long baseline interferometry (VLBI) in both cm and now mm bands. These data provide unprecedented constraints on the geometry and the dynamics of jets, such as the jet opening angle, the height and the width of knots associated with standing shock features, such as HST-1 in M87's jet. 
Recently, high-resolution VLBI observations of M87 by \citet{Hada:2016} resolved and imaged the inner core of the galaxy down to $\sim10~\rg$, revealing the innermost structure of the jet. VLBI/VLBA observations \citep{Asada:2014, Mertens:2016} constrained the bulk Lorentz factor of the jet of M87 to be mildly relativistic, i.e. $\gamma_{\rm j} \sim 1-3$. 
In the near future, with the beginning of the Event Horizon Telescope (EHT) era, observations will resolve the nearest black holes (Sagittarius $A*$ and M87) down to the event horizon scale \citep[see e.g.][]{Doeleman:2008}. This unprecedented resolution will shed new light on the immediate proximity of black holes, possibly unveiling the jet/disk connection and the mechanisms responsible for the acceleration and collimation in the first stages of jet formation.
Finally, \citet{Meyer:2013} measured the proper motions of the knots downstream of HST-1 with the Hubble Space Telescope, finding significant evidence of transverse and parallel motion with respect to the jet axis. This is evidence of a helical magnetic field beyond HST-1 and it brings important constraints on the modelling of jets at larger distances from the BH. 

Using both XRBs and AGN to obtain insight on the apparently similar jet phenomena is extremely important as demonstrated by the activity in this field of research.
However, an adequate modelling of jets, including a detailed treatment of both the radiative processes and the magnetohydrodynamics (MHD), is still far from being achieved. 

Thanks to the dramatic improvement of computational power, accretion disks and jets can be modelled with general relativistic magneto-hydrodynamic (GRMHD) simulations \citeeg{Koide:2002,McKinney:2006,HawleyKrolik:2006,Tchekhovskoy:2011,Tchekhovskoy:2016}.
Full 3D simulations allow detailed study of the stability of jets under different sets of initial conditions, and they provide a unique overview of how jets are launched, how the disk and jet interact during this phase and how they approach a stable configuration. 
However, the lack of crucial ingredients such as non-ideal processes or self-consistent radiative processes makes a direct test against observational data still a challenge. It is, however, worth noting that efforts are currently made to incorporate simplified treatments of electron microphysics in GRMHD simulations \citeeg{Ressler:2015,Moscibrodzka:2016}. 

A complementary method to simulations is given by semi-analytical models for multiwavelength emission from disk/jet systems where the geometry and the dynamics are generally fixed and simplified. Many of such models have been proposed, for instance by \citet{Romero:2003,Markoff:2005,Yuan:2005,PotterCotter:2012,Pepe:2015,ZLS:2012,ZSPS:2014} and they have been successful in reproducing the spectral energy distribution of accretion disks and jets. However, the treatment of magnetic fields is also greatly simplified and its orientation is usually not considered.
Finally, they all present a certain degree of degeneracy between combinations of input parameters that give statistically equivalent fits to the same data set. Introducing a self-consistent treatment of MHD via semi-analytical models can help reduce the freedom in the parameter space, allowing for better constraining the fits to the observations.\\
A number of such models have been developed since the 80s-90s, with pioneering works by \citet{BlandfordPayne:1982, Lovelace:1990,Li:1992, Contopoulos:1994, Sauty:1994,Bogovalov:1999}.
The MHD system of equations describing an accelerating flow is a highly nonlinear system which changes nature from elliptical to hyperbolic several times across the interval of integration. Independently of the geometry of the system, it has been shown that three critical surfaces exist in correspondence to such transitions and they determine as well the onset of magnetosonic waves of different type. These are called fast, slow magnetosonic and \alf waves. The slow magnetosonic singular surface (SMSS) appears close to the central object and upstream of the \alf surface. The fast magnetosonic singular surface (FMSS) is located downstream of the \alf point and it is suspected to be linked with a recollimation of the streamlines describing the flow \citep[see e.g.][for a complete discussion and derivation]{Meier:2012}.  The FMSS, therefore, could tentatively be identified with the jet break seen in observations of AGN and XRBs, while
the properties of the flow at SMSS could instead provide important constraints on the input parameters of radiative transfer models, such as e.g. the magnetic-to-thermal energy ratio at the base and the initial bulk velocity. 
However, the MHD equations that exhibit all the three singular surfaces must include gravity and the effect of thermal pressure in the total internal energy of the flow to properly describe the region close to the black hole. Moreover, to describe typical astrophysical jets, the equations need to allow the flow to become relativistic. 
\citet[herefter VTST00]{VTST:2000} and \citet[hereafter VK03]{VlahakisKonigl:2003} derived the MHD system of equations under the assumption of radial self-similarity, first including a simplified gravity term (kinetic term) and, later removing it, to include relativistic effects.
Only recently, \citet[hereafter, respectively PMM10, PMM13 and PMM14]{Polko:2010,Polko:2013,Polko:2014} derived the equations for a self-similar relativistic MHD flow including enthalpy and gravity, and found solutions crossing smoothly all three singular surfaces. Under the assumption of self-similarity, the singular surfaces are cones and intersect a streamline only in one point. They are usually referred to as "modified" slow and fast \emph{points} (MSP, MFP) and \alf point (AP). We will adopt this terminology from here on.
For a more detailed history of the derivation of the model, we address the interested reader to the papers cited above and references therein.\\
Since only at the AP analytical formulae can be written explicitly, locating the MFP and MSP and performing the integration of the equations through them is not an easy numerical problem to solve. These unknowns add a large degree of complexity to the problem and common integration techniques fail to retrieve solutions in most cases.
Although successful in solving such highly nonlinear set of MHD equations, \capp{} were limited in the range of jet solutions that they could retrieve due to their numerical approach. Following \cavt{}, \capp{} developed an algorithm where the integration starts from the \alf point and "shoots" towards the other two singular points (MFP and MSP). The \alf point can be regularised analytically by using the De L'H\^ opital rule on the terms that are in indefinite form $0/0$. 
The other two singular points are later found by extrapolation (for more details about the shooting method see e.g. \citealp{NRfortran:1993} and \citealp{Stoer:2013}). When the equations are integrated towards a singularity, however, the error with respect to the exact solution can be large. 
As a consequence of the limitations intrinsic to the adopted numerical scheme, \capp{} were limited to a small portion of the parameter space. \\
In this paper, we present a new method for exploring the parameter space and finding solutions to the MHD equations building on the work of the PMM papers. Using this method, we are able to explore more efficiently a much larger fraction of the parameter space. 
This will allow the model to be applicable to many other difficult flow solution problems in astrophysics, as well as potentially other fields.
Moreover, this class of models can be coupled with fairly accurate radiative transfer models \citep[see][]{Markoff:2005,Maitra:2009,Connors:2016,CCCC:2017}, which can handle spectral fitting in a reasonable computational time. \\
This paper follows the following structure: in Section~\ref{equations} we describe the system of equations that we solve and discuss modifications in the equations compared to \capp{}{} with respect to their dependence on the gravitational potential. In Section~\ref{method} we present our numerical scheme and in Section~\ref{comparison} we compare our results with \capp{}. We show that the solutions are extremely sensitive to the gravity terms and when the corrected functions of the gravitational potential are used, the self-similarity assumption is more easily broken. In Section~\ref{paramstudy} we present a partial study of the parameter space, unaccessible to previous studies.  
We included the details of our method in a series of Appendixes: in Appendix~\ref{definitions} we define the equations that we solve in explicit form with the corrected gravity terms and the new \alf regularity condition. In Appendix~\ref{derivpotential} we describe the derivation of the functions of the pseudo-potential in the gravity terms. In Appendix~\ref{sub:integ} we discuss in detail our approach in finding the locations of the unknown singular points, MFP and MSP, and how we perform the integration. Finally, in Appendix~\ref{conversion} we give the conversion of the most relevant quantities into physical units.

     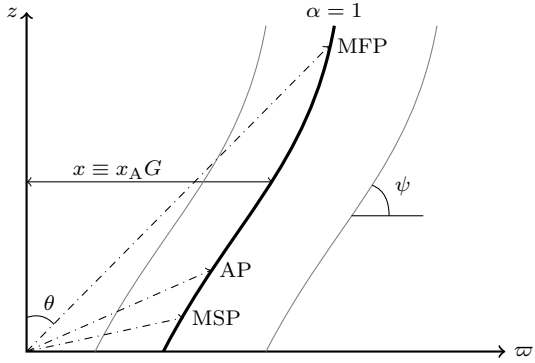
\begin{figure*}
    \centering
    \begin{tikzpicture}[scale=.9]
    \draw [thick, black,<->] (0,5) node[left] {$z$} --(0,0) -- (7,0) node[right]  {$\varpi$};
    \draw[<->] (0,2.5) node[right,xshift=0.5cm,yshift=0.16cm] {$x\equiv x\ata G$} -- (3.6,2.5);
    \draw[ very thick] (2,0) to [out=63,in=-100] (4.5,4.8) node[black,above] {$\alpha=1$};
    \draw[gray] (3.5,0) to [out=63,in=-100] (6.0,4.8);
     \draw[gray] (1.0,0) to [out=63,in=-100] (3.5,4.8);
    \draw[dashdotted,->] (0,0) -- (2.3,0.5) node[right] {MSP};
    \draw[dashdotted,->] (0,0) -- (2.7,1.2) node[right] {AP};
    \draw[dashdotted,->] (0,0) -- (4.42,4.5) node[right] {MFP};
    \draw (0,0.5) node[above,xshift=0.3cm] {$\theta$} to [out=20,in=130] (0.4,0.4);
   \draw (5.05,2.45) node[right,xshift=0.2cm] {$\psi$} to [out=-20,in=90]  (5.30,2);
   \draw (4.75,2) -- (5.8,2);
    \end{tikzpicture}\hfill
    \includegraphics[width=0.45\textwidth]{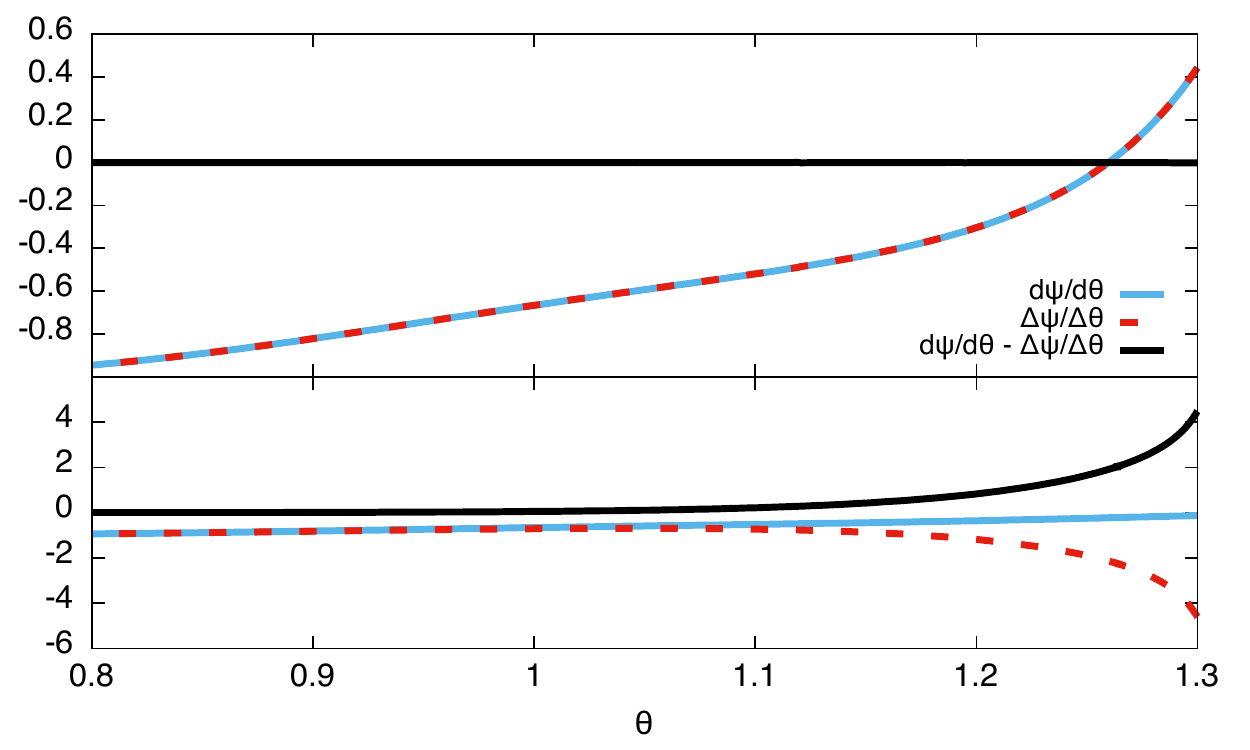}
           \caption{\textit{Left panel:} System of coordinates we adopt to describe a solution of eq.~\refp{sys:wind}-\refp{sys:xi}, which is typically the "reference" streamline identified with the label $\alpha \equiv \varpi_A^2/\varpi^2 = 1$. The four unknowns in the system~\refp{sys:wind}-\refp{sys:xi}, together with all the other quantities, are functions of $\theta$, which is the angle between a point in the streamline and the $z$-axis. The angle that the tangent to the streamline makes with the horizontal axis is $\psi$, while the distance from a point of the streamline to the $z$-axis is defined by its cylindrical radius in units of the light cylinder $x$. \textit{Right panel:} Discrepancy between the  derivative of $\psi$, calculated directly from eq.~\ref{sys:dpsidt2}, $d \psi / d\theta$, and inferred numerically at each step from eq.~\ref{sys:psi}, $\Delta \psi / \Delta \theta$ with the corrected functions of the gravitational potential (top panel) and with the one used by \capp{}{} (bottom panel). The MSP and the black hole are on the right.}
  \label{fig:angles}
   \end{figure*}

\section{Set up of the equations}
\label{equations}
Our goal is to describe an outflow launched from an accretion disk in
  the presence of a magnetic field by the Blandford-Payne mechanism \citep{BlandfordPayne:1982}. We use cylindrical coordinates
  $(\varpi,\phi,z)$ and spherical coordinates $(r, \phi, \theta)$ and
  impose axial symmetry, i.e. $\partial /\partial \phi = 0$ and assume
  the flow to be stationary, $\partial/\partial t = 0$. For the
  electric field $E$, we impose the \textit{freeze-in} condition, $E =
  -\mathbf{v} \times \mathbf{B}$, for the Ohm's law in ideal MHD and
  due to the axial symmetry $E_\phi=0$. In what follows we adopt the
  notation of \cavt{} and \capp{}.
  
 As in \cavt{}, we assume that the dependence of each variable on
  $r$ and $\theta$ is separable, and that the radial dependence can be expressed as a power law of $r$ \cpavt{see}{}. This assumption leads to self-similar solutions that obey a system of ordinary differential
  equations plus algebraic constraints. The four unknowns are the specific relativistic enthalpy $\xi
  c^2$, the poloidal \alf Mach number, $M = (\gamma V_{\rm p}/B_{\rm p}) (4\pi\rho_{\rm 0}\xi)^{1/2}$ ($V_{\rm p}$ and $B_{\rm p}$ are the poloidal components of the velocity and the magnetic field, $\rho_{\rm 0}$ is the baryon rest mass density, $\gamma$ is the bulk Lorentz factor), the cylindrical radius
  $x$ in units of the light cylinder, $x= \varpi \Omega / c$ ($c$ is
  the speed of light, $\Omega$ is the angular velocity of the flow), and the angle $\psi$ between the streamline and the
  horizontal axis. These are functions of $\theta$ only. We further substitute $x
  = x\ata G$, such that for each streamline, $G$ is the cylindrical
  radius scaled by its value at the \alf point.

The system of equations that we solve can be written in the form:
\begin{align}
\frac{d M^2}{d\theta} & = \frac{B_2C_1 - B_1C_2}{A_1B_2 - A_2B_1} \equiv \frac{\numm}{\den}, \label{sys:wind}\\
\frac{d G^2}{d\theta} & = \frac{2G^2\cos(\psi)}{\sin(\theta)\cos(\psi+\theta)}, \label{sys:dg2dt}\\
{\mu'^2 \over \xi^2}&\left\{{G^4(1-M^2-x_{\rm A}^2)^2-x^2(G^2-M^2-x^2)^2\over G^4(1-M^2-x^2)^2}\right\}  \nonumber\\
& = 1 +  {F^2\sigma_{\rm M}^2 M^4 \sin^2\theta\over\xi^2 x^4 \cos^2(\theta+\psi)}, \label{sys:psi} \\
M^2 & = q {\xi \over (\xi -1)^{1/(\Gamma-1)}}, \label{sys:xi}
\end{align}
where the functions $A_i,B_i,C_i$ with $i=1,2$ are defined in
  Appendix \ref{definitions}, while $F$ is a parameter describing the scaling of the magnetic field with respect to the radius as $B \propto r^{F-2}$. $\sigma_{\rm M} \equiv B_{\rm p}\Omega^2/(4\pi\rho_{\rm 0} V_{\rm p} c^3) $ is the magnetization parameter introduced by \citet{Michel:1969}. For more details, see Appendix \ref{definitions}, while for a complete derivation of the original equations we refer the reader to \cavt{} and
  \capp{}. Here we will provide a short description of their meaning
  and use.  
  
  Eq.~\refp{sys:wind} is the so-called \textit{wind}
  equation for the poloidal \alf Mach number, $M$, and can be derived from the Euler equation. Eq.~\refp{sys:dg2dt} describes the evolution of the
  dimensionless cylindrical radius and can be derived from the Euler equation.
  Eq.~\refp{sys:psi} is the Bernoulli equation describing the energy
  conservation along the poloidal component of the magnetic field line. From this equation we
  derive $\psi$, once $M^2$ and $G^2$ are know at each integration step. 
  Finally, Eq.~\refp{sys:xi} is the relation between the poloidal $M$, the enthalpy $\xi$ and the dimensionless adiabatic parameter $q$ (Tab. \ref{tab:modpars}), which we use to derive $\xi$ at each step.
  \begin{table*}
%\centering
\def\arraystretch{1.3}%
\setlength\tabcolsep{4.5 pt}
 \caption{Model parameters.}
 \label{tab:modpars}
 \begin{tabular}{l|c|l}
  \hline
  \thead{Input \\parameters} & &\\
  \hline
    $F$ & 			$F \equiv d \log I /d \log \varpi + 1$ with $B\propto r^{F-2}$ 	& determines the shape of the magnetic field at the base\\
  $\Gamma$ & 		$P = Q\rho_{0}^\Gamma$	& polytropic index of the gas \\
  $\sigma_{\rm M}$ &	$\sigma_{\rm M} \equiv B_{\rm p}\Omega^2/(4\pi\rho_{\rm 0} V_{\rm p} c^3) $	& magnetization parameter \\
  $\varpi_{\rm A}$ &  	$1/r = \sin\theta/(\varpi\ata G)$	& \parbox[t]{6cm}{\alf cylindrical radius, \\ used to define the gravitational potential}  \\
  $\theta_{\rm A}$ &  	(see Fig.~\ref{fig:angles})	& angular distance of the \alf point from the jet axis\\
  $\psi_{\rm A}$ &  		(see Fig.~\ref{fig:angles})	&  \parbox[t]{6cm}{inclination of the stream line \\ with respect to the horizontal axis at the AP}\vspace*{0.1cm}\\
  \hline
  \thead{Fitted \\parameters} & \\
  \hline
  $x\ata^2$ & 		(see Fig.~\ref{fig:angles})	& square of the \alf cylindrical radius in units of light cylinder \\
  $q$ &  		$q = B_{\rm 0}^2 \alpha^{F-2} x_{\rm A}^4/(4 \pi c^2 F^2 \sigma_{\rm M}^2) (\Gamma Q/(c^2(\Gamma -1)))^{1/(\Gamma-1)}$		& dimensionless adiabatic coefficient\\
  $\theta_{\rm MFP}$ &  	(see Fig.~\ref{fig:angles})	& angular distance of MFP from the jet axis  \\
  $\theta_{\rm MSP}$ &  	(see Fig.~\ref{fig:angles})	& angular distance of MSP from the jet axis  \\
  \hline
  \end{tabular}
\end{table*}

Our approach differs from that of \capp{} in how
  the gravity term is treated.  Gravity enters through the pseudo-potential
  $P_{\rm g}$ (see Appendix \ref{definitions}).  \capp{}{, following  \cite{Meier:2012},} considered $P_{\rm g}$ to be
  small and therefore approximated terms like $1 / ( 1 - P_{\rm g} )$
  as $( 1 + P_{\rm g} )$ or used other similar approximations. $P_{\rm g}$
  enters the $C_i$ terms in Eq. \eqref{sys:wind} and the
  analogous equation for $\psi$ 
  \beq 
  \frac{d\psi}{d\theta}=\frac{A_1C_2 - A_2C_1}{A_1B_2 - A_2B_1} \equiv \frac{\nump}{\den},
  \label{sys:dpsidt2}
  \eeq
  which in principle we do not use, since we can exploit the much more
  tractable and accurate Bernoulli equation
  (Eq. \ref{sys:psi}).\\ However, we noticed that the rate of change
  of $\psi$ with respect to $\theta$ as derived from the system of
  equations \eqref{sys:wind} - \eqref{sys:xi} was not consistent with the
  prediction of Eq. \eqref{sys:dpsidt2}, with a substantial discrepancy upstream of the \alf point (see right panel of Fig.~\ref{fig:angles}).

  The reason can be understood as follows: \cite{Meier:2012}
      obtained the Bernoulli and the transfield equations neglecting
      terms $\propto P_{\rm g}^2$ (see equations F.16 and F.18 in
      \citealp{Meier:2012}). Our Eq. \eqref{sys:psi} corresponds to
      \citeauthor{Meier:2012}'s equation F.16. In order to obtain Eq.
      \eqref{sys:wind}, we take derivatives of the Bernoulli equation
      Eq. \eqref{sys:psi} and therefore of $P_{\rm g}$ (see Appendix
      \ref{derivpotential}). However, \emph{we also keep the Bernoulli
        equation in its original form}. If we were to approximate the
      term $1 / ( 1 - P_{\rm g} )$ in the derivatives, the resulting
      equation would not be consistent with Eq. \eqref{sys:psi}
      anymore, because we would obtain a different version of the Bernoulli equation when integrating them back. Of course, this
      difference would be more pronounced upstream of the \alf{} point
      where $P_{\rm g}$ is greater, explaining the discrepancy we
      measured.

   We use Eq. \eqref{sys:psi} in order to evaluate quickly
     $\psi$ and therefore we need to retain in $C_1$ the full term $1
     / (1 - P_{\rm g})$ and calculate full derivatives of $P_{\rm g}$
     with respect to $\theta$ when needed\footnote{We use $d P_{\rm
         g}/d \theta = \partial P_{\rm g}/\partial \theta + \partial
       P_{\rm g}/\partial G \cdot d G/d \theta$.}. This approach
     brings back self-consistency between Eqs. \eqref{sys:wind}-\eqref{sys:xi} and Eq. \eqref{sys:dpsidt2}.

   Analogously, the gravity correction term to $C_2$ comes
     from derivatives of $P_{\rm g}$ in the transfield equation
     written to the same order of approximation as the Bernoulli
     equation (equation F.18 of \citealp{Meier:2012}). For keeping the same
     consistency as mentioned before, we keep the full terms of the
     derivatives of $P_{\rm g}$ in $C_2$ as well (Appendix
     \ref{derivpotential}).

It is worth noting that some of the constants of motions along the streamlines are affected by the inclusion of gravity within a general relativistic formalism as done by \capp{} following \cite{Meier:2012}. 
The specific energy $\mu$, defined in \cavt{} (eq. 13d), is not a constant of motion anymore, while the following one is:
\beq
(1-P_{\rm g})\left[\gamma+\gamma(\xi - 1) + S \right] =  (1-P_{\rm g}) \mu \equiv \mu'. \label{berneq}
\eeq
The first term on the left-hand side is the kinetic energy, the second is the internal energy, the third is the Poynting energy ($S = - \varpi \Omega B_\phi/ \Psi c^2$, with $\Psi = 4\pi\rho\zero \gamma V_{\rm p}/B_{\rm p}$ is the mass-to-magnetic flux ratio\footnote{Here we do not use the italic subscript "A" for $\Psi$ to avoid confusion with the roman subscript meaning that a quantity has been calculated at Alfv\`en, but it is the equivalent to Eq. 13b in \cavt{}.}), the multiplicative factor is the contribution of gravity with $P_{\rm g}$ defined as in eq.~\refp{potential}.
Similarly, the constant specific angular momentum (see \cavt{}, Eq. 13c), becomes now 
\beq
(1-P_{\rm g})\left[\xi \gamma \varpi V_\phi - \frac{\varpi B_\phi}{\Psi }\right] =  (1-P_{\rm g}) L \equiv L'. \label{momeq}
\eeq
All the other constants of motion remain unchanged.
We can recast Eq. \refp{berneq} and \refp{momeq} in the following compact notation:
\begin{align}
\mu' & = \mu'_{\rm HD} + \mu'_{\rm M}, \label{berncomp} \\
L' & = L'_{\rm HD} + L'_{\rm M},  \label{momcomp}
\end{align}
where 
\begin{align}
\mu'_{\rm HD} & = (1-P\gr)\xi\gamma , \qquad  L'_{\rm HD} =  (1-P\gr) \xi\gamma\varpi V_\phi, \\
\mu'_{\rm M} & = (1-P\gr)S  , \qquad\quad~   L'_{\rm M} =    (1-P\gr)\varpi B_\phi/\Psi,
\end{align}
are the hydrodynamical and magnetic components of the total energy and the angular momentum which will be used in Sec. \ref{paramstudy}.

\section{Numerical method}
\label{method}
\begin{figure*}
\centering
\includegraphics[width=0.5\linewidth]{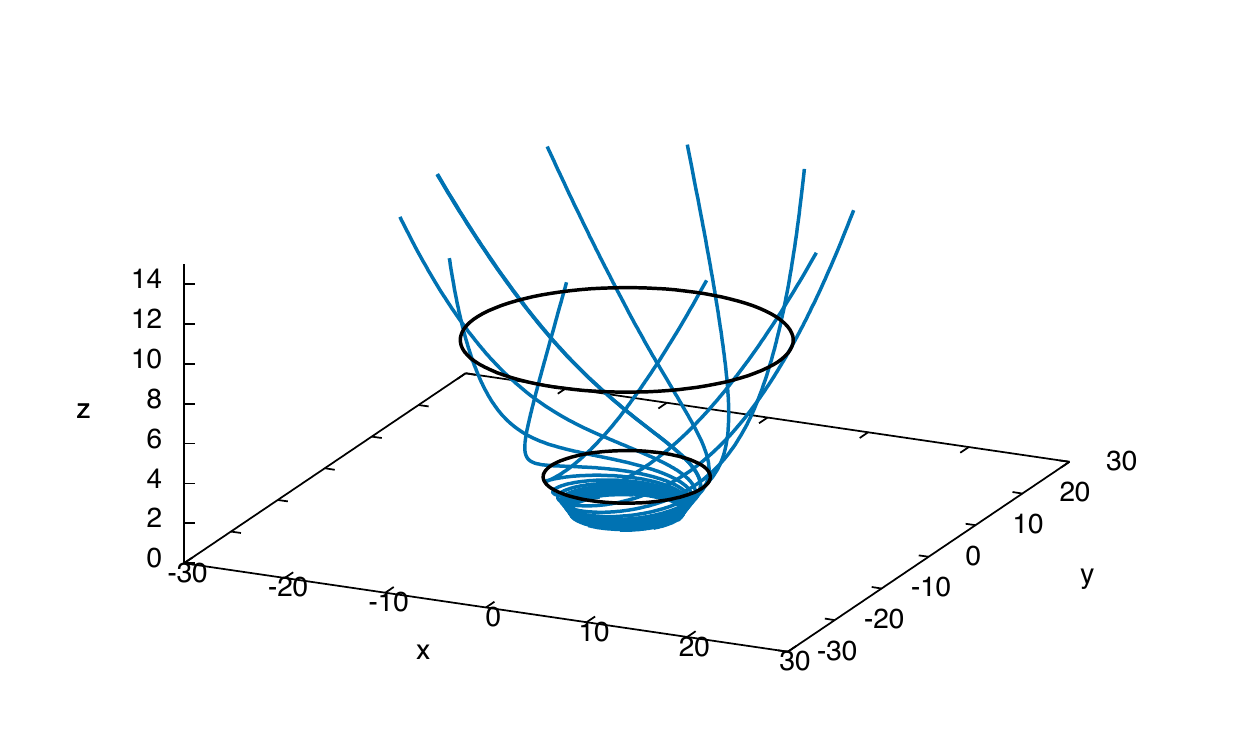}\hfil
\includegraphics[width=0.5\linewidth]{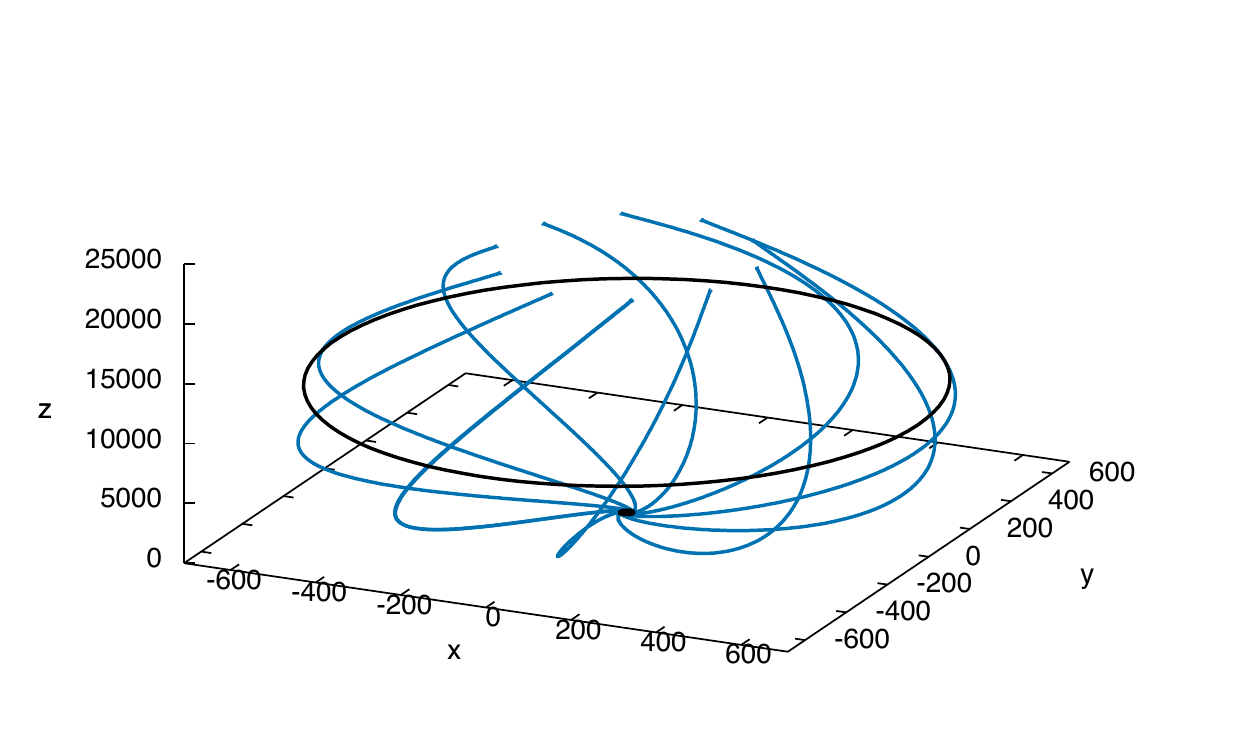}
\caption{3D plot of the streamlines corresponding to model (6) in Tab.~\ref{tab:pars} (see Fig.1a in \citet{Contopoulos:1995}  for a comparison). \textit{Left:} Zoom at the base (MSP and AP black circles, starting from the bottom respectively). \textit{Right:} MFPs.}
\label{fig:streamlines}
\end{figure*}

The \syst{} allows for different families of solutions which may or may not cross the singular point(s), similar to the case for hydrodynamic \citep{Parker:1958} or magnetohydrodynamic \citep{WeberDavis:1967} solar winds. In the case of the \syst{} and under the assumption of self-similarity, there can be up to three singular points: the \alf point AP, the modified fast magnetosonic point MFP and the modified slow magnetosonic point MSP (see Fig.~\ref{fig:streamlines}).  There is only one family of solutions that crosses all the points while it accelerates away from the black hole (see Fig. 1 and 2 of \citealt{WeberDavis:1967}, solution $u_{\alpha 1}$). Therefore, we shaped our approach in such a way that we automatically select solutions having this topology.

Each solution is characterised by the 10 parameters described in Table
  \refp{tab:modpars}. In particular, $F$ and
  $\Gamma$ describe the magnetic field configuration and the kind of plasma in the system, therefore they can be
  considered as defined for a given class of outflows. 
 The magnetisation parameter, $\sigma_{\rm M}$, gives the efficiency with which matter is pulled out of the rotating plasma at the base of the jet. 
  As shown in Table \refp{tab:modpars}, it is a function of the angular velocity of the streamlines, the magnetic field distribution and mass load. Hence, for a jet to be efficiently launched, it has to be strongly magnetised, rapidly rotating and/or have little mass load, as pointed out by \cite{Fendt:2004}. The $\sigma_{\rm M}$ parameter could be used as a fitted parameter, like we did to reproduce \capp{}{}'s results, but in general we keep this fixed, assuming again that this parameter will be defined by specific applications. Following a similar logic, we usually fix the position and the inclination of the field line, and the radial distance from the axis of the \alf point with $\theta\ata,\psi\ata$ and $\varpi\ata$. This last parameter is related to the strength of the gravitational potential as described in Appendix \ref{definitions}. We will discuss in detail the role of the fixed parameters in Section \ref{paramstudy}.
  In principle, the rest of the parameters should
  follow from the integration of the \syst{} once a suitable set of
  initial conditions is given for $M^2$, $G^2$ and $\theta$
  \cpavt{see}{}.

The integration is not simple, however, because the initial location of the MSP and MFP is not known. Therefore, some of the
  unknown parameters must be fixed, while the remaining ones are being
  retrieved as part of the solution process. The constraint we have on
  the solution, i.e. that all three singular points must be crossed,
  determines how many parameters will be found and how many need to be
  fixed a priori. The procedure that we follow is then an iterative one: \textit{1)} we fix a set of parameters ($F$, $\Gamma$, $\sigma_{\rm M}$,
  $\masa$, $\thea$ $\psia$),  \textit{2)} we make an educated guess for the remaining ones ($\thes$,
  $\thef$, $q$, $\xa$),  \textit{3)} we derive initial conditions for the
  integration and integrate the equations, and finally \textit{4)} we evaluate the "goodness"
  $\fits$ of the solution, improve the guesses and integrate again
  until a high enough $\fits$ is achieved. We now justify and describe
  in more detail our approach for each of these steps.

\subsection{Fixed and guessed parameters}
\label{sub:fixedguessed}

As a basic requirement we impose that all our solutions cross smoothly all the three
  critical points: MSP, AP, MFP (see Fig. \ref{fig:streamlines}).  At
  the AP analytical conditions are known \cpavz{see}{}. Given
  \emph{all} the parameters\footnote{$\thea$, $\masa$ and $\xa$
    determine the position of the AP, but $F$, $\Gamma$, $q$, $\sigma_{\rm M}$ and
    $\psia$ are needed for $M^2$ and $G^2$ and their
    derivatives.}, these conditions allow the determination of $G^2\ata$,
  $M^2\ata$, $d G^2 / d \theta|\ata$, and $d M^2 / d \theta|\ata$ through the \alf regularity condition (ARC, see
    Appendix \ref{definitions}), thus allowing integration away from the AP.

The regularity conditions at the modified
  magnetosonic points MFP and MSP are not known analytically. They are both of
  the same kind: the denominator $\den$ and the the numerators $\numm$
  and $\nump$ of Eqs. \eqref{sys:wind} and \eqref{sys:dpsidt2} must
  vanish, while $d M^2 / d \theta$ and $d \psi / d \theta$ remain
  finite. Other authors, like \cavz{}, \cavt{}, \capt{}, \capp{},
  found their solutions by using a shooting method to integrate from the \alf point upstream
  towards the MSP and downstream towards the MFP. This method has major caveats, however, since it is very
  hard to numerically integrate towards singular points, in which the numerators
  and the denominator approach zero simultaneously, while keeping the
  accuracy of the solution. It is however numerically stable to
  integrate away from a singularity. This inspired us to explore a different approach. 
  
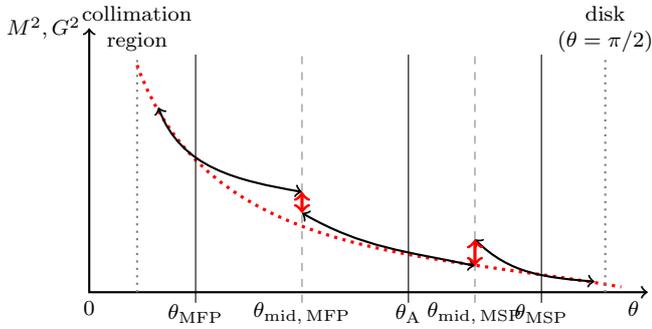
\begin{figure}
    %\centering
    \begin{minipage}{0.2\textwidth}
    \begin{tikzpicture}[scale=0.7]
    \draw [black,thick,<->] (-0.50,5) node[left] {$M^2,G^2$} -- (-0.5,0)  node[below] {$0$} -- (10,0) node[below left] {$\theta$};
    \draw[red, very thick, dotted] (0.4,4.3) to [out=-70,in=170] (9.5, 0.1); 
    \draw[thick,<->] (0.8,3.5) to [out=-70,in=170] (3.5, 1.9); 
    \draw[thick,<->] (3.5, 1.5) to [out=-30,in=168] (6.75,0.5);
    \draw[thick,<->] (6.75,1.0) to [out=-45,in=175] (9,0.2);
     \draw[thick, gray,dotted] (0.4,0.) -- (0.4,4.5);
     \node[thick] at (0.4,5) {\makecell[c]{collimation \\ region}};
    \draw[black] (1.5,-0.2) -- (1.5,4.5);
    \node[thick] at (1.5,-0.4) {$\thef$};
    \draw[gray,dashed] (3.5,-0.2) -- (3.5,4.5);
     \draw[red, very thick,<->] (3.5,1.5) -- (3.5,1.9);
    \node[thick] at (3.5,-0.4) {$\themf$};
    \draw[black] (5.5,-0.2) -- (5.5,4.5);
    \node[thick] at (5.5,-0.4) {$\theta_{\rm A}$};
    \draw[gray,dashed] (6.75,-0.2) -- (6.75,4.5);
     \draw[red, very thick,<->] (6.75,0.5) -- (6.75,1);
    \node[thick] at (6.75,-0.4) {$\thems$};
    \draw[black] (8,-0.2) -- (8,4.5);
    \node[thick] at (8,-0.4) {$\thes$};
    \draw[thick, gray,dotted] (9.2,0.) -- (9.2,4.5);
    \node[thick] at (9.2,5) {\makecell[c]{disk \\ $(\theta = \pi/2)$}};
    \end{tikzpicture}
    \end{minipage}
    \hspace{1.4cm}
    \caption{Schematic of the method to derive the form of the functions $M^2(\theta)$ or $G^2(\theta)$. The solid black lines are the branches of integration with the corresponding direction shown with arrows. A typical situation where the given set of input parameters plus the guesses for the $\theta_{\rm MFP}$ and $\theta_{\rm MSP}$ and $x\ata^2$ and $q$ at these points do not provide a smooth solution through all the singular points and we see large offsets at the midpoints (red solid arrowed lines), while the closest \textit{good} solution looks like the red dotted line. Note that the last integration branches towards the disk and downstream of MFP are not used in the evaluation of the fitness, but calculated at later times.}
    \label{fig:sol}
\end{figure}

  We first \emph{guess the locations of the critical points,
    $\thef$ and $\thes$} and derive values for
  $M^2$, $G^2$ and their derivatives with respect to $\theta$ based on the
  regularity conditions $\numm = \nump = \den = 0$ evaluated at the
  $\theta_{\rm MFP/MSP}$ of choice (see Appendix \ref{sub:integ} for our numerical
  technique). We then are able to integrate \textit{away} from each initial guess for the
  modified magnetosonic points and avoid numerical inaccuracies.
At the same time, we integrate away from the AP towards both magnetosonic points
  and consider how good the match is for the values of $M^2$ and $G^2$ of the various solutions at the
  midpoints $\themf = (\thea + \thef)/2$ and $\thems = (\thea +
  \thes)/2$ (see Fig. \ref{fig:sol}). These are in total four
  conditions, which imply four free parameters. $\thef$ and $\thes$
  are two necessary ones and we are left with the freedom to choose
  two other parameters. We chose to leave $\xa$ and $q$ free and fix
  the others. This choice is the most natural and convenient one: $\xa$
  immediately determines $M^2\ata$ as per Eq. \ref{alf:m}, while knowing $q$ and the position of the AP allows our
  algorithm to derive $\left.d M^2 / d \theta\right|\ata$ very quickly.
  
\subsection{Initial conditions at the singular points}
\label{sub:initcond}

As mentioned above, the AP is completely determined. As for the two modified magnetosonic points, there is no analytical condition that can be used to regularize the equations. We use a combination of root-finding techniques that allow us to find the values of $M^2$ and $G^2$ that give $\numm=\nump=\den=0$. A similar procedure can be applied to both the MFP and the MSP. 
Once the values of $M^2$ and $G^2$ at the singular points are found, the last step before starting the integration is finding the derivative of $M^2$. By making use of Eq. \ref{sys:dg2dt} and of $M^2$ and $G^2$ at the points of interest, we determine $d M^2 / d \theta$ by finding the root of the function \ref{FM2_1}. 
The integration from all the three singular points can now start.
This step is very important, serving as a \emph{numerical regularity condition}, but it is somewhat laborious, so that for a more detailed discussion, we refer the interested reader to the Appendix \ref{sub:integ}. \\
At this stage we have both initial values \emph{and} derivatives for $M^2$ and $G^2$, evaluated at $\theta_{\rm MFP/MSP}$, and we can finally start the integration inwards towards the AP. Since we integrate \emph{away} from singular points, a standard adaptive step Runge-Kutta scheme is sufficient to integrate Eqs. \refp{sys:wind} and \refp{sys:dg2dt} (giving $M^2$ and $G^2$), while we retrieve $\psi$ from Eq. \refp{sys:psi} and $\xi$ from Eq. \refp{sys:xi}.
Thanks to this procedure it is also possible to integrate downstream from the MFP further away from the black hole and upstream from the MSP towards the equatorial plane of the disk: we simply repeat the same procedure \emph{on the other side} of the critical points.
    
\subsection{Goodness of solution and solution finding}
\label{sub:fitnes}

In order to find the best parameters ($\thes$, $\thef$, $q$, $\xa$) for a
  given problem set ($F$, $\Gamma$, $\sigma_{\rm M}$, $\masa$, $\thea$ $\psia$) which minimise the offsets at each midpoint simultaneously (see Fig. \ref{fig:sol}),
  we make use of the open-source Bayesian inference algorithm \mny{}
  \citep{Multinest:2008,Multinest:2009,Multinest:2013}. \mny{} is a
  very robust software package which is also fast due to MPI
  parallelisation. It also has the advantage of not requiring
  derivatives of the fitted function, like for example the
  Newton-Raphson method. This makes it even better suited for our case,
  because the derivatives can only be calculated numerically and such an approach would
  increase the numerical error and would prevent most attempts
  at finding a good fit. We only use the fitting algorithm of
  \mny{}. In this case, the only information that needs to be given to the algorithm is the \fitn{} of a given calculation. It
  then proceeds to maximise such a function by exploring the given
  domain of the free parameters\footnote{We set a flat prior for all
    parameters.}. 

To quantify the \fitn{} of a single integration, we measure the
  mismatch at the offsets between the values of $G^2$ and $M^2$ from the
  integration from AP and the modified magnetosonic points, summing
  the relative differences of all variables:
  \begin{equation}
    \fits = \left[f_{\rm G, F}^2 + f_{\rm M, F}^2 + f_{\rm G, S}^2 +
      f_{\rm M, S}^2\right]^{-1/2} \label{fitness2}
  \end{equation}
  where
  \begin{equation}
    f_{\rm G, *}  = \left.\frac{2(G_{L}^2 - G_{R}^2)}{(G_{L}^2 + G_{R}^2)}\right|_{\themg{*}},
    \quad
    f_{\rm M, *}  = \left.\frac{2(M_{L}^2 - M_{R}^2)}{(M_{L}^2 + M_{R}^2)}\right|_{\themg{*}}
  \end{equation}
  and the subscripts $L$ and $R$ stand for left and right, such that
  at $\themf$ $L$ is the result of the integration from the MFP and
  $R$ the results of the integration from AP; while the opposite holds
  at $\thems$.  In case one of the various preparatory steps described
  in Appendix \ref{sub:integ} fails, we are unable to calculate $\fits$
  and simply return zero, discarding the corresponding point in the parameter space. 
  %We instructed \mny{} to ignore such points
  %when calculating the ellipsoid in the parameter space. 
  We accept a
  solution and stop the iterations when we reach a fitness value\footnote{$\fits$ is the inverse RMS of all the fitting errors.  Double precision in a computer gives up to $\sim10^{-14} - 10^{-16}$ machine roundoff error, but roundoff errors can build up in any calculation. So, if the chosen $1/\fits$ is too small, convergence could suffer.  We found $10^{-9}$ to be a good compromise choice.} of
  $\fits \geq 10^9$. During the process of finding the solutions we noted that there is a
  highly non-linear relation between the amount of change of the
  different parameters $q$, $\xa$, $\thef$ and $\thes$ and the
  resulting change in the fitness function Eq. \eqref{fitness2}, such
  that each parameter should be known with at least six significant
  digits to make a good fit.

\begin{table*}
%\centering
 \caption{Parameter study of the solutions recovered from \capp{}. The classes of models \textit{I} and \textit{II} are the "reference" and "first" solutions in Tab.1 of \capp{}. Sub-classes of models differ for the adiabatic index $\Gamma$, the gravitational potential, $P_{\rm g}$ (Newtonian=N or \pw=PW), and whether the functions $f_i(P_{\rm g})$ are approximated (A) or corrected (C) (see Appendixes \ref{definitions} and \ref{derivpotential}). The fitted parameters are shown in italic and with 6 significant digits.}
 \label{tab:pars}
 %\begin{threeparttable}
 \def\arraystretch{1.5}%
 \setlength\tabcolsep{4.5 pt}
 \begin{tabular}{l|cccccccccccc}
%  \hline
%   & \multicolumn{4}{c|}{Fitted} & \multicolumn{6}{c}{Fixed} \\
%  \hline
  Model & $x\ata^2$ &  $q$ &  $\theta_{MFP}$ &   $\theta_{MSP}$ &   $\theta_{A}$ &   $\psi_{A}$ &    $\sigma_{M}$ &   $\varpi_{A}$ &  $\Gamma$ &  F & Style & $P_{g}$\\
  \hline 
   \hline
   Reference & \textit{0.145330}     & \textit{2.4184E-2}   &  \textit{0.118635}  &  \textit{1.26022}  &  60   & 40  &   0.02  & 15  &  5/3  &  0.75  &  A  & PW \\
   \hline
    \textit{Ia} & \textit{0.145329}     & \textit{2.41844E-2}   & \textit{0.118608}          & \textit{1.26383}  &  60   & 40  &   0.02  & 15  &  5/3  &  0.75  &  A  & PW \\  
    \textit{Ib} & \textit{0.158777}     & \textit{2.74016E-2}   & \textit{0.116215}	    & \textit{1.19411}  &  60   & 40  &   0.02   & 15  &  5/3  &  0.75 &   C & PW \\ 
    \textit{Ic} & \textit{0.143936}     & \textit{1.86592E-2}   & \textit{0.118492}	    & \textit{1.31034}  &  60   & 40  &   0.02   & 15  &  5/3  &  0.75 &   C & N \\ 
 \hline
    \textit{Id} & \textit{0.228007}     & \textit{1.19385E-3}   & \textit{0.104885}	    & \textit{1.16340}  &  60   & 40  &   0.02   & 15  &  4/3  &  0.75 &   A & PW \\
    \textit{Ie} & \textit{0.264180}     & \textit{7.43368E-4}   & \textit{0.104566}	        & \textit{1.15714}  &  60   & 40  &   0.02   & 15  &  4/3  &  0.75 &   C & PW \\  
    \textit{If} & \textit{0.225918}     & \textit{8.34464E-4}   & \textit{0.105647}	    & \textit{1.19239}  &  60   & 40  &   0.02   & 15  &  4/3  &  0.75 &   C & N \\
  \hline
   \hline
  First solution & 0.01     & \textit{1.4359E-2}   &   \textit{0.120427}   &  \textit{1.18682} &  60   & 45  &   \textit{7.85798E-4}  & 18.2088  &  5/3  &  0.75  &  A  & PW \\
  \hline
  \textit{IIa} & 0.01 		& \textit{1.43588E-2}	& \textit{0.114838}	& \textit{1.19846}  &  60   & 45  & \textit{7.85794E-4 }  & 18.2088 &   5/3 &   0.75  &  A & PW \\
  \textit{IIb} & 0.01 		& \textit{1.66386E-2}   & \textit{0.115231}	& \textit{1.12314}  &  60   & 45  & \textit{5.94572E-4}   & 18.2088 &   5/3 &   0.75  &  C & PW \\  
  \textit{IIc} & 0.01 		& \textit{1.25403E-2}   & \textit{0.119471}       & \textit{1.20381}  &  60   & 45  & \textit{7.86151E-4}   & 18.2088 &   5/3 &   0.75  &  C & N \\     
\hline
  Others & &  &   &  &   &   &   &   &  &  &  & \\
   \hline
  (1) & \textit{0.643674}   &    \textit{6.13069E-3}   &     \textit{8.06108E-2}   &   \textit{1.32117}  &   60 &  45 &  0.50  & 15 &  4/3 &  0.75 & C & PW \\ %middle 
  (2) & \textit{0.324834}   &    \textit{6.91188E-3}   &     \textit{8.37261E-2}   &   \textit{1.28111}   &  60 &   46 &  0.10  & 15 & 4/3  &  0.75  &  C & PW \\ %org
  (3) & \textit{0.258726}   &    \textit{4.90328E-3}   &    \textit{9.73984E-2}    &   \textit{1.22659}   &  60 &  44 &  0.05  &  15 & 4/3  &  0.75  &  C & PW  \\ %psiA 26
  (4) & \textit{0.475936}   &    \textit{7.87961E-3}   &    \textit{9.30176E-2}    &    \textit{1.27947}   & 60  &  44 &  0.20  &   15 & 4/3  & 0.75  &  C & PW  \\ %psiA 29
  (5) & \textit{0.816325}   &    \textit{1.70893E-5}   &    \textit{7.02122E-2}    &   \textit{1.27569}  &  60 & 44  &  0.75  &   15 & 4/3  &  0.75   &  C & PW\\ %psiA 40
  (6) & \textit{0.864334}   &    \textit{ 6.86982E-6}   &   \textit{6.27254E-2}    &   \textit{1.33771}   &  60 & 47 &  1.45   &  15 & 4/3  &  0.75  &  C & PW \\ %psiA 110
  (7) & \textit{0.503771}   &    \textit{ 2.85668}        &    \textit{2.58765E-2}    &   \textit{1.38088}   &  57.5 & 47 &  1.45  &  15 & 4/3  &  0.85 & C & PW  \\ %wall 27
  (8) & \textit{0.470045}   &    \textit{ 2.53731E-5}        &    \textit{0.10165}    &   \textit{1.15853}   &  60 & 39 &  0.05  &  15 & 4/3  &  0.75 & C & PW  \\ %psiA 1 
\hline
 \end{tabular}
% \begin{tablenotes}
%    \item[\textit{f}] Fitted parameter.
%   \item[\textit{*}] Values taken from Tab.1 of \capp{}. \textbf{NP = not published. Peter!?}
%  \end{tablenotes}
%\end{threeparttable}
\end{table*}

\section{Comparison with previous solutions}
\label{comparison}

We start our parameter study by recovering the solutions presented in Table 1 of \capp{}. We list our parameters corresponding to \capp{}'s \emph{reference} and \emph{first} solutions in Tab.~\ref{tab:pars}, as models \textit{I} and \textit{II}. We identify two factors that can explain the discrepancies in the parameters: 1) differences in the numerical scheme and  2) definition of the functions of the gravitational potential in the gravity terms (Eq.~\ref{fPgs} and Appendix \ref{derivpotential}). 
Comparing then the parameters published in \capp{} (first solution) with model \textit{IIa}, both listed in Tab.\ref{tab:pars}, we noticed that whilst there is not an appreciable difference in $q$ and $\sigma_{\rm M}$, which are the fitted parameters for \capp{}, the locations of both MFP and MSP change by a few $\%$. 
% However, if we then compare the height of the MFP given in \capp{} ($4298~r_{\rm g}$) with the one we find ($4500~r_{\rm g}$), we see that the relative error is effectively larger ($\Delta z/z \gtrsim 4 \%$). This could be important when fitting to very high-resolution observations of nearby sources, such as M87. 

\begin{figure*}
\centering
    \includegraphics[width=0.5\textwidth]{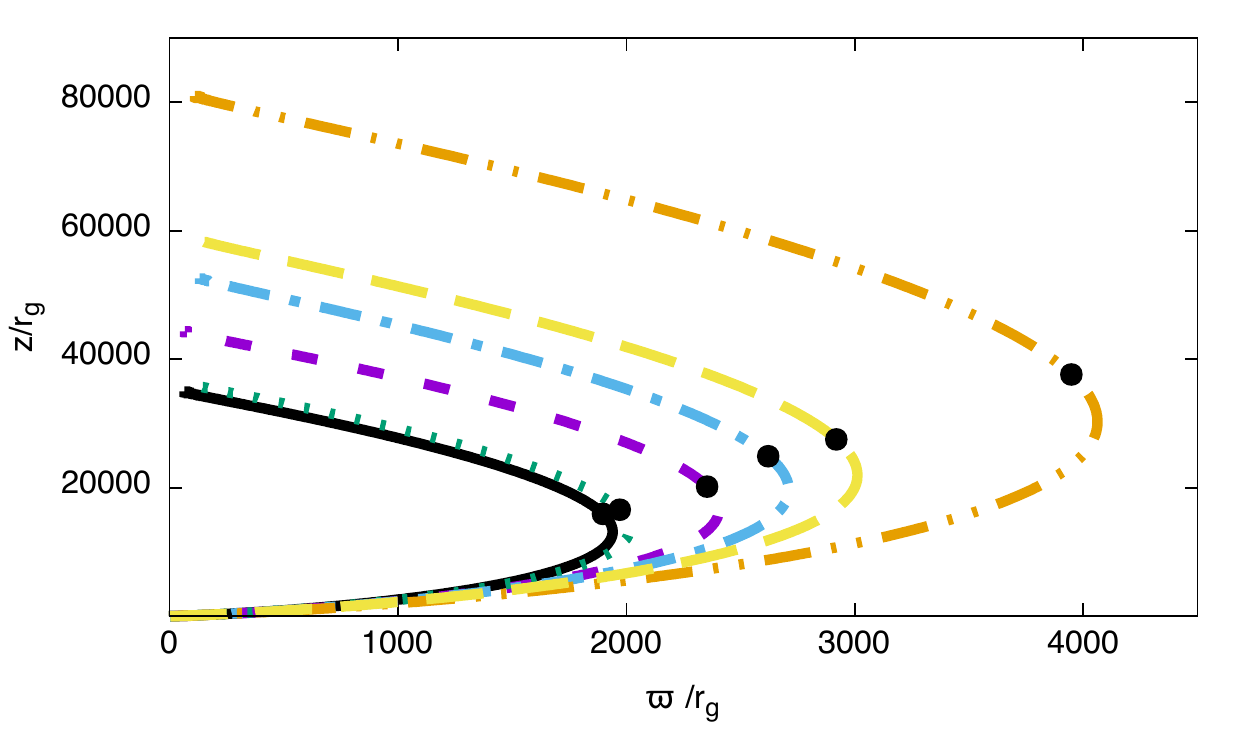}\hfil
    \includegraphics[width=0.5\textwidth]{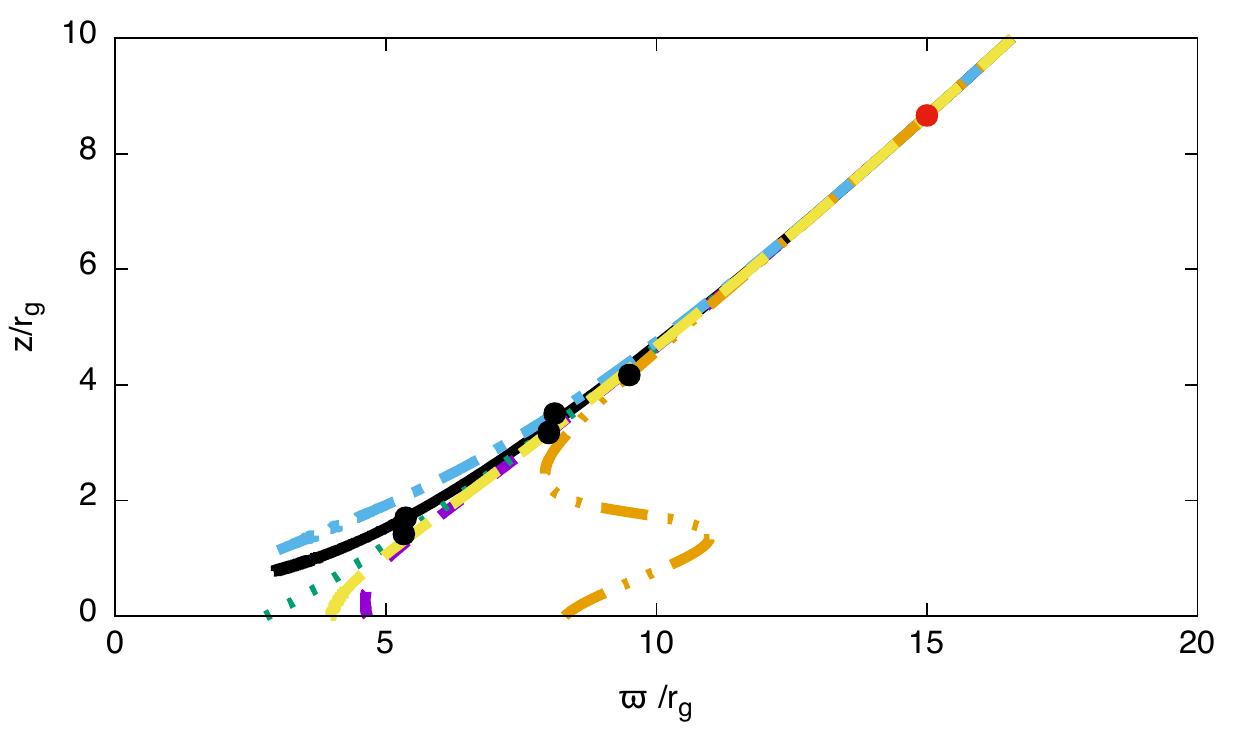}
\caption{Comparison of stream lines of models \textit{Ia} to \textit{If} from Tab.~\ref{tab:pars}. Colours correspond to same model in both panels: black solid line is model \textit{Ia}, purple short-dashed line is model \textit{Ib}, dotted-green line is model \textit{Ic}, dot-dashed blue line is model \textit{Id}, orange dot-dot-dashed line is model \textit{Ie}, and yellow long-dashed line is model \textit{If}. The models \textit{Ia}, \textit{Ib}, and \textit{Ic} have adiabatic index of $5/3$ and differ for the potential, $P_{\rm g}$,  used (N or PW) and whether the functions $f_i(P_{\rm g})$ are approximated (A) or corrected (C). The models \textit{Id}, \textit{Ie}, and \textit{If} are the same but with $\Gamma=4/3$. \textit{Left:} MFP region, where the black dots mark the height of the MFP for the corresponding streamline. \textit{Right:} Zoom on the AP and MSP. The AP (red dot) is fixed in all the solutions, therefore the streamlines converge to the same point with the same derivative. The black dots define the location of the MSP of each streamlines.}
\label{fig:polko}
\end{figure*}

As discussed in Section \ref{equations}, if we remove the approximation of a small potential in the gravity terms, we find substantial differences in our solution parameters (see models \textit{Ia}-\textit{Ib}, \textit{Id}-\textit{Ie} and \textit{IIa}-\textit{IIb}), that are clearly seen when drawing the corresponding streamlines: in Fig.~\ref{fig:polko} we plot the projected streamlines in the $z\varpi$-plane for the models \textit{Ia} to \textit{If}. Model \textit{Ia} is our reference solution, which differs from \capp{} only for the numerical method used to solve the equations \refp{sys:wind}-\refp{sys:xi} (black solid line). Model \textit{Ib} is the reference solution with the corrected potential functions (purple dashed line) and model \textit{Ic} is the reference solution with the corrected potential functions, but with Newton potential (green dotted line). It is evident that introducing the approximation of small $P_{\rm g}$ in the derivatives of $P\gr$ itself (see discussion in Sec. \ref{equations} and Appendix \ref{derivpotential}) reduces the effect of gravity close to the BH (right panel in Fig.~\ref{fig:polko}). Larger discrepancies at MSP appear when we consider a relativistic gas with adiabatic index $4/3$ (see models \textit{Id} (blue dot-dashed line), \textit{Ie} (orange dot-dot-dashed line) and \textit{If} (yellow long-dashed line)). Generally, the effect of a small \pw\, potential is very similar to a Newtonian potential.  
We define the last recollimation point (LRP) as the last point of the streamline where the integration downstream of MFP stops.
In the left panel of Fig.~\ref{fig:polko}, we note that the positions of the MFP and the LRP lie far apart from each other when comparing solutions with corrected or approximated functions of the gravitational potential. 

\begin{figure*}
\centering
   \subfloat[]{ \includegraphics[width=0.47\linewidth]{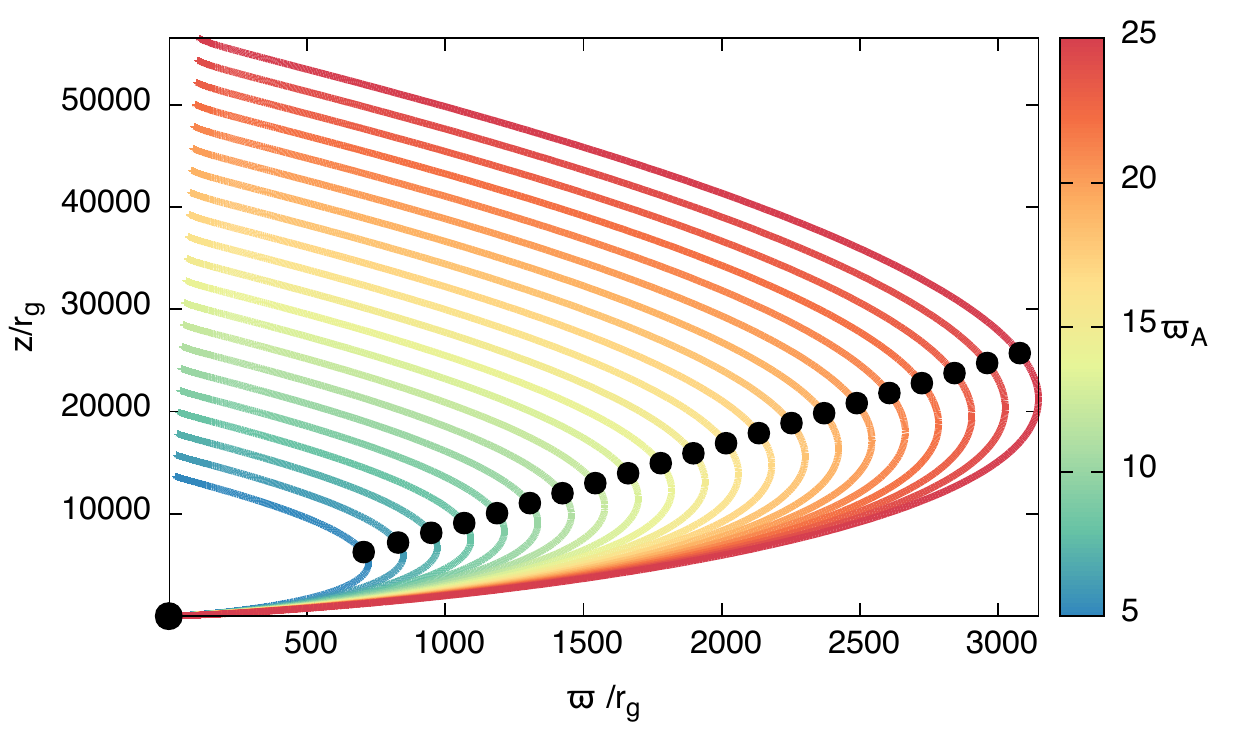}}\hfil
    \subfloat[]{\includegraphics[width=0.47\linewidth]{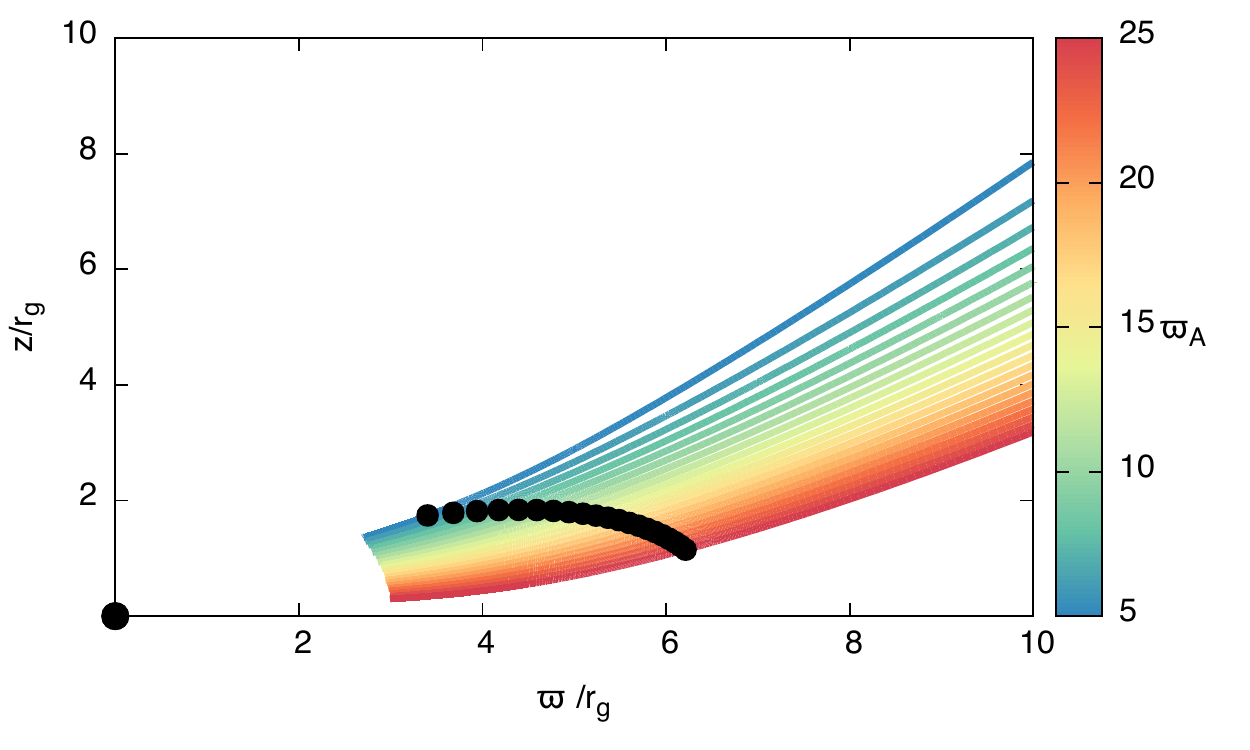}}\par\medskip
    \subfloat[]{\includegraphics[width=0.47\linewidth]{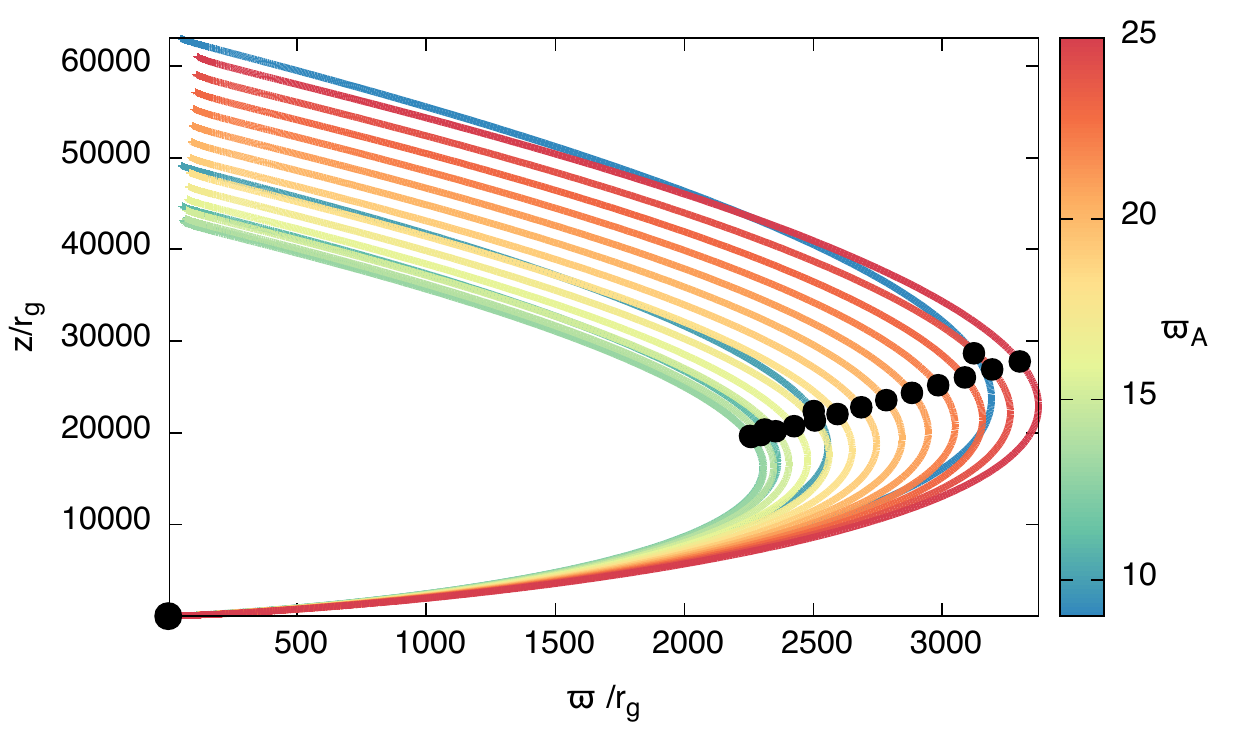}}\hfil
    \subfloat[]{\includegraphics[width=0.47\linewidth]{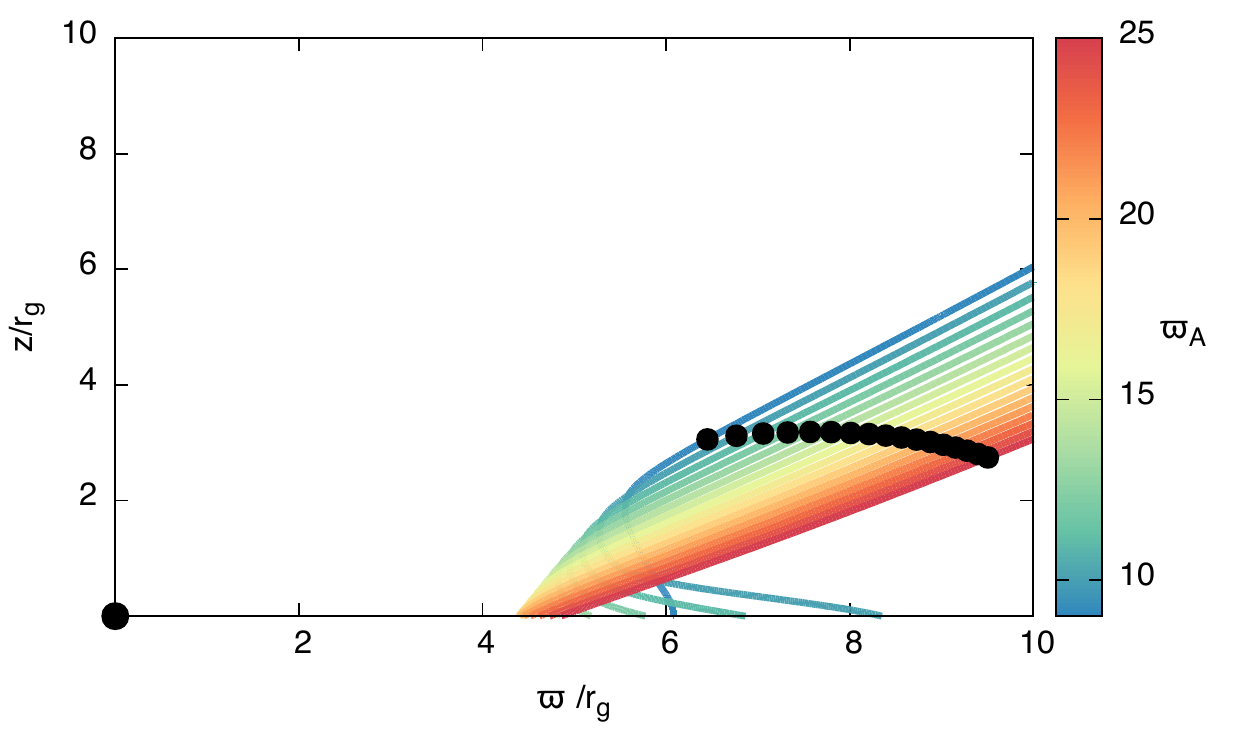}}
\caption{Family of solutions obtained by varying the cylindrical radius of the \alf point, $\varpi_A$. The reference solution ($\varpi_A=15$) for the upper panels is model \textit{Ia} and for the bottom panels is \textit{Ib} in Tab.~\ref{tab:pars}. \textit{Upper panels:} We vary $\varpi_A$ in the interval $5-25$, as the lines go from blue to red. On the left, the black dots on the curves mark the position of the MFP,  while on the right they are showing the MSP. The bigger black dot at the origin of the axis is the BH. \textit{Lower panels:} Same but the interval is restricted to $9-25$.}
\label{fig:wa}
\end{figure*}

\subsection{The self-similarity assumption}
\label{selfsim}

As pointed out by \capp{}, the self-similarity assumption is a serious limitation intrinsic in the derivation of the equations. The inclusion of gravity further complicates the matter.  Radial self-similarity, even without the inclusion of gravity, introduces a few geometrical constraints.
Ultimately, to properly quantify these issues, our solutions will be benchmarked against GRMHD simulations in future works. In this section, we focus on determining the effect of gravity on the self-similarity assumption. In Fig.~\ref{fig:wa} we show two families of solutions obtained varying $\varpi\ata$ from the reference models \textit{Ia} using eq. \ref{peterfPgs} and \textit{Ib} (Tab.~\ref{tab:pars}) using eq. \ref{fPgs}. 
All the solutions found with the the corrected functions of the gravitational potential (eq. \ref{fPgs}) can be integrated down to the disk midplane ($z=0$). Each field line will have a characteristic angular velocity, $\Omega(\varpi,z)$, which is a constant of motion along the field line, but it will vary from one field line to another (see Appendix \ref{conversion}). As expected from relativistic self-similar models, e.g. \citet{Li:1992} and \cavt{}{}, $\Omega(\varpi,z=0)$ follows a profile which is $\propto R^{-1}$. However the gravity terms add some perturbation in the proximity of the black hole. The maximum deviation between the angular velocity of a single field line at $z=0$ compared to $\Omega_{\rm K} = (R/R_{\rm g})^{-3/2}$ ($G=c=M=1$) is a factor of 2, with our field lines rotating at lower speed than the keplerian one.
All streamlines should look as scaled copies of the same shape; however, where gravity is strong enough, for example upstream of the MSP, the field lines bend and cross each others until $\varpi_A$ is larger than $\sim19~r_{\rm g}$ (panel {(\emph d)} in Fig.~\ref{fig:wa}). The MFP is heavily affected as well for smaller values of $\varpi\ata$ ($\lesssim19$). It is worth noting that for the family of solutions with model \textit{Ib} as reference, we could not obtain solutions for $\varpi_A<9$ due to the stronger gravitational potential, while for solutions found around model \textit{Ia}, we were able to go down to $\varpi_A=5$. 

\begin{figure*}
\centering
   \subfloat[]{ \includegraphics[width=0.47\linewidth]{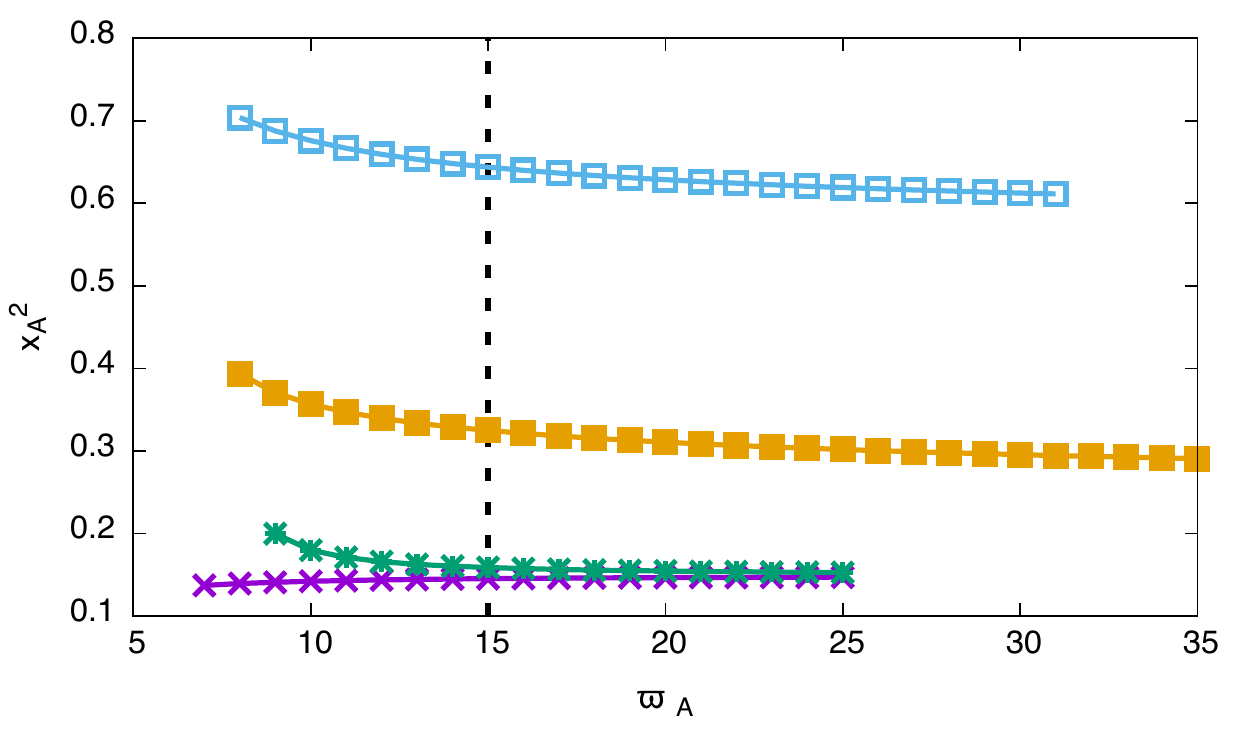}}\hfil
   \subfloat[]{ \includegraphics[width=0.47\linewidth]{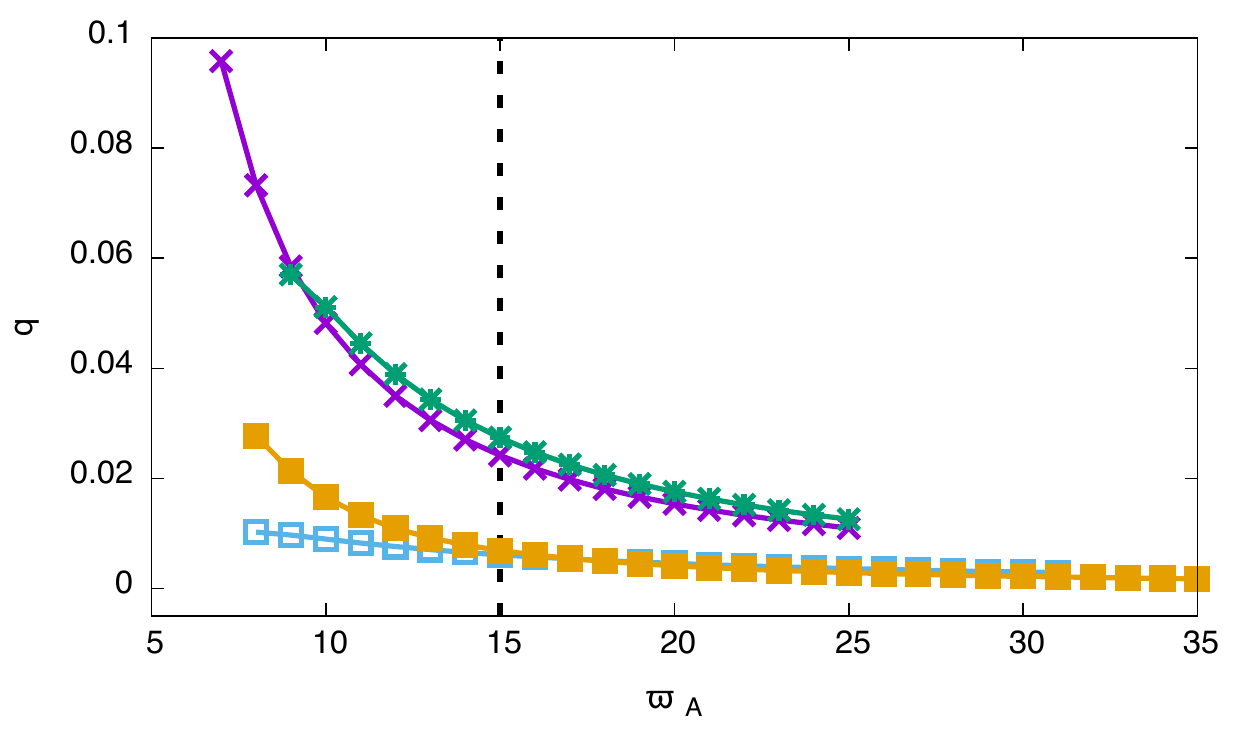}}\par\medskip
    \subfloat[]{\includegraphics[width=0.47\linewidth]{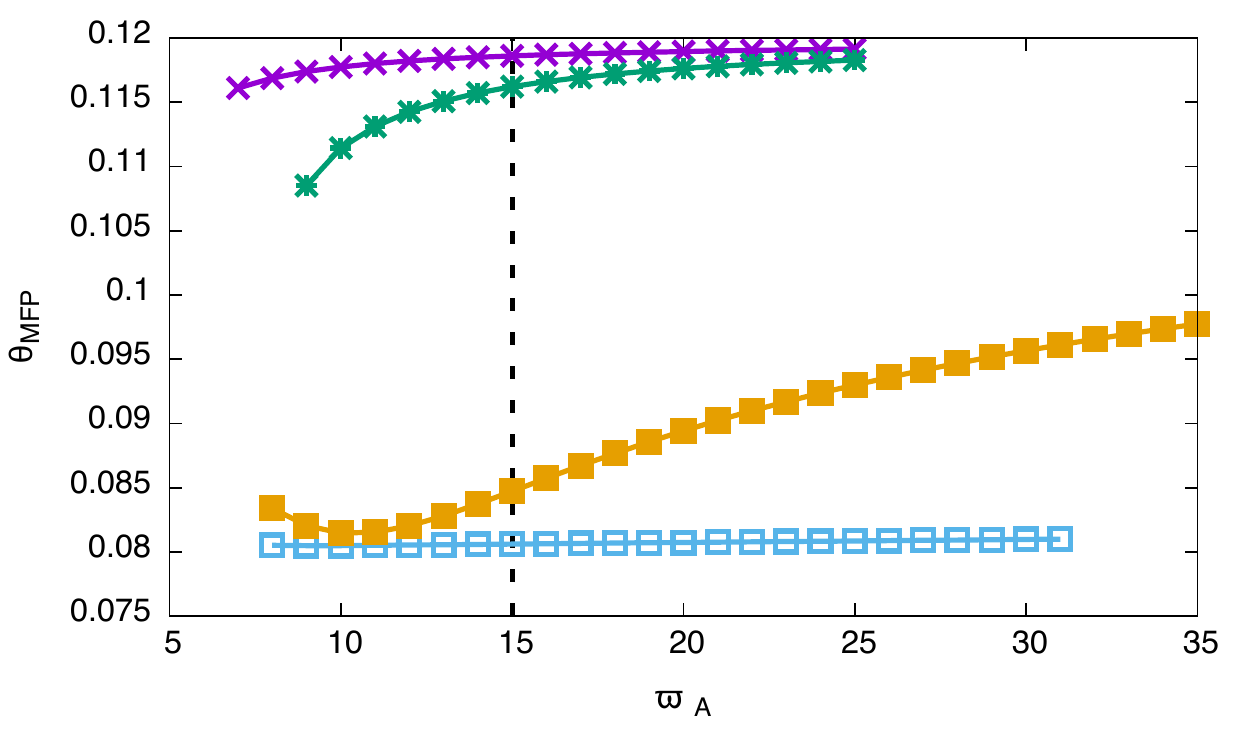}}\hfil
    \subfloat[]{\includegraphics[width=0.47\linewidth]{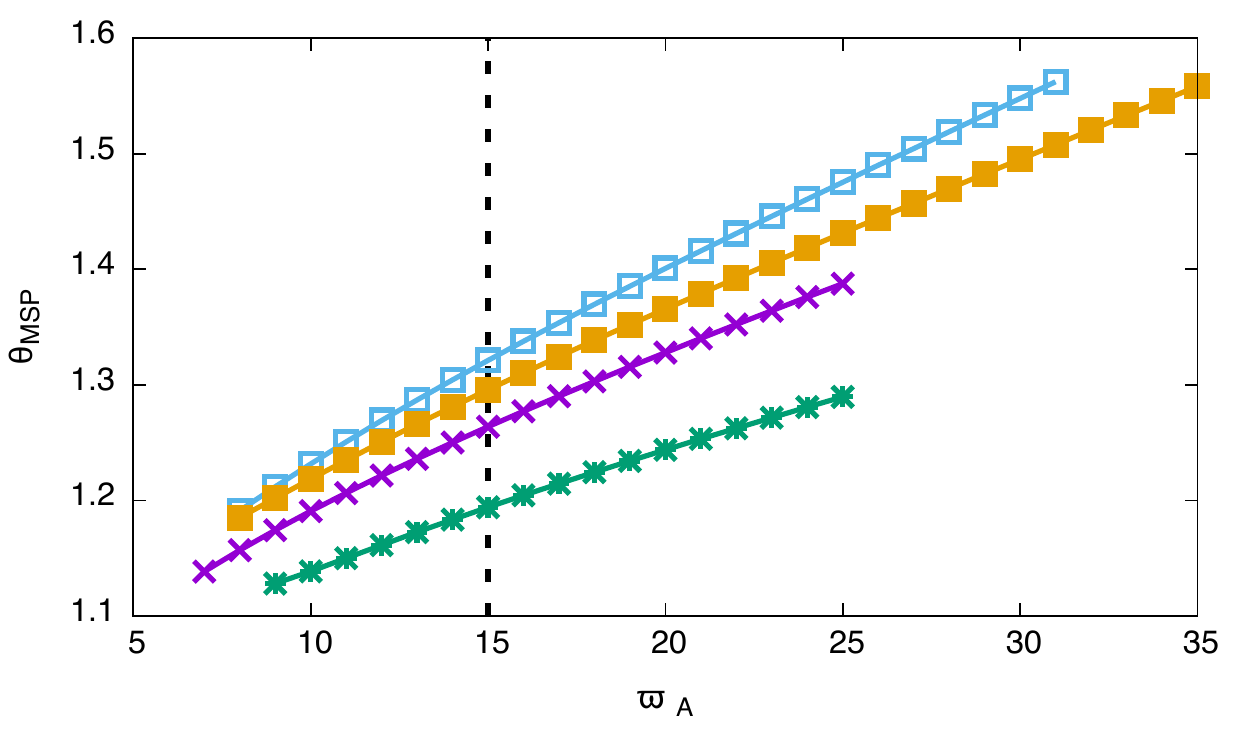}}
\caption{Fitted parameters ($x\ata^2,q,\theta_{\rm MFP},\theta_{\rm MSP}$) of the same families of solutions as function of the \alf cylindrical radius, $\varpi_A$. The purple line with crosses and the green line with stars represent the families of solutions corresponding to the reference parameters of the model \textit{Ia} and \textit{Ib} in  Tab.~\ref{tab:pars}, respectively. The blue line with hollow squares and the orange line with filled squares are derived from the reference solutions given in Tab.~\ref{tab:pars} as model (1) and (2) respectively. The vertical dashed lines mark the reference $\varpi_A$.}
\label{fig:multipar_wA}
\end{figure*}

The degree in which self-similarity is affected by gravity depends on the other parameters, such as $\sigma_{\rm M}$ and $\psi\ata$: we show this in Fig.~\ref{fig:multipar_wA}. The four panels in the figure show the values of the fitted parameters of various families of  solutions, as a function of $\varpi\ata$. In the absence of gravity, self-similarity would ensure that $\xa$ , $q$, $\theta_{\rm MFP}$ and $\theta_{\rm MSP}$ be constant within the same family of solutions, independent of any variation in  the value of the cylindrical radius $\varpi\ata$. In Fig.~\ref{fig:multipar_wA} we show families that correspond to four models of Tab.\ref{tab:pars}:  \textit{Ia} (purple), \textit{Ib} (green), \textit{1} (light blue), \textit{2} (orange).

Solutions corresponding to model \textit{Ia} and \textit{Ib} have lower $\psi\ata$ ($=40\degree$) and lower $\sigma_{\rm M}$ ($=0.02$) with respect to model (1) and (2) which have $\psi\ata=45\degree,~ \sigma_{\rm M}=0.5$ and $\psi\ata=46\degree, ~\sigma_{\rm M}=0.1$, respectively. When we introduce gravity, we are introducing a disturbance in the radial scaling of the streamlines, that becomes more pronounced as we move closer to the BH, i.e. $\varpi_A\lesssim 15$. Gravity breaks self-similarity for all the parameters at play, but its effect is more dramatic on $\theta_{\rm MSP}$.  Nonetheless, the median of the absolute deviations from the results' median (\textit{mad}=median$|X_i-{\rm median}(X)|$) for the parameters $x\ata^2,~q,~\theta_{\rm MFP}, ~\theta_{\rm MSP}$ over the full interval of $\varpi_A$ is at most $\sim10\%$. This gives us the confidence in the use of this method. 

\section{Parameter space study}
\label{paramstudy}

\begin{figure*}
\centering
    \includegraphics[width=0.47\linewidth]{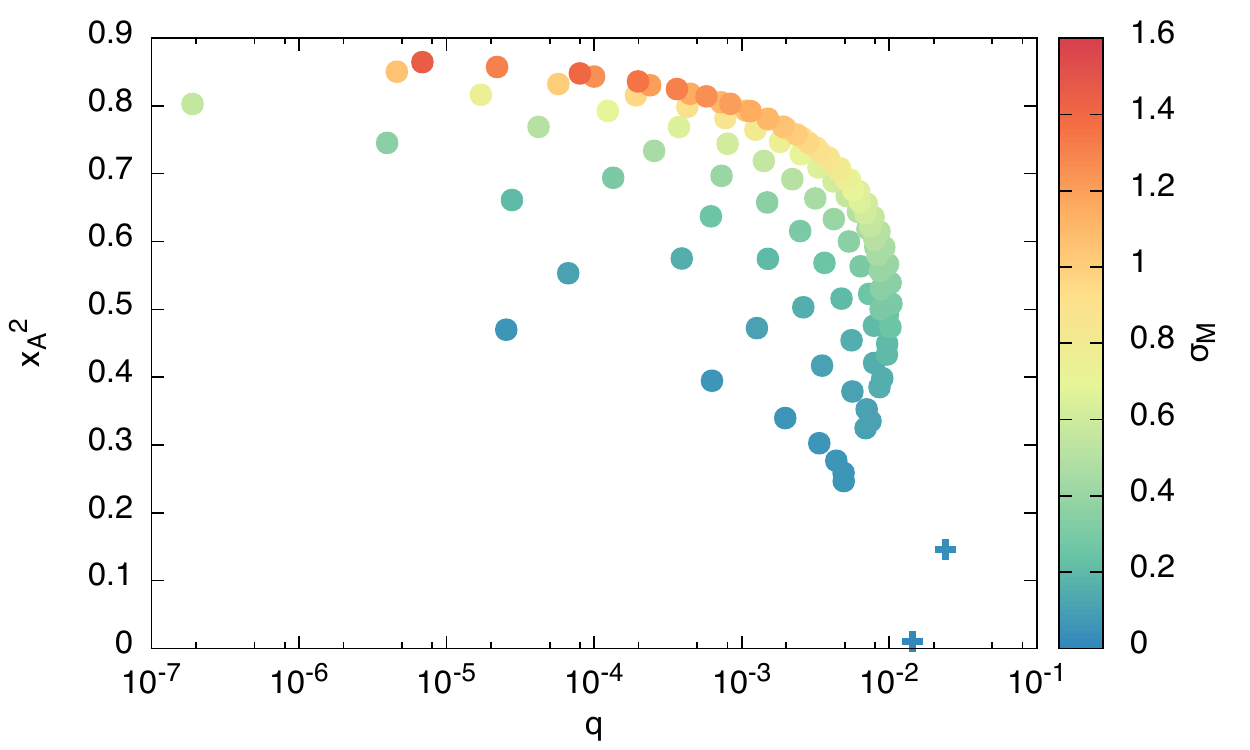}\hfil
    %\par\medskip
%     \includegraphics[width=0.47\linewidth]{lowq-msp}\hfil
      \includegraphics[width=0.47\linewidth]{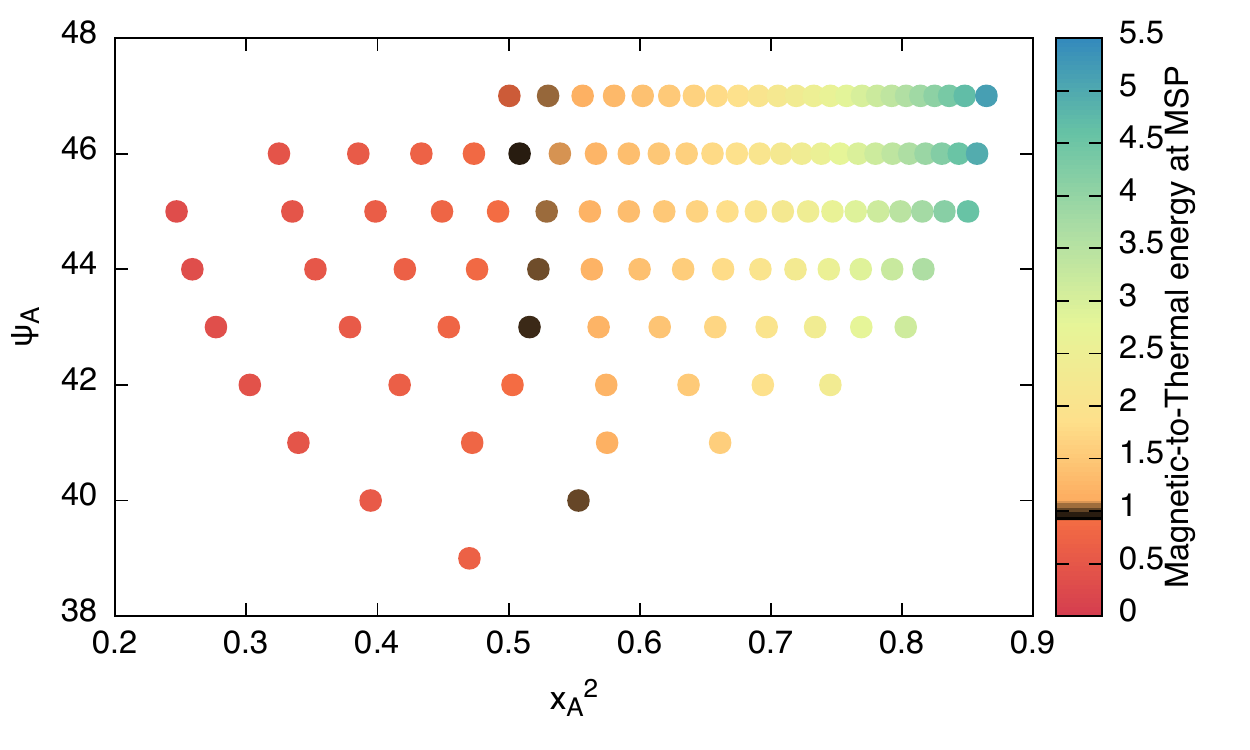}
\caption{Two-dimensional plots ($x\ata^2$ versus $q$ in the left panel and $\psi\ata$ versus $x\ata^2$) of the parameter space we explored by making discrete steps in the parameters $\psi\ata$ and $\sigma_{\rm M}$, while fitting for $x\ata^2,q,\theta_{\rm MFP}$ and $\theta_{\rm MSP}$. The third dimension ($\sigma_{\rm M}$ in the left panel and $S/\xi\gamma$ with $S = - \varpi \Omega B_\phi/ \Psi c^2$ at the MSP in the right panel) is expressed with the colour scheme on the right of each panel. The two crosses in the left panel are the first and reference solutions of \capp{} with corrected gravity terms. In the right panel, we show which solutions of the grid are magnetically or thermally dominated at the base. The black dots are the solution that have roughly equipartition between magnetic and thermal energy at the base. On the left of these all solutions are thermally powered jets at the base, while on the right there are solutions that are Poynting dominated at the base..}
\label{fig:parspace}
\end{figure*}

We present here a more detailed analysis of an ensemble of solutions found in a $(\psi\ata,\sigma_{\rm M})$ slice of the parameter space, for a specific choice of a subset of fixed parameters, namely  $F=0.75, ~\theta\ata=60\degree$, $\Gamma=4/3$ and $\varpi\ata=15$.
Although we fixed it for all runs, $\varpi\ata$ has been changed when testing self-similarity for models \textit{Ia}, \textit{Ib}, (1) and (2) of Tab.~\ref{tab:pars}. 
Changing $\varpi\ata$ changes the radius of the stream line, allowing one to check self-similarity of a given outflow with the parameters $F, \theta\ata$, $\sigma_{\rm M}$ and $\Gamma$.
We populate a two-dimensional grid of solutions by varying $\psi\ata$ and $\sigma_{\rm M}$, while fitting for  $x\ata^2,~q,~\theta_{\rm MFP}$ and $\theta_{\rm MSP}$ and keeping all the other parameters fixed. In Tab.~\ref{tab:pars}, models (3)-(6) belong to this grid. We use them to present features also found in other solutions in the grid. Models (7) and (8) are solutions found in other areas of the parameter space and we include them into the discussion to show their peculiar characteristics.
Our primary goal in such exploration is to expand the pool of solutions found by \capp{}{} and search for a wider range of relativistically boosted jet solutions, i.e. with bulk Lorentz factors in the range of $2-10$, in order to find suitable solutions for future applications to astrophysical objects. Considering the difficulties of starting from a random initial position, we started by changing the polytropic index to $4/3$ and keeping the other fixed parameters as in the reference solution in \capp{} (first row in Tab.~\ref{tab:pars}), then slowly increasing $\sigma_{\rm M}$ and $\psi\ata$ until it was possible to find solutions. 
As we can see from Fig.~\ref{fig:parspace}, the parameter space is not a continuous volume and has patches where no solution exists. Due to the high dimensionality of this space, it is often difficult to know \textit{a priori} where solutions can be found. We will focus on the nature of these boundaries later on in this section.

The solutions found in the $(\psi\ata,\sigma_{\rm M})$ slice presented here, lie on a three-dimensional curve which extends on a limited range of the interested parameters.  
Although solutions with $\sigma_{\rm M} < 0.05$ could be recovered, we dedicated little time to the exploration of this class of solutions since they appear to have very slow bulk velocities, $\gamma_{\rm MFP}\gtrsim 1$. 

Just from the solutions in this sparse 2-dimensional grid, we see already a variety of flow shapes and dynamics. For instance, we show in Fig.~\ref{fig:energy} six examples of the dynamical evolution of the energy terms (Eq.~\ref{berneq}) along the jet streamlines corresponding to the solutions (3) to (8) in Table \ref{tab:pars}.  The panels (\emph{a})-(\emph{c}) differ in the value of $\sigma_{\rm M}$, and indeed the Poynting energy at the base increases going from (\emph{a}) to (\emph{c}), while the enthalpy and kinetic energy remain unchanged. At the launching site, the jet can be powered by a different source of energy depending on the parameters, being first thermally-dominated ($0.05 \geq \sigma_{\rm M} < 0.2$, panel (\emph{a})), crossing equipartition between magnetic and thermal energy at $\sigma_{\rm M} = 0.2$ (panel (\emph{b})) and later becoming magnetically-dominated ($0.2 <\sigma_{\rm M} \leq 0.75$, panel (\emph{c})).
However, we note that all these solutions have very little thermal energy overall, so they are still relatively "cold" jets.
When $\sigma_{\rm M}$ increases even further (panel (\emph{d})), the Poynting energy completely overtakes all the other energy contributions until it converts entirely into kinetic energy at $\sim 1000 ~ r_{\rm g}$. 
A common feature of all our jet solutions is the conversion of the primary source of energy (enthalpy or Poynting energy) at the base into kinetic energy at some distance from the BH between the \alf point and the MFP, as it is to be expected for relativistic flows \citep[][]{Komissarov:2010}. The jet is always \emph{kinetically-dominated} by the time it approaches the MFP.
There is another channel for the energy exchange, however, which we discuss below. 

\subsection{Counter-rotation in hot jets}
\label{counterrot}

\begin{figure*}
\centering
    \subfloat[]{\includegraphics[width=0.47\linewidth]{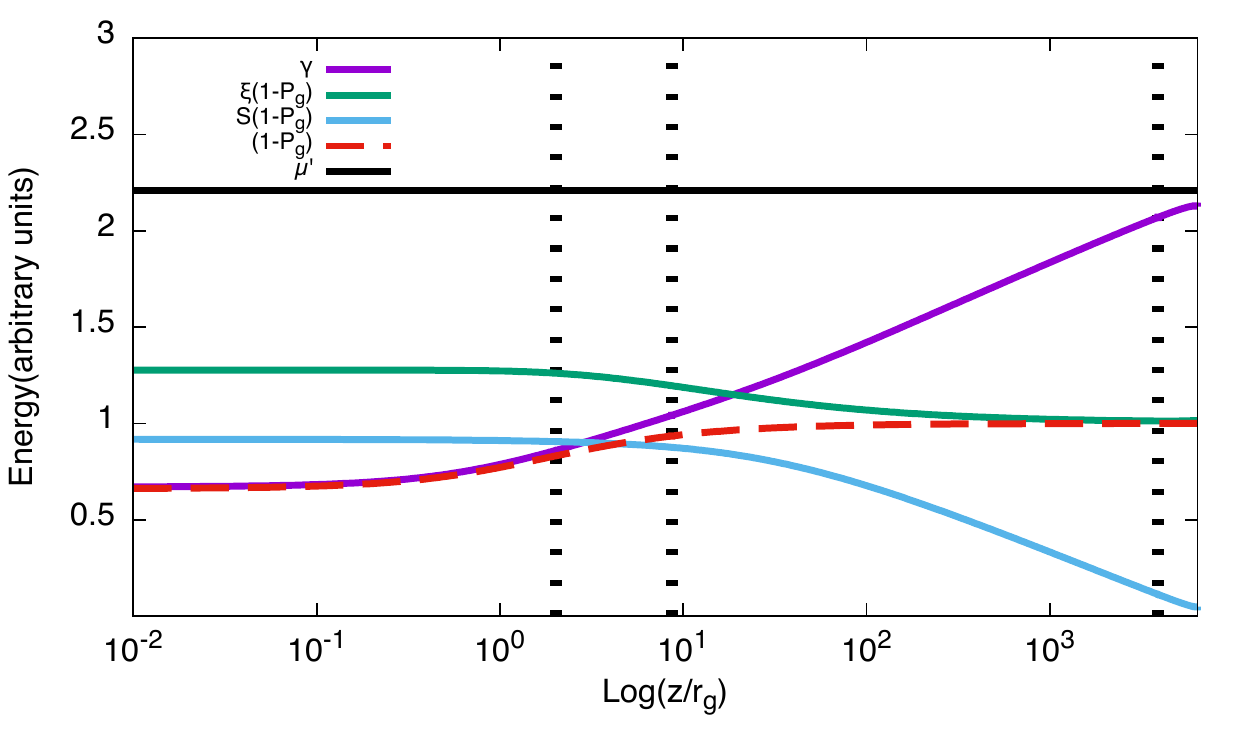}}\hfil
    \subfloat[]{\includegraphics[width=0.47\linewidth]{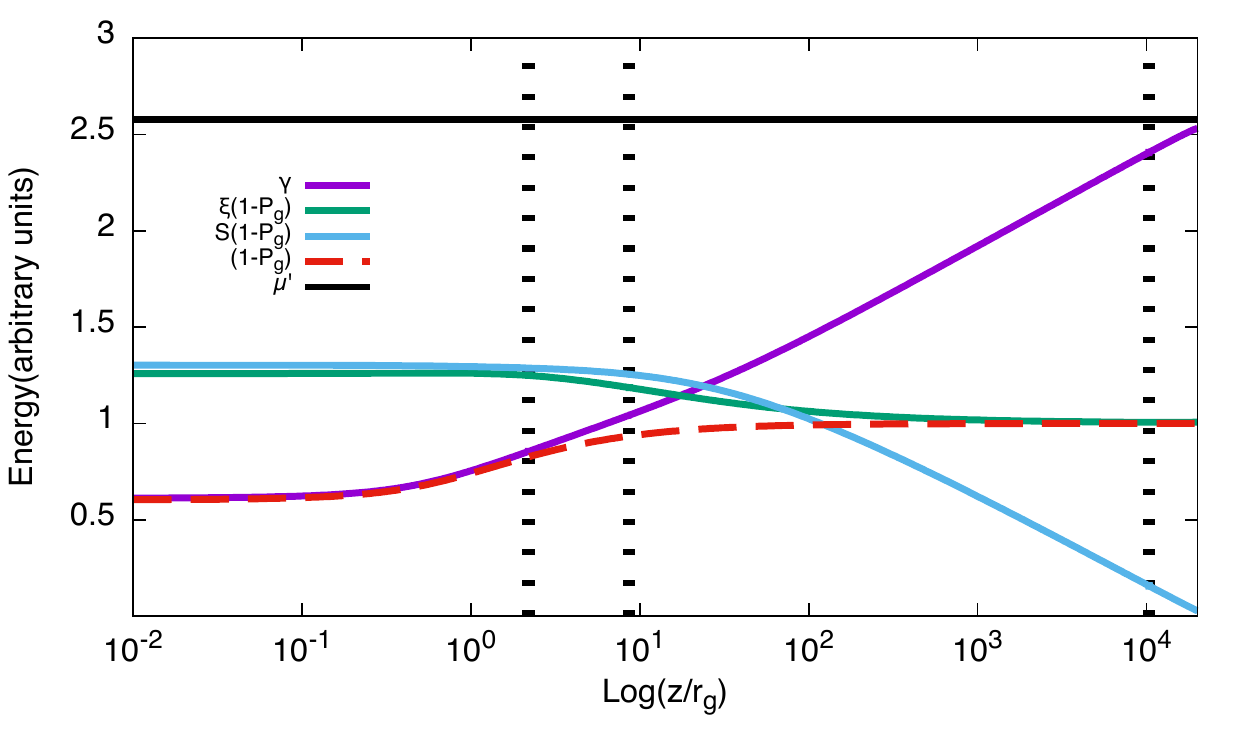}}\par\medskip
    \subfloat[]{ \includegraphics[width=0.47\linewidth]{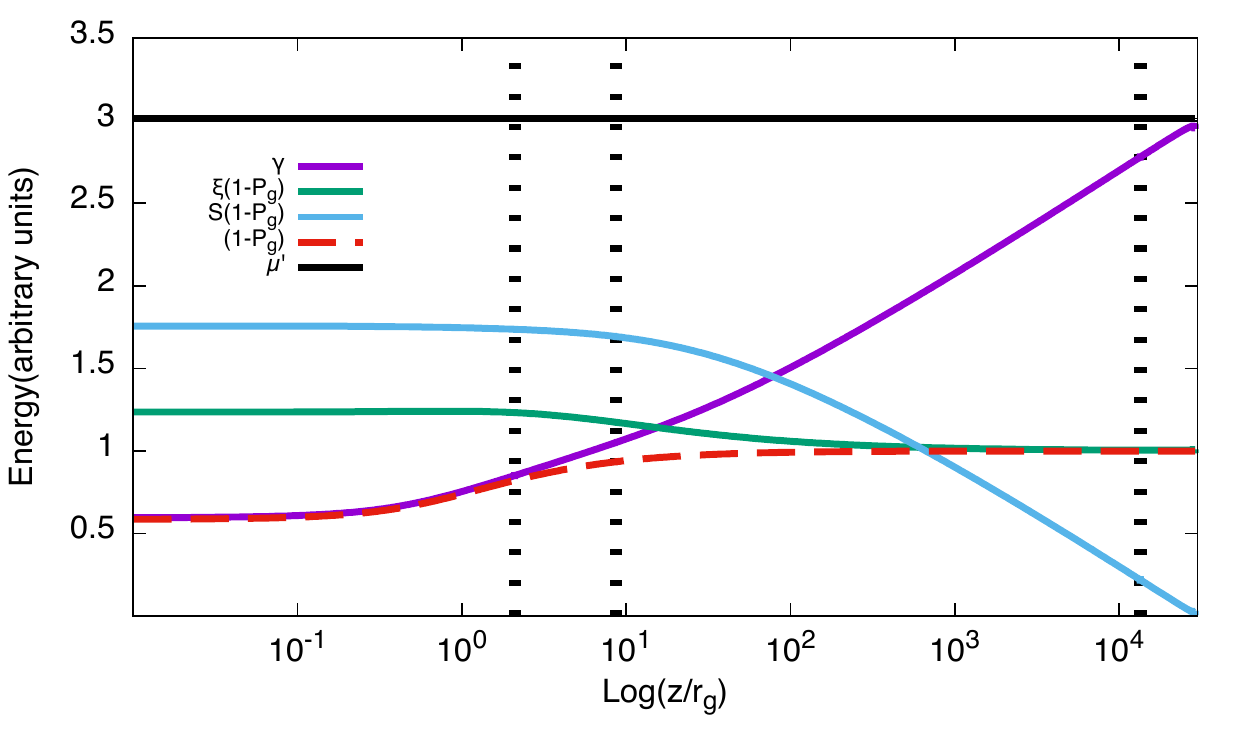}}\hfil
     \subfloat[]{ \includegraphics[width=0.47\linewidth]{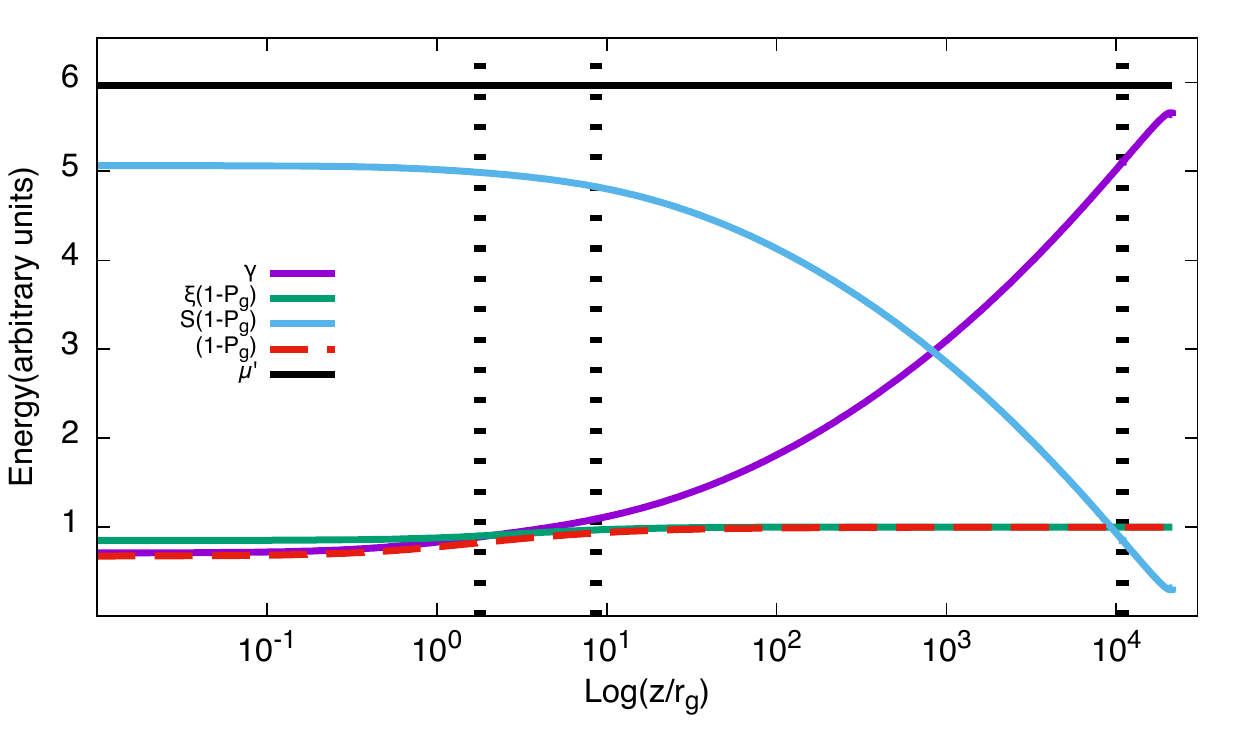}}\par\medskip
     \subfloat[]{ \includegraphics[width=0.47\linewidth]{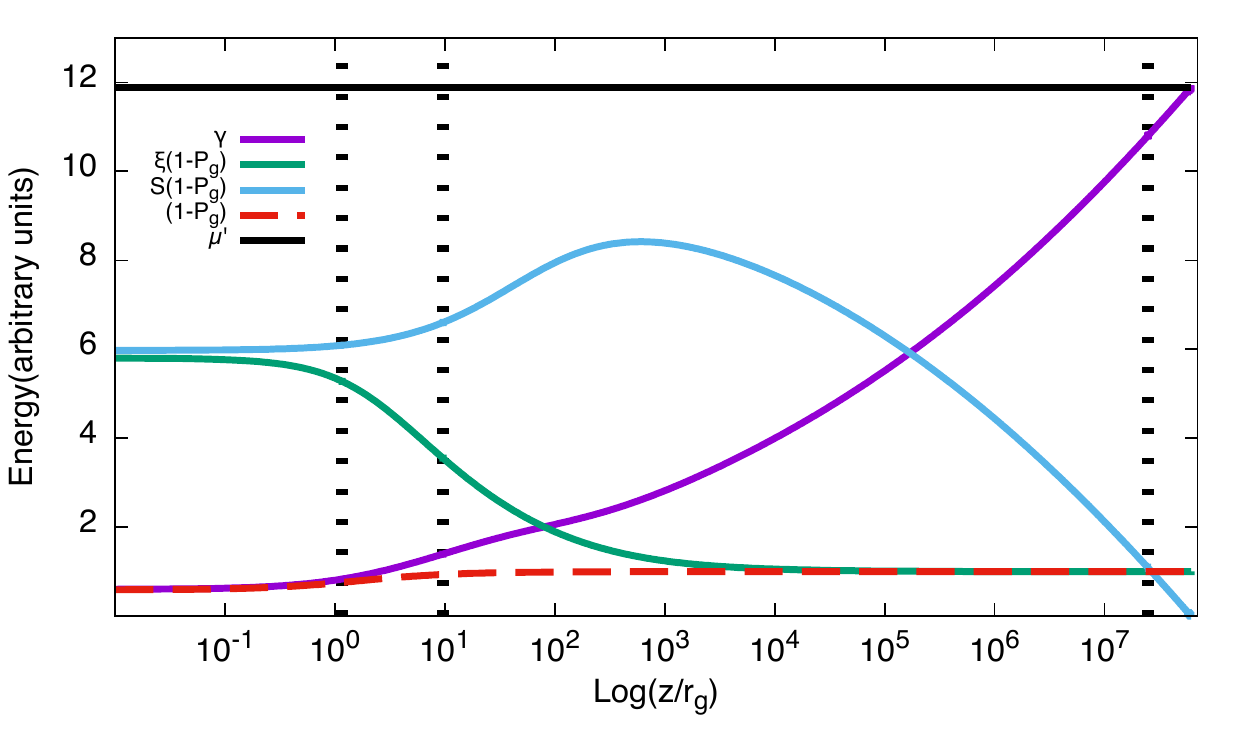}}\hfil
      \subfloat[]{\includegraphics[width=0.47\linewidth]{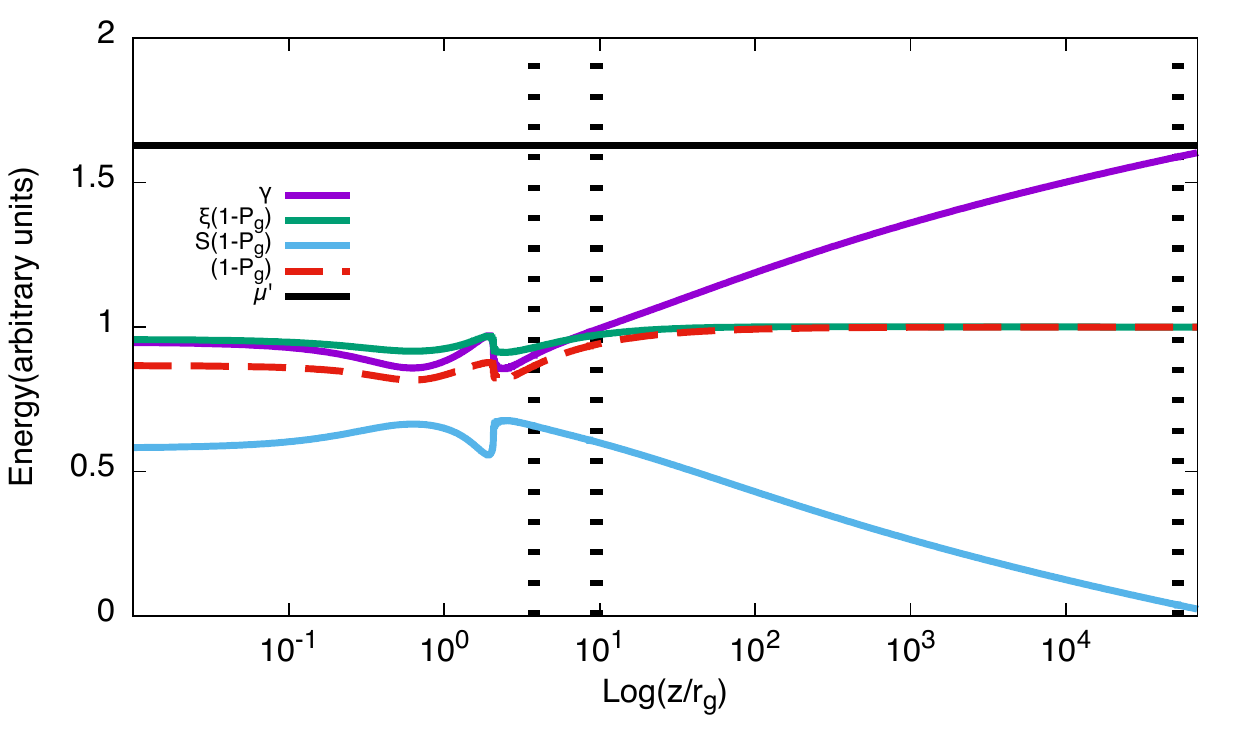}}
\caption{Evolution of the energy components along the jet streamlines. The black dotted lines indicate the position of the three singular points (from left to right, MSP, AP and MFP). The black solid line is the total energy, $\mu'$, the green line is the specific enthalpy, $\xi \gamma$, the light blue line is the energy carried by the magnetic field $-\varpi \Omega B_\phi/\Psi$, the purple line is the kinetic energy $\gamma$ and the dashed red line is the function $(1-P_{\rm g})$ of the gravitational potential. As described in Section \ref{equations} (Eq.~\ref{berneq}), each component is multiplied by this function to account for gravity in the energy balance. 
Panel (\emph{a}) shows the energy balance for model (3) with $\sigma_{\rm M}=0.05$, $F=0.75$, panel (\emph{b}) is model (4) with $\sigma_{\rm M}=0.20$, $F=0.75$, panel (\emph{c}) is model (5) with $\sigma_{\rm M}=0.75$, $F=0.75$, panel (\emph{d}) is model (6) with $\sigma_{\rm M}=1.45$, $F=0.75$, panel (\emph{e}) is model (7) with $\sigma_{\rm M}=1.45$, $F=0.85$ and panel (\emph{f}) is model (8) with $\sigma_{\rm M}=0.05$, $F=0.75$.}
\label{fig:energy}
\end{figure*}

\begin{figure*}
\centering
    \includegraphics[width=0.47\linewidth]{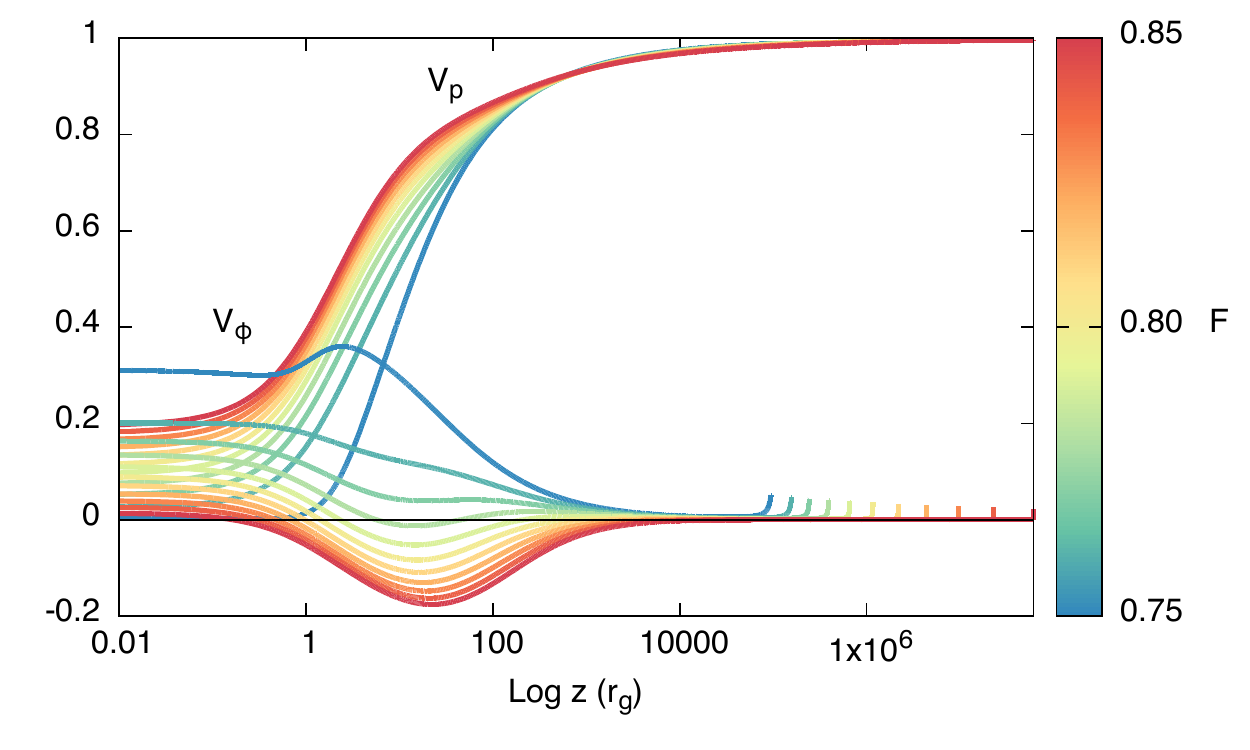}\hfil
     \includegraphics[width=0.47\linewidth]{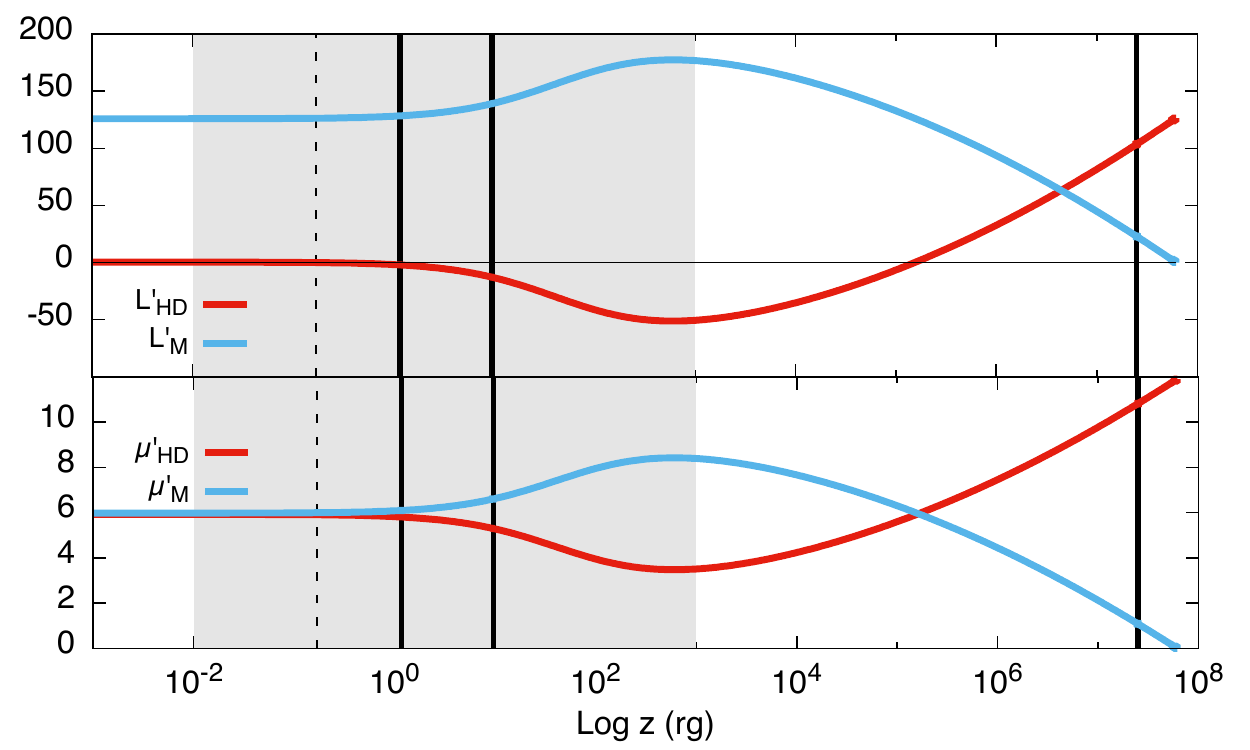}
\caption{ \textit{Left panel:} Toroidal and poloidal velocities in units of $c$ for a series of solutions obtained for increasing $F$. The solution of this series with higher $F$ is model (7) in Tab.~\ref{tab:pars}. \textit{Right panel:}. Angular momentum and energy components defined in equations \ref{berncomp} and \ref{momcomp} for model (7). The vertical solid lines are the MSP, AP and MFP of this solution, while the dashed line marks the height of the jet where $L'_{\rm HD}$ becomes negative. The shaded grey areas are shown as a comparison with the range considered in figure 2 of \citet{Cayatte:2014}.} 
\label{fig:vel_counterrot}
\end{figure*}

Panel (\emph{e}) of Fig. ~\ref{fig:energy} shows the energy components of model (7). 
This jet is roughly at equipartition at its foot point, with both enthalpy and Poynting energy being large. This is a \textit{hot, magnetized} jet. We see that, starting from the MSP, the Poynting energy increases as a consequence of the transfer of a fraction of the thermal energy into the magnetic field, while the kinetic energy increases at lower pace. The fraction of energy transferred between the different components is regulated by the conservation of the total energy $\mu'$. This third channel of energy transfer from the kinetic and thermal components to the magnetic component has never been seen before in a semi-analytical model, although present in simulations \citep[see model B2H and Section 5.5 in][]{Komissarov:2009} and discussed analytically by \citet{Sauty:2012} and \citet{Cayatte:2014}. This result is thus important because it demonstrates that our semi-analytical framework produces the full range of flows seen also in MHD simulations, and is not limited to the simplest scenarios. 
The increase of the Poynting energy also results in an increase in the magnetic component of the angular momentum $L'_{\rm M}$ in Eq. \refp{momeq} that corresponds to $L'_{\rm HD}$ becoming negative (see right panel Fig. \ref{fig:vel_counterrot} and figure 2 in \citealp{Cayatte:2014}).
The change in sign of $L'_{\rm HD}$ is due to $V_{\phi}$ becoming negative (see left panel Fig. \ref{fig:vel_counterrot}). 
In the cold regime, we normally see both components of the velocity, $V_{\rm p}$ and $V_\phi$, always above zero. The jet starts off with a larger toroidal velocity that then decreases in correspondence to the poloidal component taking the lead.  
In the case of hot magnetized solutions, the toroidal velocity can become negative before the canonical behaviour of a cold jet is restored at larger distances from the black hole. This inversion of sign of the toroidal component of the velocity is interpreted as a counter-rotation of the jet with respect to the disk.
\citet{Sauty:2012} and \citet{Cayatte:2014} claim that the counter-rotation in jets is the signature of the magnetization of the jet and it can be due to several effects: deceleration of the flow, steep gradients of the magnetic field and/or energy transfer from enthalpy to the magnetic field. Model (7) belongs to the latter case. In Fig.~\ref{fig:vel_counterrot} we start from model (7) and progressively decrease F. In the left panel, we see how for higher $F$ the solutions become hotter and their toroidal velocity minima decrease and they exhibit counter-rotation.

Similar interplay between energy terms can also result in almost oscillatory behaviours in the evolution of the energy components, seen for instance when the initial amount of enthalpy in the system is close to its minimum ($=1$, i.e. $q \sim 10^{-5} - 10^{-9}$) and $\sigma_{\rm M} < 1$ (panel (\emph{f}) of Fig.~\ref{fig:energy}, depicting solution (8) of Tab. \ref{tab:pars}).
Here we see two episodes where the energy is transferred to the magnetic field from the thermal energy and the flow is decelerating. However, the toroidal component of the velocity does not reverse sign, therefore no counter-rotation is established. 
This behaviour is usually associated to a highly wound-up streamline below MSP. The radial distance of the streamline during this phase is not constant, inducing a small change in the gravitational potential. Usually, after an initial acceleration powered by the leading force, either thermal or magnetic, the flow starts decelerating while approaching the jet axis, then the gravitational potential becomes relevant again, leading to an acceleration of the flow. Such interplay of forces can happen a few times and it results in complete winding up of the field line before the jet gains enough poloidal velocity to be launched outwards after crossing MSP. If both $q$ and $\sigma_{\rm M}$ are small, upstream of the MSP the jet has little poloidal component in its velocity, therefore it keeps slowly spiralling upwards, until it gains enough poloidal velocity to be slung out  of the disk. Solutions with lower $\psi\ata$ could not be found because the jet will not have enough initial energy to acquire sufficient poloidal velocity to then be launched out.

\subsection{On the effect of $\sigma_{\rm M}$ and $\psi\ata$}
\label{sigMpsiA}

\begin{figure*}
\centering
    \includegraphics[width=0.47\linewidth]{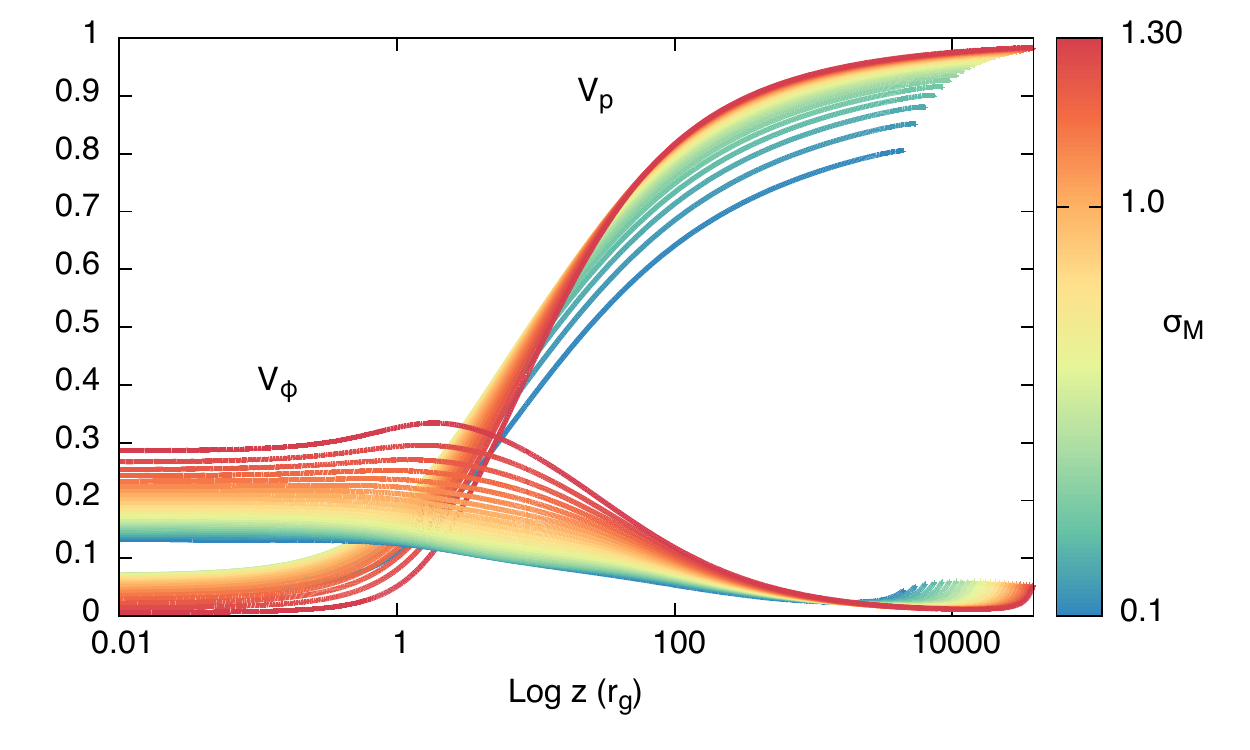}\hfil
     \includegraphics[width=0.47\linewidth]{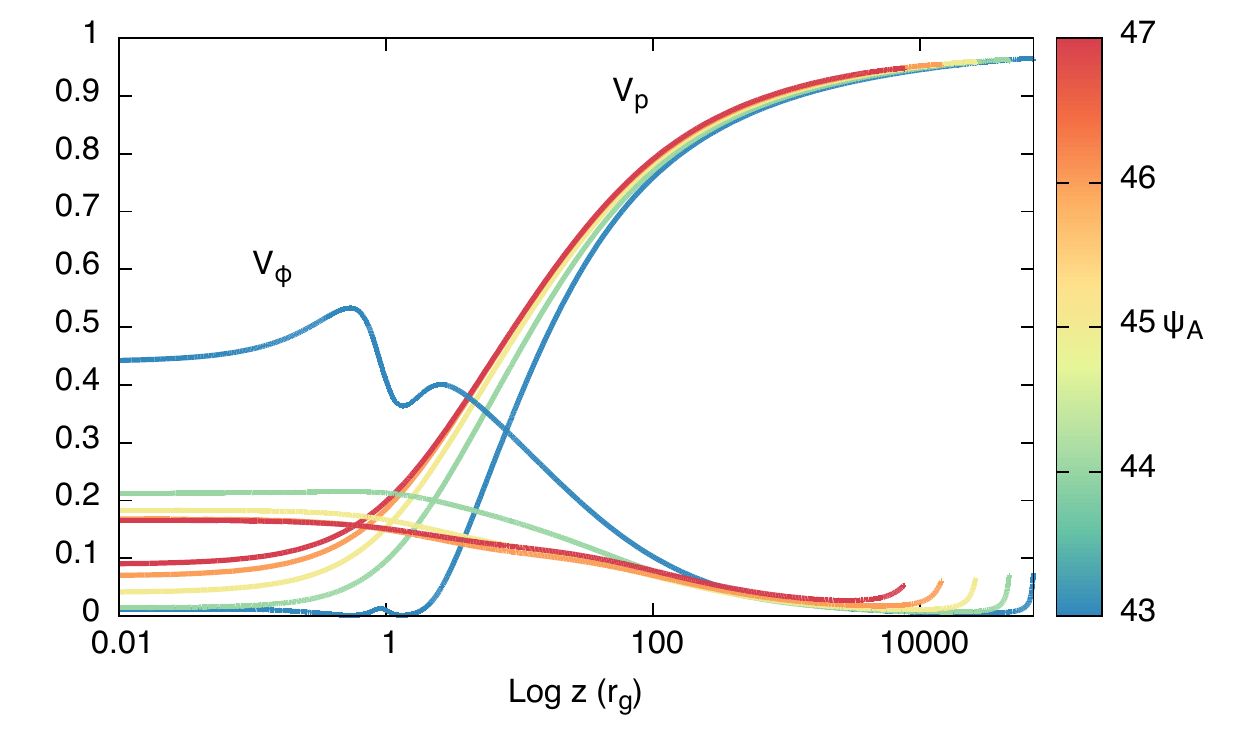}
\caption{ \textit{Left panel:} Toroidal and poloidal velocities in units of $c$ for a series of solutions obtained increasing $\sigma_{\rm M}$ at fixed $\psi\ata=46\degree$. \textit{Right panel:} Same figure as the left panel, but for increasing $\psi\ata$ at fixed $\sigma_{\rm M}=0.55$. The other fixed parameters for both plots are $\theta\ata=60\degree,\, F= 0.75,\,\Gamma=4/3$ and $\varpi\ata=15$.} 
\label{fig:velocities}
\end{figure*}

We show two examples of the evolution of the toroidal and poloidal components of the velocity for two sets of solutions in Fig.~\ref{fig:velocities}, obtained by varying $\sigma_{\rm M}$ and $\psi\ata$.  
We note that at small $z$ (upstream of the MSP) as $\sigma_{\rm M}$ increases (left panel of Fig.~\ref{fig:velocities}) the toroidal component of the velocity increases, while the poloidal component first becomes larger and then it starts decreasing. For increasing $z$ the toroidal component decreases towards the MFP, but, as $\sigma_{\rm M}$ increases, a maximum appears around the MSP. 
The poloidal velocity increases with distance from the black hole and accelerates, rapidly approaching the speed of light close to the MFP/LRP. Indeed, we cannot find solutions with a larger magnetisation parameter for the given set of input parameters ($\psi\ata=46\degree,\,\theta\ata=60\degree,\, F= 0.75,\,\Gamma=4/3$ and $\varpi\ata=15$). 
In the right panel of Fig.~\ref{fig:velocities}, we show a similar plot of the velocity components for a series of solutions obtained by increasing $\psi\ata$ and fixing $\sigma_{\rm M}=0.55,\,\theta\ata=60\degree,\, F= 0.75,\,\Gamma=4/3$ and $\varpi\ata=15$. We see that varying either $\psi\ata$ or $\theta\ata$ individually restricts the search to a small range in such parameters. From the ARC (See Appendix~\ref{definitions}, Eq.~\ref{ARC}), we can put a constraint on the sum of these two angles in order to select solutions that have a negative derivative of the poloidal Mach number, $d M^2/d\theta$ (Eq.~\ref{sys:wind}), at the \alf point ($90\degree<\theta\ata+\psi\ata<180\degree$). This interval for the sum of $\theta\ata$ and $\psi\ata$ ensures that the flow is accelerating while moving away from the black hole. From the evolution of the velocity components with respect to this sum we see that the range where we can find solutions, for a given set of fixed parameters, is much smaller than the one inferred from the sign of $d M^2/d\theta$ at \alf and that the solution can be radically different between the lower end and the upper end of the range. We also note that the larger $\psi\ata$ becomes, the closer the MFP moves towards the black hole. We cannot find solutions for $\psi_{\rm A} > 47\degree$ probably because the LRP happens before the MFP, while our method is focused on finding solutions that pass through all the three singular points.
When $\psi\ata$ decreases, we see that suddenly the jet has a much larger toroidal velocity, while the poloidal component is almost zero. The last solution is therefore the one for which the jet can still be launched from beyond MSP, while it keeps circulating upstream of this point.  
 
\subsection{Exploring other regions of the parameter space}
\label{sigMpsiA}

\begin{figure*}
\centering
     \includegraphics[width=0.47\linewidth]{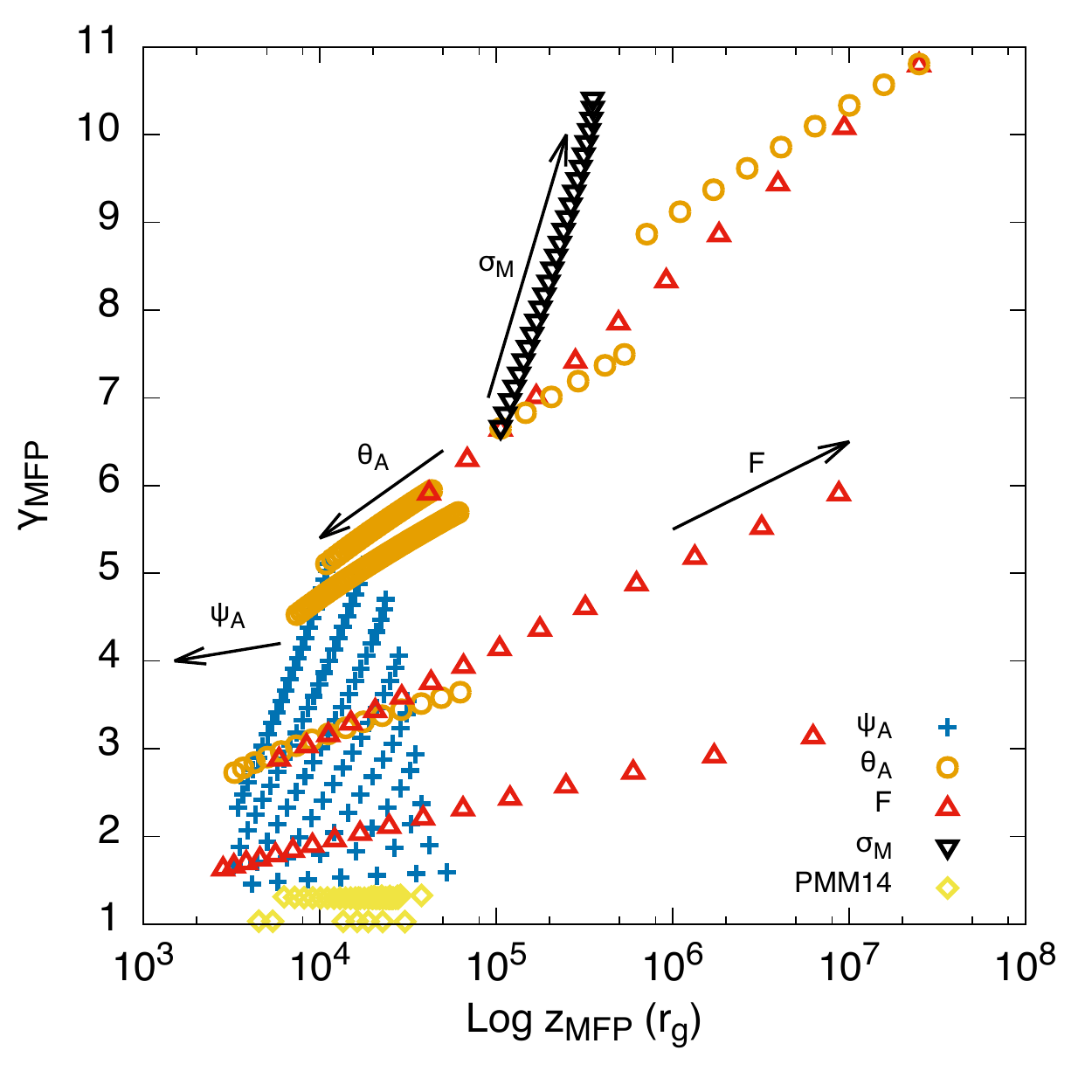}
     \includegraphics[width=0.47\linewidth]{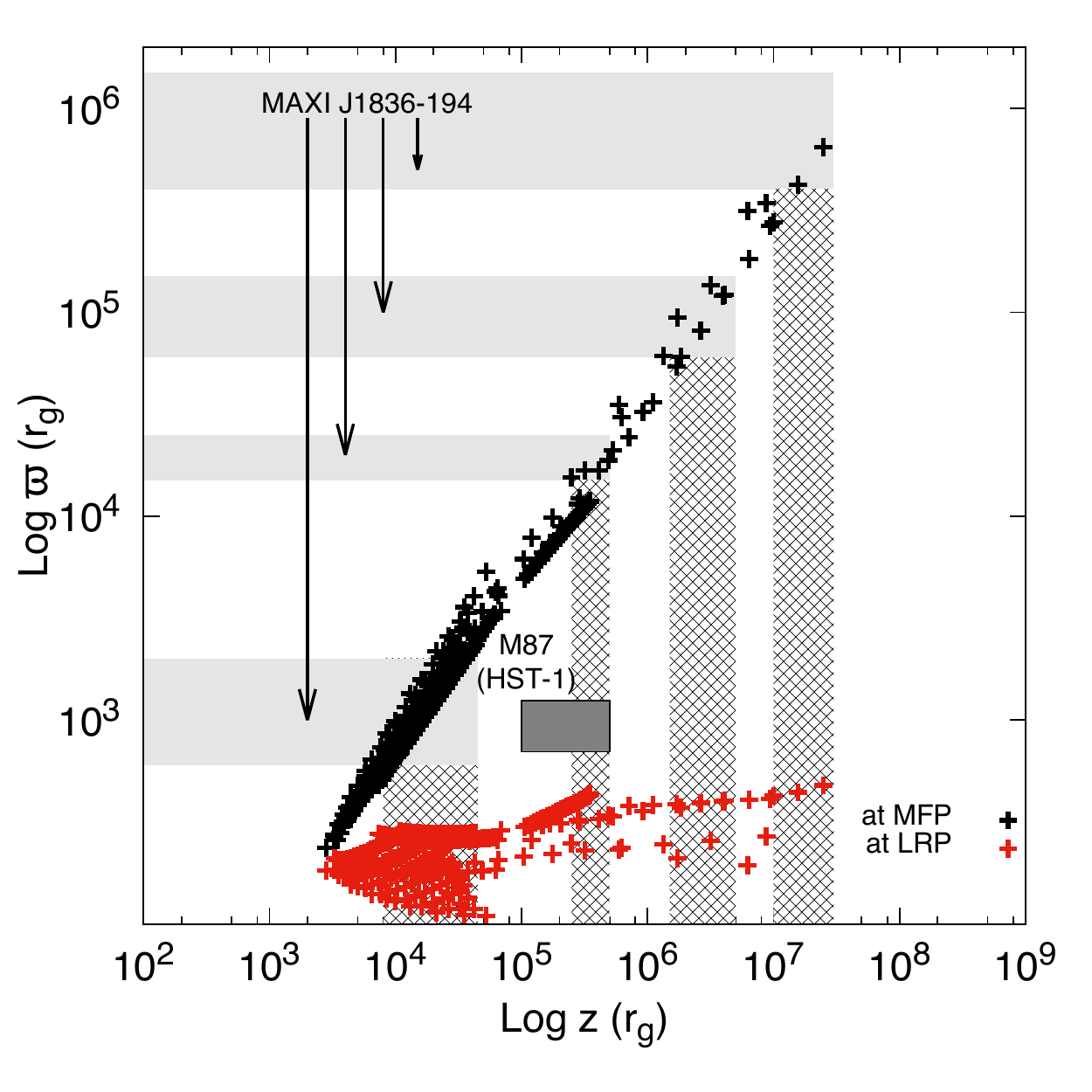}
\caption{\textit{Left panel:} Distribution of the Lorentz factors at MFP of all solutions, as a function of the height of the MFP. The blue crosses are solutions belonging to the 2-dimensional search varying $\psi_{\rm A}$ for given values of $\sigma_{\rm M}$. 
The other symbols are for solutions found by varying $\theta\ata$ (orange hollow circles),  $\sigma_{\rm M}$ (black hollow downward triangles), $F$ (red hollow triangles) and \capp{} with correct gravity terms (yellow hollow diamonds). The arrows mark the direction of increasing values of the given parameter. 
\textit{Right panel:} Distribution of the same collection of solutions shown in the the left panel with respect to cylindrical radius versus height of the MFP (black crosses) and of the last recollimation point (LRP, red crosses). The shaded light grey areas corresponds to the radial coordinates of the jet at the jet break with error bars reported by \citet{Russell:2014} for the XRB MAXI J1836-194. Solutions are present within each region where the jet break has been seen (grey areas). The areas with the oblique-line pattern denotes the predicted heights for the jet break. The dark grey rectangle is the area where the  knot HST-1 in the jet of M87 is estimated to be by \citet{Asada:2012}. The authors pointed out that the jet cross section at HST-1 is smaller than the one predicted by conical and parabolic jet models. We note that HST-1 could be tracing an intermediate location between the MFP and LRP, where the jet is rapidly collimating towards the axis.}
\label{fig:collimation}
\end{figure*}

We already presented here a few solutions that were found outside the initial 2-dimensional grid found by varying $\psi\ata$ and $\sigma_{\rm M}$.
Driven by the need to better understand which parameters lead to a shift in the observable values, such as the Lorentz factor of the jet at the MFP/collimation region, the height of the MFP, the energy balance at the launching site, we chose to try to explore different regions of the parameter space, i.e. varying previously fixed parameters. In the left panel of Fig.~\ref{fig:collimation} we show some of the alternative directions that we pursued. 
We note that each of the parameters used has a substantial effect in determining the height of the MFP compared to the relatively small steps we adopted in this search. Also, the variation of some of them (particularly $\sigma_{\rm M}$) result in large changes in the bulk Lorentz factor of the jet at MFP, e.g. a $\Delta\,\sigma_{\rm M} \sim 1.2$ around $\sigma_{\rm M} = 0.70$ corresponds to $\Delta\,\gamma_{\rm MFP} \sim 4$.

We find that increasing $F$, thereby changing the scaling of the magnetic field ($B \propto r^{F-2}$), is the most efficient way of moving into a different area of the parameter space, letting us touch terminal Lorentz factors of about 11. However, as discussed e.g. in \citet{BlandfordPayne:1982}, \citet{Contopoulos:1995} and \cavt{}{} , the higher $F$ the more the MFP is moving to larger distances from the BH, eventually to infinity.  Although in principle there is no constraint in our algorithm, with the exception of loss of numerical accuracy, on how high the MFP can lie with respect to the BH, we indeed have not yet found solutions with $F \gtrsim 0.9$ for the parameters explored so far due to the aforementioned numerical accuracy issues, therefore all our solutions are in the so-called return-current regime, i.e. the current decreases with radius.

As a further general feature shared by all solutions, we point out that they are suddenly terminated soon after the MFP, while rapidly recollimating  towards the jet axis (see also Fig.~\ref{fig:streamlines}). 
From a numerical point of view that happens because the denominator of the wind equation (Eq.~\ref{sys:wind}) and the equation for $\psi$ (Eq.~\ref{sys:dpsidt2}) goes to zero, $D \to 0$. We see that the streamlines are rapidly becoming vertical ($\psi \to \pi/2$).
At the moment it is difficult to say whether this is due to physical effects such as compression of the gas or just due to the ``polar axis singularity'' imposed by the self-similarity assumption, which makes the equations degenerate for $\theta\to0$.

\section{Comparison with observations}

The goal of the parameter search presented here has been primarily to populate an initial portion of the parameter space where solutions that are good candidates for the application to real sources reside. 
As shown in Fig.~\ref{fig:collimation}, we retrieve solutions that have bulk Lorentz factors at the MFP ranging from $1-11$ and MFP's height that spans 4 orders of magnitude ($\sim10^3-10^7 \rg$). Moreover, we find jets with different initial conditions exhibiting a large range of magnetic-to-thermal energy ratios and different degrees of winding at the base. 
The set of initial conditions, the positions of the singular points, velocity, density and magnetic field profiles can be either directly compared to observations or can be used to provide constraints for the calculation of the emission and the polarisation degree expected from a particular jet configuration. 

For instance, the MFP is the location where the jet streamlines start to recollimate and the flow downstream of this region loses causal contact with the upstream flow. These are favourable conditions for the onset of a shock, while maintaining the structure upstream intact \citep{Polko:2010}. 
If, furthermore, we assume that there is a correspondence between the MFP and the frequency of the jet break, the self-absorption turnover in the synchrotron spectrum from the region where particle acceleration initiates in the jet \citep{Markoff:2001,Markoff:2005,Markoff:2010}, we can use observational constraints on the position of the jet break to determine the best jet solution within the set that we have.

In AGN, the equivalent jet break can be at large offsets from the BH ($> 10^4\rg$) and there could be substantial contamination in the determination of the jet break by the interaction of the jet with the environment \citep{Russell:2015} or the scenario could be further complicated by a multiple-flow dynamics (see \citealp{Meier:2003} for a theoretical motivation of \emph{structured} jets and e.g. \citealp{Giroletti:2004,Harris:2006,Meyer:2013} for observational evidence on the radial structure of the jets of M87). 
High-resolution observations can now resolve the details of the jet structure in AGN down to a few gravitational radii. 
Radio interferometry and high-resolution optical and X-ray observations unveiled bright subfeatures, usually referred to as \emph{knots}, which are identified with shocks and particle acceleration. 
Knots can occur everywhere along the jet. Some of them are attributed to internal shocks
caused for example by the interactions of the jet with a dense external cloud \citep[][]{MendozaLongair:2001}, variations in the injected mass and velocity \citep[][]{ReesMeszaros:1994,Malzac:2014} or magnetic recollimation \citep{Markoff:2010}. 
However, knots that occur at larger distances from the black holes can also be produced by self-collimation of the streamlines \cite[][]{Polko:2014} or a change in the ISM pressure profile \citep[][]{Nakamura:2013}. \\
The jet of M87, for example, is one of the main targets for this type of study. It exhibits a complex pattern of knots with different values and orientations of the proper motion. In particular, the HST-1 knot has received significant interest over time. HST-1 is located approximately at $2-5\times 10^5 \rg$, which also corresponds to the region where the Bondi radius resides ($r_{\rm B} \sim 3.5\times 10^5 \rg$ for $M_{\rm BH}=6\times10^9 M_\odot$).  
The Bondi radius marks the volume where the gravitational potential of the black hole dominates over the thermal energy of the gas contained within this volume. The density profile within and outside the Bondi radius can be substantially different, hence the jet could experience a steep gradient in the external pressure which could induce a recollimation shock (for observational evidence see \citealp{Russell:2015} and for recent simulations see \citealp{BarniolDuran:2016}).

\citet{Asada:2012} and \citet{Asada:2014} show that the M87 jet maintains a parabolic profile up to HST-1. The authors show that this is evidence of the extension of  the acceleration and collimation region of the jet of M87 up to the end of the sphere of influence of the black hole. 

If we assume then that the HST-1 knot in the jet of M87 coincides with a recollimation shock and acceleration region in the jet produced by self-collimation, we can identify it with the MFP or LRP. In all the solutions presented in this paper we see that the jet continues recollimating downstream of the MFP and we possibly see conditions that could lead to a shock due to the compression of the gas in correspondence of the last recollimation point. The flow could become causally disconnected but still remain smooth through the MFP, then undergoing a shock only when it is over-compressed.  The distance between the MFP and LRP is relatively large in AGN so that it can be resolved by current high-resolution observations.\\
Using the interval given by \citet{Asada:2012} for the width of the jet at HST-1 (dark grey area in Fig.~\ref{fig:collimation}, right panel), we see that the estimated position of HST-1 falls in between the MFP and the last recollimation point of the same jet solutions. HST-1 has a smaller cross-section with respect to a canonical paraboloidal jet, probably indicating that the streamlines are converging towards the jet axis and a shock could take place soon \emph{after} the MFP.  \\
Although the identification of HST-1 with a self-collimation shock is tempting, also the change in the external pressure could influence the collimation of the jet and the shock formation due to the proximity of the Bondi radius.
There is also increasing evidence that a first recollimation occurs within the first few tens of $\rg$ \citep{Prieto:2016,Hada:2016}, suggesting a more complex scenario. 

In XRBs, the position of the jet break has been identified in a few cases in the range of of $10-1000~\rg$. The turnover has been directly observed only in three cases (GX339-4 by \citealp{CorbelFender:2002}, MAXI J1836-194 by \citealp{Russell:2014} and 4U 0614+091 by \citealp{Migliari:2006}). 
\citet{Russell:2014} show that the jet break shifts by $\sim 3$ orders of magnitude in frequency during state transition in the source MAXI J1836-194. The estimated width of the acceleration region also presents large variations during the transition of the source, decreasing as the jet break moves closer to the BH. With the solutions that we have collected so far, we cover the full range of the jet widths seen in MAXI J1836-194 (see right panel of Fig.~\ref{fig:collimation}). 
Solutions found within the grey areas in Fig.~\ref{fig:collimation} not only can reproduce the inferred widths of the jet at the location of the shock, but they give a full set of predicted physical properties of the corresponding jet, such as the height of the shock (identified with the MFP), the bulk velocity of the flow and the magnetic field morphology at any point of the jet.  

Empirically constraining some of the jet properties, such as the Lorentz factor of the flow, indirectly from observations is very difficult.  
For instance, XRBs are thought to have mildly relativistic jets with $2 <\gamma_{\rm j}< 5$, while the Lorentz factor of AGN could span a larger range. However, recent works have shown that the range of indirectly inferred Lorentz factors for XRBs could be biased towards the lower end \citep{Fender:2003,JMFN:2006} and in fact be the same of AGN $(\gamma_{\rm j}\sim 1-50)$. With the solutions that we present here, the velocity profile is determined as part of a jet solution, and can be constrained together with other observables.

With the increasing attention on polarisation studies and the growing number of current and upcoming devoted facilities, particularly the first X-ray polarimeter IXPE \citep[Imaging X-ray Polarimetry Explorer, ][]{Weisskopf:2016}, it is crucial to properly model the jet magnetic field and the emission produced by such configuration of fields and particles. 
Indeed, as shown by various works, e.g. \citet[][]{Perlman:1999,Avachat:2016,RussellShahbaz:2014}, multiwavelength polarisation studies can help constrain the magnetic field morphology over large scales. The parameter degeneracy in the evaluation of the spectral energy distribution of the given source is also greatly reduced by polarisation measurements. The need of a more detailed MHD treatment of the modelling of jets, which could be used to infer their emission is therefore compelling.

\section{Summary and conclusion}

We presented here a new numerical scheme to solve the equations for stationary, axisymmetric, radially self-similar, relativistic MHD jets. 
Such class of jet models have been studied extensively over the last three decades and adopted in numerical studies and simulations of the acceleration/collimation region of jets with various applications, from gamma-ray bursts to young stellar objects.
In the case of semi-analytical models of \capmp{},\capt{},\capp{}, the numerical approach adopted for the integration of the equations describing the outflow was heavily affecting the efficiency of the parameter space search. 
Moreover, in the most recent development, the gravity terms in \capp{} were calculated with excessive approximations, which led to extra inaccuracies in the results. 
We avoid inconsistent approximations and also use the correct derivatives of the gravitational potential with respect to $\theta$: in
this way we use fully self-consistent gravity terms.
With this new setup, we are able to recover and compare the solutions found in \capp{} as a test for our algorithm. The consistency between the equations affects the parameters of the solutions and, consequently, the observables that we derive from them. 
Once the extra approximations are removed, the contribution of gravity is stronger and this leads to a more restricted range of radii within which self-similarity holds for a given set of outflow parameters. 

We were able to find solutions corresponding to single streamlines in large parts of the parameter space. The multiple jet configurations retrieved from the initial search presented in this paper are shown to be diverse in geometrical and dynamical properties, such as magnetic-to-thermal energy ratio at the jet foot point, bulk velocity, morphology of the streamlines and the position of the MFP. 
However, the parameter space is not continuous and the exploration along a specific direction, i.e. varying the value of a given parameter, can be interrupted due to the approach of physical limits or the break-down of our assumptions.
Solutions can cease to exist because of the jet velocity approaching the speed of light around the MFP or due to the impossibility of launching the jet because of a too small energy reservoir at the base. In the latter case, the flow keeps circulating around the jet axis and never gains enough poloidal velocity to be slung out.
Alternatively, solutions cannot be found when the LRP takes place before the MFP, while our method is constructed so to cross all three singular points.

By studying the evolution of the energy and the velocity components for all solutions, we encounter both cold and hot relativistic jets, the latter configurations exhibiting counter-rotation at distances  $\sim 0.1-10^5 \, \rg$. 
With the exploration of the parameter space that we have conducted, we are able to cover a large area of the parameter space and to move towards a desired direction, e.g. higher bulk Lorentz factors, by making use of the trends that we observed so far.

Taking into account the large variety of jet configurations and dynamics that we can obtain with our numerical scheme, the extension to other accreting objects, e.g. young stellar objects, is within the capabilities of this model.
Future steps will be taken to increase the density of the solution grid and cover larger volumes of the parameter space. 
This mapping of the parameter space is ultimately aimed at providing a dynamic and flexible base for future coupling with a radiative code. That, in turn, will
eventually allow data-fitting of both XRBs and AGN, which will be the subject of forthcoming papers.
We are also aware of the need to benchmark the results of our method to fully numerical simulations to properly quantify the limits of our approach. 

\section{Acknowledgement}

CC is supported by the Netherlands Organisation for Scientific Research (NWO), grant Nr. 614.001.209.
SM acknowledges support from NWO VICI grant Nr. 639.043.513.
YC is supported by the European Union's Horizon 2020 research
and innovation programme under the Marie Sklodowska-Curie Global Fellowship grant agreement No 703916.

\onecolumn
\appendix

\section{Definitions of the equations}
\label{definitions}

We present here the \syst{} in its full form as given in \capp{}{} and we report the changes in the function of the gravitational pseudo-potential. \\
The Bernoulli equation and the transfield equation assume the same form when we write them down in terms of the functions $A_i,B_i,C_i$ with $i=1,2$
\begin{align}
A_1 \frac{d M^2}{d\theta} + B_1 \frac{d\psi}{d\theta} & = C_1 \label{origBern}\\
A_2 \frac{d M^2}{d\theta} + B_2 \frac{d\psi}{d\theta} & = C_2 \label{origTrans}
\end{align} 
where
\begin{align}
 A_1=&\frac{\cos^3\argtp}{\sin^2\argt\sin\argtp}\left[ \left(\frac{\mu x\ata^2}{F \sigma_{\rm M}}\right)^2\frac{x\ata^2M^2}{G^2}\frac{(1-G^2)^2}{\fmx^3}-\left(\frac{\xi x\ata^2}{F\sigma_{\rm M}}\right)^2\frac{(\Gamma-1)(\xi-1)}{(2-\Gamma)\xi+\Gamma-1}\frac{1}{M^2}\right]+\frac{M^2}{G^4}\frac{\cos\argtp}{\sin\argtp} \\
 B_1=&\frac{M^4}{G^4}\\
 C_1=&\frac{\cos(\psi)\cos^2\argtp}{\sin^3\argt\sin\argtp}\left\{\left(\frac{\mu x\ata^2}{F \sigma_{\rm M}}\right)^2\frac{(1-M^2-x\ata^2)^2}{\fmx^2}-\left(\frac{\xi x\ata^2}{F\sigma_{\rm M}}\right)^2+\left(\frac{\mu x\ata^2}{F \sigma_{\rm M}}\right)^2\frac{2x^2}{G^4\fmx^3}\left[G^4(1-M^2-x\ata^2)^2\right.\right. + C_1^+\nonumber\\
 &\qquad\left.\left.-x^2(G^2-M^2-x^2)^2-\fmx G^2(1-x\ata^2)(G^2-M^2-x^2)\right]\right\}\nonumber\\
 A_2=&\sin\argtp\cos\argtp\left[\left(\frac{\mu x\ata^2}{F \sigma_{\rm M}}\right)^2\frac{x^2 (1-G^2)^2}{\fmx^3}-\left(\frac{\xi x\ata^2}{F\sigma_{\rm M}}\right)^2\frac{(\Gamma-1)(\xi-1)}{(2-\Gamma)\xi+\Gamma-1}\frac{G^4}{M^4}\right]\\
 B_2=&\sin^2\argt\left[\frac{(1-x^2)}{\cos^2\argtp}-M^2\right]\\
 C_2=&\frac{\cos(\psi)\sin\argtp}{\sin\argt}\frac{G^4}{M^2}\left[\left(\frac{\mu x\ata^2}{F \sigma_{\rm M}}\right)^2\frac{(1-M^2-x\ata^2)^2}{\fmx^2}-\left(\frac{\xi x\ata^2}{F\sigma_{\rm M}}\right)^2+\left(\frac{\mu x\ata^2}{F \sigma_{\rm M}}\right)^2\frac{2x^2}{G^2\fmx^3}M^2(1-G^2)(1-M^2-x\ata^2)\right]\nonumber\\
 &+\frac{2}{M^2}\left(\frac{x^2}{F\sigma_{\rm M}}\right)^2\frac{(\Gamma-1)(F-2)\xi(\xi-1)}{\Gamma}+\left[2x^2+\fmx\right]\frac{\cos(\psi)\sin\argt\sin\argtp}{\cos^2\argtp}\nonumber\\
 &+\frac{\sin^2\argt}{\cos^2\argtp}(F-2-Fx^2+x^2)+\left(\frac{\mu x\ata^2}{F \sigma_{\rm M}}\right)^2\frac{x^2\left[M^2(1-G^2)^2(F-1)-(G^2-M^2-x^2)^2\right]}{M^2\fmx^2} + C_2^+.
\end{align}
where $\mu = \mu'/(1-P\gr)$.
The last terms in the $C_i$ are the gravity terms and they can be written schematically as
\beq
C_1^+   = f_1(P\gr) ~C_1^{+,noP}, \qquad C_2^+   = f_2(P\gr) ~C_2^{+,noP} \label{newCplusses}
\eeq
where
\begin{align}
C_1^{+,noP}  & = {\mu^2 x\ata^4\over F^2 \sigma_{\rm M}^2} {\cos^2\argtp\over\sin^2\argt}\left[\frac{G^2\fmxA^2-x\ata^2\fgmx^2}{G^2\fmx^2}\right]\\
C_2^{+,noP}  & = \left\{\left({x^4\over F^2\sigma_{\rm M}^2}\right)\left[\frac{\mu^2}{M^2}{\fmxA^2\over\fmx^2} + {\mu^2x^2\over2G^4}{\fg^2\over\fmx^2}-{\Gamma-1\over\Gamma}{\xi(\xi -1)\over M^2}\right]+  \frac12 (1+x^2){\sin^2\argt\over\cos^2\argtp}\right\}\cos^2\argtp
\end{align}
The functions $f_i(P\gr)$ ($i=1,2$) of the potential can be evaluated in a Newtonian or a \pw~scenario. The choice of using the \pw~pseudo-potential is easily motivated. The \pw~ potential diverges at the Schwarzschild radius, $r_{\rm s} = 2 r_{\rm g}$, so it mimics general relativity with the advantage of maintaining a newtonian formalism. If we define the potential as
\begin{equation} 
P\gr =  \left\{
\begin{aligned}
& ~\,\quad \Phi,\qquad \text{if} \quad Newton \\
&{\Phi \over (1-2\Phi)}, ~ \text{if} \quad \text{\pw} \label{potential}
\end{aligned}
\right. 
\end{equation}
where $\Phi = \sin(\theta)/(\varpi_A G)= r_{\rm g}/r$, $\varpi\ata$ is in units of $\rg = \mathcal{G M}/c^2$. Defining the following function 
\beq \mathcal F(P\gr) =  \left\{
\begin{aligned}
&1, \qquad \quad \text{if} \quad Newton \\
&1 + 2 P\gr,  ~\text{if} \quad \text{\pw}, \label{fpg}
\end{aligned}
\right. 
\eeq
the gravity terms appear in the convenient form \refp{newCplusses} with 
\beq
f_1(P\gr)  = -P\gr \frac{\mathcal F(P\gr)}{(1 - P\gr)},  \qquad f_2(P\gr)  =  P\gr~\mathcal F(P\gr). 
\label{fPgs}
\eeq
It is worth noting that the equations \refp{fPgs} differ from \capp{} which in our notation are
\beq
 f_1(P\gr) = -P\gr ,  \qquad f_2(P\gr)  =  P\gr . 
\label{peterfPgs}
\eeq
Rearranging the terms in Eq. \refp{origBern}-\refp{origTrans}, we obtain two differential equations, \refp{sys:wind}  and \refp{sys:dpsidt2}, for $M^2$ and $\psi$.

At the \alf point, all the equations in the \syst~ can be regularized with De L'H\^opital rule, starting from the quantities $\theta = \theta\ata, \psi = \psi\ata, x = x\ata$ and $\xi = \xi_A$, as follows:
\begin{align}
& G\ata = 1, \qquad M\ata^2 = 1 - x\ata^2 \label{alf:m} \\
&\sigma = {x\ata^2 - x^2 \over 1-M^2-x\ata^2}  \quad \longrightarrow \quad \sigma\ata =  {2x\ata^2\cos\psi\ata \over p\ata \sin\theta\ata\cos(\theta\ata + \psi\ata)}, \label{alf:sigma}\\
& \left({1-M^2-x\ata^2\over1-M^2-x}\right)\ata = {1\over \sigma\ata +1}, \qquad \left({1-G^2\over1-M^2-x}\right)\ata = { \sigma\ata/x^2\ata\over \sigma\ata +1}, \qquad \left({G^2-M^2-x^2\over1-M^2-x}\right)\ata = {x^2\ata - (1 - x^2\ata) \sigma\ata \over x^2\ata(\sigma\ata +1)}, \label{alf:ratios}  \\
& \left.{dM^2\over d\theta}\right|\ata = p\ata, \qquad  \left.{dG^2\over d\theta}\right|\ata = {2 \cos\psi\ata\over\sin\theta\ata\cos(\theta\ata + \psi\ata)} \label{alf:derivs}
\end{align}
where $p\ata$ is given by the \alf regularity condition (ARC) as described in \cite{Polko:2010} and \cite{Polko:2014}. The ARC is obtained by calculating the wind equation at the \alf point using the De L'H\^opital rule (Eq. \ref{alf:m}-\ref{alf:derivs}). Therefore we have
\beq
\left.\frac{d M^2}{d\theta}\right|\ata = \frac{B_{2,\rm A}C_{1,\rm A} - B_{1,\rm A}C_{2,\rm A}}{A_{1,\rm A}B_{2,\rm A} - A_{2,\rm A}B_{1,\rm A}}, 
\eeq
that, after recasting terms, becomes
\begin{align}
0 &= -\mathscr{G}\ata + 2 \frac{\Gamma - 1}{\Gamma} \frac{F - 2}{F^2 \sigma_{\rm M}^2} \xi\ata (\xi\ata - 1) (1 - x\ata^2) x\ata^4 + \frac{\sin^2(\theta\ata) (1 - x\ata^2)^2}{\cos^2(\psi\ata + \theta\ata)} \left[ (1 - x\ata^2) (F - 1) - 1 \right] \nonumber \\
&\qquad + \frac{\mu^2 x\ata^2}{F^2 \sigma_{\rm M}^2} (F - 1) \sigma\ata^2 \frac{(1 - x\ata^2)^2}{(\sigma\ata + 1)^2} - \frac{\mu^2 x\ata^2}{F^2 \sigma_{\rm M}^2} \frac{1 - x\ata^2}{(\sigma\ata + 1)^2} \left[ x\ata^2 - \sigma\ata (1 - x\ata^2) \right]^2 \nonumber \\
&\qquad + 2 \frac{\cos(\psi\ata) \sin(\theta\ata) \sin(\psi\ata + \theta\ata)}{\cos^2(\psi\ata + \theta\ata)} x\ata^2 (1 - x\ata^2)^2 \frac{\sigma\ata + 1}{\sigma\ata} \label{ARC}
\end{align}
where the gravity term is given by
\begin{align}
\mathscr{G}\ata = C_{1,\rm A}^+ B_{2,\rm A} - C_{2,\rm A}^+ B_{1,\rm A} & =  f_{1,\rm A}(P\gr)(1-x\ata^2)\sin(\theta\ata)^2 \tan(\psi\ata+\theta\ata)^2\left\{\left(\frac{\xi\ata x\ata^2}{F\sigma_{\rm M}}\right)^2\frac{\cos^2(\psi\ata+\theta\ata)}{\sin^2(\theta\ata)} + (1-x\ata^2)^2\right\} \nonumber\\
&\quad - f_{2,\rm A}(P\gr)(1-x\ata^2)^2 \cos^2\argtpa\left\{\left({\mu\ata x\ata^2 \over F\sigma_{\rm M}}\right)^2 {2x\ata^2+(1-x\ata^2)\sigma\ata^2\over 2x\ata^2(1-x\ata^2)(\sigma\ata+1)^2}\right. \nonumber\\
&\qquad\qquad \left.- {x\ata^4\over(F\sigma_{\rm M})^2}  {\Gamma-1\over\Gamma} {\xi\ata(\xi\ata-1)\over 1-x\ata^2} + {1+x\ata^2\over2} {\sin^2(\theta\ata)\over\cos^2(\psi\argtpa)}\right\}
\end{align}
with $f_{1,\rm A}(P\gr)$ and $ f_{2,\rm A}(P\gr)$ are calculated with $\Phi\ata =  \sin(\theta\ata)/\varpi\ata$.

\section{Derivation of the pseudo-potential functions in the gravity terms}
\label{derivpotential}

Here we give the details of the derivation of the functions \ref{fPgs}. First and foremost, we need to carry out the derivative of the gravitational potential to include them into the system \refp{origBern}-\refp{origTrans}. 
We take the derivative with respect to $r$ (since the gravitational potential depends solely on $r$), and then, applying the assumptions of axisymmetry and self-similarity, we obtain the corresponding forms in $\theta$. Since we want to keep the flexibility of changing between the Newtonian potential and the \pw~ pseudo-potential, we differentiate both $P_{\rm g}$s as follows
\begin{equation} 
 \pder{P\gr}{r} =  \left\{
\begin{aligned}
&- \frac{P\gr}{r}, \qquad \qquad \text{if} \quad Newton \\
&- \frac{P\gr}{r}(1+2P\gr), ~ \text{if} \quad \text{\pw} \label{potential}
\end{aligned}
\right. 
\end{equation}
or, in a more compact way 
\beq
\pder{P\gr}{r} = - \frac{P\gr}{r}\mathcal F(P\gr)
\eeq
where $\mathcal F(P\gr)$ is defined as in \ref{fpg}. 
Now, we need to calculate again the gravity terms in both \refp{origBern}-\refp{origTrans}. Let's start from the gravity term in the transfield equation in the general form equivalent to Eq. 8 of \capp{}{}
\beq
-(\gamma\rho_{\rm 0} + \mathcal E/c^2) c^2 \nabla P\gr \cdot \hat n \label{gravtrans_r}
\eeq
where $ \mathcal E = \gamma (\gamma -1) \rho_{\rm 0} c^2+ P (\gamma^2 \Gamma/(\Gamma -1) -1) + (B^2 +E^2)/(8\pi) $ is the energy density. Since we are calculating the derivative with respect to $r$ and $\hat n = \cos\argtp \hat r -\sin\argtp \hat \theta$ ($\hat n$ is the unit vector  perpendicular to the field line, towards the jet axis, as defined in Section 2.1 and 3.1 of \cavt{}{}), then $\nabla  P\gr \cdot \hat n = \cos\argtp\partial_r  P\gr $. The derivative of $P\gr$ with respect to $r$ is given by Eq. \eqref {potential}.
  
Recasting the term in parenthesis in Eq. \refp{gravtrans_r} in dimensionless units (see Appendix \ref{conversion} and \cavt{}{}), applying self-similarity ($\partial/\partial \phi=0$) and using $\varpi = r \sin\theta$ and $G\equiv \varpi/\varpi\ata,$ we have 
\beq
\nabla P\gr \cdot \hat n = \cos\argtp {\sin\theta \over \varpi} P\gr \mathcal F(P\gr) . 
\eeq
therefore the gravity term in the transfield equation is
\beq
-\left\{{B\zero^2 \alpha^{F-2}\over 4\pi \varpi G^4} {\sin\argt \over \cos\argtp}\right\}\left[{\cos^2\argtp}P\gr \mathcal F(P\gr)\right] {C_2^{+,noP}\over \cos^2\argtp}. \label{gravtrans}
\eeq 
Similarly, we take the derivative of the Bernoulli equation (Eq. 4 in \capp{}{}) with respect to $r$ and, using Eq. \ref{berneq}, we find
\beq
\tder{\mu}{r} \equiv \mu'^{2} \tder{}{r}\left[1\over (1-P\gr)^2\right] = - 2 \mu^2{P\gr\over(1-P\gr)} \mathcal F(P\gr) {1\over r}
\eeq
which once being included in the derivative of the Bernoulli equation under the self-similar assumption becomes
\beq
\left\{ -2 \tan\argtp G^6F^2\sigma_{\rm M}^2 (1-M^2 -x^2)^2 \sin^2\theta\right\} \left[ - 2 \mu^2{P\gr\over(1-P\gr)} \mathcal F(P\gr) \right] C_1^{+,noP}. \label{gravbern}
\eeq
It is worth noting that the scaling of both equations \ref{gravtrans} and \ref{gravbern} (terms in curly brackets) can be simplified from all functions $A,B,C$, as done by \capp{}{}.

\section{Initial conditions}
\label{sub:integ}

Once we find the initial conditions for $M^2$ and $G^2$ and their
  derivatives at the three critical points, we perform the integration
  from each point with an adaptive stepsize Runge-Kutta scheme as
  described in \citet[hereafter NR93]{NRfortran:1993}, until we reach
  the midpoints $\themf$ and $\thems$. However, setting up the initial
  conditions is both the real challenge and the base for the
  robustness of our scheme.

% let's start easy: AP

As we mentioned in the Sect.\ref{method}, given the full set of
  parameters, the initial conditions at the AP are readily
  calculated. The only numerical step is finding $d M^2 / d
  \theta$. The derivative of $M^2$ at \alf is commonly referred to as $p\ata$ and can be found from the \alf regularity condition as described in \cite{Polko:2010,Polko:2014} and in eq. \ref{ARC}. 
  % \textbf{At \alf also $\xi\ata$ is know from Eq.~\ref{sys:xi} and using the constrain for $M^2$ at this point, i.e. $M\ata^2 = 1 - \xa$.}
  Now the integral of motion $\mu'$ can also be
  calculated (see Eq. \ref{berneq}) and all the other quantities can be analytically calculated at the \alf point. 

% now the big guys!: MFP & MSP
  
The situation is more complicated at the modified magnetosonic
  points, since no analytical condition is known. Our solution is a
  sequence of root finding routines to find the zeroes of specific
  functions. The method is almost the same at both MSP and MFP,
  therefore we will describe only the case of the MFP, highlighting
  the few differences when they exist.

Given the value of $\thef$, we need to find the values of $M^2$ and $G^2$
that gives $\numm = \nump = \den = 0$. In order to do this, we
minimize the \emph{\cri} function
 \begin{equation}
\crif_{\rm F} = \nump^2 +\den^2.
\label{fmfp}
\end{equation}
Even if it can be shown that if any two of $\numm$, $\nump$, $\den$
are zero, the third one must be zero as well, we found that for numerical
reasons using $\nump$ and $\den$ is more robust. We also found that at
MSP it is better to use all the three numbers
\begin{equation}
  \crif_{\rm S} =
  \numm^2 + \nump^2 +\den^2.
  \label{fmsp}
\end{equation}
That has to do with the topology of the zeroes of the three numbers,
which in the regimes we explored turn out to have deep narrow valleys
running almost parallel to each other, so that it is difficult
to find their intersection with enough precision.

In order to find the minimum of the critical function, we create
  a grid in the $M^2$ and $G^2$ space with boundaries dictated by the
  location of the \alf point. We then use the three points which have
  the lowest values of the critical function $\crif$ as the starting nodes of a global
  simplectic minimising algorithm such as \texttt{amebsa}
  \cpan{see}{for a description}. When the minimum has been found, we try
  to polish the solution with a further local \texttt{powell}
  minimising routine \cpan{}{}. At the end, we further verify that the
  absolute values of all the three numbers $\numm$, $\nump$ and $\den$
  are close enough to zero (< $10^{-10}$). While this procedure of
  polishing the first solution and then cross checking again seems
  redundant and CPU consuming, it is our experience that this process
  avoids having spurious starting points which would contaminate the
  results of the next steps in our solution finding scheme.

Although we now know the initial values of $M^2$ and $G^2$, we
  cannot yet start the integration, since the derivative $d M^2 / d
  \theta$ is still unknown: i.e. Eq. \eqref{sys:wind} would still be zero
  over zero. In order to recover the value of the derivative, we
  search the roots of another function.
   
For a point $\theta$, very close to the critical point at $\thef$, we can Taylor expand
$G^2$ and $M^2$ to first order:
\begin{align}
  G^2(\thef) =& G^2(\theta) +  \left.\frac{d G^2}{d \theta} \right|_{\theta} (\thef - \theta) \label{approxG}\\ 
  M^2(\thef) =& M^2(\theta) + \left.\frac{d M^2}{d \theta} \right|_{\theta} (\thef - \theta) \label{approxM}
\end{align}
%\begin{align*}
%  G(\thef) =& G(\theta) + \left.\frac{d G}{d \theta} \right|_{\theta} (\thef - \theta)\\ 
%  M(\thef) =& M(\theta) + \left.\frac{d M}{d \theta} \right|_{\theta} (\thef - \theta)
%\end{align*}
Combining the two equations to eliminate $\thef - \theta$, we define
the following function
\begin{equation}
  Y\left(M^2(\theta)\right) = [M^2(\theta) - M^2(\thef)] - [G^2(\theta) - G^2(\thef)]
  \frac{\left.\frac{d M^2}{d \theta}\right|_{\theta}}{\left.\frac{d G^2}{d \theta}\right|_{\theta}}
  \label{FM2_1}
\end{equation}
The final step is finding the zeroes of this function through an iterative method. 
Note that only $M^2(\theta)$ is a free variable.
We approximate $G^2(\theta)-G^2(\thef)= \left.d G^2/d \theta \right|_{\thef} (\theta - \thef)$.
Once a guess of $M^2(\theta)$ is tried, we can calculate $d M^2 / d
\theta$ and $d G^2 / d
\theta$ at $\theta$ where the first one is not singular.  At MFP we take $\theta
= \thef(1 + \delta)$ and at MSP we take $\theta = \thes(1 - \delta)$,
where $\delta=10^{-7}$.

In most cases we find two roots for $M^2(\theta)$: one smaller and
  one larger than $M^2(\thef)$. Based on the physical picture of the accelerating jet and that $\left . dM^2 / d\theta \right |_A < 0$, we assume
  that the derivative is negative also at the critical
  points. Therefore at MFP we choose $M^2(\theta) < M^2(\thef)$ and at MSP
  we take $M^2(\theta) > M^2(\thes)$.  We cannot justify this assumption
  based on general principles, and cannot provide a definitive proof,
  but we performed various tests allowing for the alternative choices,
  and never managed to find a solution. Now that we have the initial conditions we can integrate.

\section{Conversion to physical quantities}
\label{conversion}

When a solution is found, we convert and calculate all the relevant quantities in physical units to be comparable with observables. The conversions are taken from \cavt{}{}.
A few additional input quantities need to be provided in order to transform back a solution to physically measurable quantities. These are the magnetic field strength at the base, $B_{\rm 0}$, the ratio between radiation and matter pressure, $P_{\rm M}/P_{\rm R}$ (section 4, Eq. 29 in \cavt{}{}) and the streamline label $\alpha\equiv \varpi^2/\varpi\ata^2$ (Section 2.1, Eq. 17 in \cavt{}{}). The quantities $\alpha$ and $B_{\rm 0}$ provide the scaling for the given solution. 

\begin{align}
\varpi & = \varpi\ata G, \qquad
z	 = {\varpi\ata G \over \tan(\theta)}, \qquad
\gamma  = \frac{\mu'}{(1-P_{\rm g})}{1 \over \xi}  {(1-M^2-x\ata^2)\over(1-M^2-x^2)} \\
%T_e  & = \sqrt{x\ata^2 {B\zero\over F \sigma_{\rm M}}} \left( {(\xi-1)^{\Gamma/(\Gamma-1)}  \alpha^{F-2} \over q (1+P_{ratio})} {3(\Gamma-1)\over4 \pi a~\Gamma}\right)^{1/4}, \qquad P_{\rm M}/P_{\rm R} = 0~(\Theta_e \leq 1) \quad \mathrm{or} \quad P_{\rm M}/P_{\rm R} = 1.85~(\Theta_e \geq 1)\\
T_e  & = \sqrt{x\ata^2 {B\zero\over F \sigma_{\rm M}}} \left( {(\xi-1)^{\Gamma/(\Gamma-1)}  \alpha^{F-2} \over q\, (1+P_{\rm M}/P_{\rm R})} {3(\Gamma-1)\over4 \pi a~\Gamma}\right)^{1/4}, \qquad 
 P_{\rm M}/P_{\rm R} = \left\{
\begin{aligned}
&0, \qquad \quad\, \theta_{\rm i } \leq  1\\
&1.85, \qquad \theta_{\rm i } \geq  1
\end{aligned}
\right., \qquad \textrm{with} \quad \theta_{\rm i } = k T_{\rm e} /m_{\rm e} c^2 \\
V_{\rm p}    & = -{c F \sigma_{\rm M}  M^2  \sin(\theta) \over \gamma \xi x^2 \cos(\psi+\theta)}, \qquad
V_{\rm \phi}  =  {c x\ata\over \gamma \xi G} \frac{\mu'}{(1-P_{\rm g})}{G^2-M^2-x^2\over 1-M^2-x^2}, \qquad
V_{\rm tot}  = \sqrt{V_{\rm p}^2 + V_{\rm\phi}^2} \\
B_{\rm p}    &  = - {B\zero \sin(\theta) \alpha^{(F-2)/2} \over G^2 \cos(\psi+\theta)}, \qquad
B_{\rm \phi}     = -\frac{\mu'}{(1-P_{\rm g})} x\ata^4  {1-G^2\over 1-M^2-x^2} {B\zero \alpha^{(F-2)/2} \over F \sigma_{\rm M} x}, \qquad
B_{\rm tot}  = \sqrt{B_{\rm p}^2 + B_{\rm \phi}^2} \\
\rho\zero & =  {x\ata^4 \xi B\zero^2\alpha^{(F-2)} \over 4\pi M^2 (c F \sigma_{\rm M})^2}, \qquad
P     = {B\zero^2 \alpha^{(F-2)} \over 4\pi} {\Gamma-1 \over \Gamma} {x\ata^4 \over (F\sigma_{\rm M})^2} {\xi (\xi-1) \over M^2 }\\
h    &= \xi\gamma, \qquad
S    = -  {x F\psi\ata B_\phi \over x\ata^2  B\zero \alpha^{(F-2)/2}} \qquad
\mu' = E_{tot} = (h+S)(1-P_{\rm g}) = \mu (1-P_{\rm g})\\
\Omega  & = \frac{1}{\varpi}\left(V_{\rm \phi} - B_{\rm \phi}{V_{\rm p}\over B_{\rm p}}\right)
\label{energies}
\end{align}
where $a$ is the Stefan-Boltzmann constant.

\bibliographystyle{mnras}
\bibliography{mybiblio}

\begin{thebibliography}{}
\makeatletter
\relax
\def\mn@urlcharsother{\let\do\@makeother \do\$\do\&\do\#\do\^\do\_\do\%\do\~}
\def\mn@doi{\begingroup\mn@urlcharsother \@ifnextchar [ {\mn@doi@}
  {\mn@doi@[]}}
\def\mn@doi@[#1]#2{\def\@tempa{#1}\ifx\@tempa\@empty \href
  {http://dx.doi.org/#2} {doi:#2}\else \href {http://dx.doi.org/#2} {#1}\fi
  \endgroup}
\def\mn@eprint#1#2{\mn@eprint@#1:#2::\@nil}
\def\mn@eprint@arXiv#1{\href {http://arxiv.org/abs/#1} {{\tt arXiv:#1}}}
\def\mn@eprint@dblp#1{\href {http://dblp.uni-trier.de/rec/bibtex/#1.xml}
  {dblp:#1}}
\def\mn@eprint@#1:#2:#3:#4\@nil{\def\@tempa {#1}\def\@tempb {#2}\def\@tempc
  {#3}\ifx \@tempc \@empty \let \@tempc \@tempb \let \@tempb \@tempa \fi \ifx
  \@tempb \@empty \def\@tempb {arXiv}\fi \@ifundefined
  {mn@eprint@\@tempb}{\@tempb:\@tempc}{\expandafter \expandafter \csname
  mn@eprint@\@tempb\endcsname \expandafter{\@tempc}}}

\bibitem[\protect\citeauthoryear{{Asada} \& {Nakamura}}{{Asada} \&
  {Nakamura}}{2012}]{Asada:2012}
{Asada} K.,  {Nakamura} M.,  2012, \mn@doi [\apjl]
  {10.1088/2041-8205/745/2/L28}, \href
  {http://adsabs.harvard.edu/abs/2012ApJ...745L..28A} {745, L28}

\bibitem[\protect\citeauthoryear{{Asada}, {Nakamura}, {Doi}, {Nagai}  \&
  {Inoue}}{{Asada} et~al.}{2014}]{Asada:2014}
{Asada} K.,  {Nakamura} M.,  {Doi} A.,  {Nagai} H.,   {Inoue} M.,  2014,
  \mn@doi [\apjl] {10.1088/2041-8205/781/1/L2}, \href
  {http://adsabs.harvard.edu/abs/2014ApJ...781L...2A} {781, L2}

\bibitem[\protect\citeauthoryear{{Avachat}, {Perlman}, {Adams}, {Cara}, {Owen},
  {Sparks}  \& {Georganopoulos}}{{Avachat} et~al.}{2016}]{Avachat:2016}
{Avachat} S.~S.,  {Perlman} E.~S.,  {Adams} S.~C.,  {Cara} M.,  {Owen} F.,
  {Sparks} W.~B.,   {Georganopoulos} M.,  2016, \mn@doi [\apj]
  {10.3847/0004-637X/832/1/3}, \href
  {http://adsabs.harvard.edu/abs/2016ApJ...832....3A} {832, 3}

\bibitem[\protect\citeauthoryear{{Barniol Duran}, {Tchekhovskoy}  \&
  {Giannios}}{{Barniol Duran} et~al.}{2016}]{BarniolDuran:2016}
{Barniol Duran} R.,  {Tchekhovskoy} A.,   {Giannios} D.,  2016, preprint, \href
  {http://adsabs.harvard.edu/abs/2016arXiv161206929B} {} (\mn@eprint {arXiv}
  {1612.06929})

\bibitem[\protect\citeauthoryear{{Belloni} \& {Motta}}{{Belloni} \&
  {Motta}}{2016}]{BelloniMotta:2016}
{Belloni} T.~M.,  {Motta} S.~E.,  2016, \mn@doi [Astrophysics of Black Holes:
  From Fundamental Aspects to Latest Developments]
  {10.1007/978-3-662-52859-4_2}, \href
  {http://adsabs.harvard.edu/abs/2016ASSL..440...61B} {440, 61}

\bibitem[\protect\citeauthoryear{{Blandford} \& {Payne}}{{Blandford} \&
  {Payne}}{1982}]{BlandfordPayne:1982}
{Blandford} R.~D.,  {Payne} D.~G.,  1982, \mn@doi [\mnras]
  {10.1093/mnras/199.4.883}, \href
  {http://adsabs.harvard.edu/abs/1982MNRAS.199..883B} {199, 883}

\bibitem[\protect\citeauthoryear{{Bogovalov} \& {Tsinganos}}{{Bogovalov} \&
  {Tsinganos}}{1999}]{Bogovalov:1999}
{Bogovalov} S.,  {Tsinganos} K.,  1999, \mn@doi [\mnras]
  {10.1046/j.1365-8711.1999.02413.x}, \href
  {http://adsabs.harvard.edu/abs/1999MNRAS.305..211B} {305, 211}

\bibitem[\protect\citeauthoryear{{Cayatte}, {Vlahakis}, {Matsakos}, {Lima},
  {Tsinganos}  \& {Sauty}}{{Cayatte} et~al.}{2014}]{Cayatte:2014}
{Cayatte} V.,  {Vlahakis} N.,  {Matsakos} T.,  {Lima} J.~J.~G.,  {Tsinganos}
  K.,   {Sauty} C.,  2014, \mn@doi [\apjl] {10.1088/2041-8205/788/1/L19}, \href
  {http://adsabs.harvard.edu/abs/2014ApJ...788L..19C} {788, L19}

\bibitem[\protect\citeauthoryear{{Connors} et~al.,}{{Connors}
  et~al.}{2016}]{Connors:2016}
{Connors} R.~M.~T.,  et~al., 2016, preprint, \href
  {http://adsabs.harvard.edu/abs/2016arXiv161200953C} {} (\mn@eprint {arXiv}
  {1612.00953})

\bibitem[\protect\citeauthoryear{{Contopoulos}}{{Contopoulos}}{1994}]{Contopoulos:1994}
{Contopoulos} J.,  1994, \mn@doi [\apj] {10.1086/174590}, \href
  {http://adsabs.harvard.edu/abs/1994ApJ...432..508C} {432, 508}

\bibitem[\protect\citeauthoryear{{Contopoulos}}{{Contopoulos}}{1995}]{Contopoulos:1995}
{Contopoulos} J.,  1995, \mn@doi [\apj] {10.1086/176170}, \href
  {http://adsabs.harvard.edu/abs/1995ApJ...450..616C} {450, 616}

\bibitem[\protect\citeauthoryear{{Corbel} \& {Fender}}{{Corbel} \&
  {Fender}}{2002}]{CorbelFender:2002}
{Corbel} S.,  {Fender} R.~P.,  2002, \mn@doi [\apjl] {10.1086/341870}, \href
  {http://adsabs.harvard.edu/abs/2002ApJ...573L..35C} {573, L35}

\bibitem[\protect\citeauthoryear{{Crumley}, {Ceccobello}, {Connors}  \&
  {Cavecchi}}{{Crumley} et~al.}{2017}]{CCCC:2017}
{Crumley} P.,  {Ceccobello} C.,  {Connors} R.~M.~T.,   {Cavecchi} Y.,  2017,
  preprint, \href {http://adsabs.harvard.edu/abs/2017arXiv170302842C} {}
  (\mn@eprint {arXiv} {1703.02842})

\bibitem[\protect\citeauthoryear{{Doeleman} et~al.,}{{Doeleman}
  et~al.}{2008}]{Doeleman:2008}
{Doeleman} S.~S.,  et~al., 2008, \mn@doi [\nat] {10.1038/nature07245}, \href
  {http://adsabs.harvard.edu/abs/2008Natur.455...78D} {455, 78}

\bibitem[\protect\citeauthoryear{{Fabian}}{{Fabian}}{2012}]{Fabian:2012}
{Fabian} A.~C.,  2012, \mn@doi [\araa] {10.1146/annurev-astro-081811-125521},
  \href {http://adsabs.harvard.edu/abs/2012ARA%26A..50..455F} {50, 455}

\bibitem[\protect\citeauthoryear{{Falcke}, {K{\"o}rding}  \&
  {Markoff}}{{Falcke} et~al.}{2004}]{Falcke:2004}
{Falcke} H.,  {K{\"o}rding} E.,   {Markoff} S.,  2004, \mn@doi [\aap]
  {10.1051/0004-6361:20031683}, \href
  {http://adsabs.harvard.edu/abs/2004A%26A...414..895F} {414, 895}

\bibitem[\protect\citeauthoryear{{Fender}}{{Fender}}{2003}]{Fender:2003}
{Fender} R.~P.,  2003, \mn@doi [\mnras] {10.1046/j.1365-8711.2003.06386.x},
  \href {http://adsabs.harvard.edu/abs/2003MNRAS.340.1353F} {340, 1353}

\bibitem[\protect\citeauthoryear{{Fender}, {Belloni}  \& {Gallo}}{{Fender}
  et~al.}{2004}]{Fender:2004}
{Fender} R.~P.,  {Belloni} T.~M.,   {Gallo} E.,  2004, \mn@doi [\mnras]
  {10.1111/j.1365-2966.2004.08384.x}, \href
  {http://adsabs.harvard.edu/abs/2004MNRAS.355.1105F} {355, 1105}

\bibitem[\protect\citeauthoryear{Fendt \& Ouyed}{Fendt \&
  Ouyed}{2004}]{Fendt:2004}
Fendt C.,  Ouyed R.,  2004, The Astrophysical Journal, 608, 378

\bibitem[\protect\citeauthoryear{{Feroz} \& {Hobson}}{{Feroz} \&
  {Hobson}}{2008}]{Multinest:2008}
{Feroz} F.,  {Hobson} M.~P.,  2008, \mn@doi [\mnras]
  {10.1111/j.1365-2966.2007.12353.x}, \href
  {http://adsabs.harvard.edu/abs/2008MNRAS.384..449F} {384, 449}

\bibitem[\protect\citeauthoryear{{Feroz}, {Hobson}  \& {Bridges}}{{Feroz}
  et~al.}{2009}]{Multinest:2009}
{Feroz} F.,  {Hobson} M.~P.,   {Bridges} M.,  2009, \mn@doi [\mnras]
  {10.1111/j.1365-2966.2009.14548.x}, \href
  {http://adsabs.harvard.edu/abs/2009MNRAS.398.1601F} {398, 1601}

\bibitem[\protect\citeauthoryear{{Feroz}, {Hobson}, {Cameron}  \&
  {Pettitt}}{{Feroz} et~al.}{2013}]{Multinest:2013}
{Feroz} F.,  {Hobson} M.~P.,  {Cameron} E.,   {Pettitt} A.~N.,  2013, preprint,
  \href {http://adsabs.harvard.edu/abs/2013arXiv1306.2144F} {} (\mn@eprint
  {arXiv} {1306.2144})

\bibitem[\protect\citeauthoryear{{Gallo}, {Fender}, {Kaiser}, {Russell},
  {Morganti}, {Oosterloo}  \& {Heinz}}{{Gallo} et~al.}{2005}]{Gallo:2005}
{Gallo} E.,  {Fender} R.,  {Kaiser} C.,  {Russell} D.,  {Morganti} R.,
  {Oosterloo} T.,   {Heinz} S.,  2005, \mn@doi [\nat] {10.1038/nature03879},
  \href {http://adsabs.harvard.edu/abs/2005Natur.436..819G} {436, 819}

\bibitem[\protect\citeauthoryear{{Giroletti} et~al.,}{{Giroletti}
  et~al.}{2004}]{Giroletti:2004}
{Giroletti} M.,  et~al., 2004, \mn@doi [\apj] {10.1086/379663}, \href
  {http://adsabs.harvard.edu/abs/2004ApJ...600..127G} {600, 127}

\bibitem[\protect\citeauthoryear{{Hada} et~al.,}{{Hada}
  et~al.}{2016}]{Hada:2016}
{Hada} K.,  et~al., 2016, \mn@doi [\apj] {10.3847/0004-637X/817/2/131}, \href
  {http://adsabs.harvard.edu/abs/2016ApJ...817..131H} {817, 131}

\bibitem[\protect\citeauthoryear{{Harris} \& {Krawczynski}}{{Harris} \&
  {Krawczynski}}{2006}]{Harris:2006}
{Harris} D.~E.,  {Krawczynski} H.,  2006, \mn@doi [\araa]
  {10.1146/annurev.astro.44.051905.092446}, \href
  {http://adsabs.harvard.edu/abs/2006ARA%26A..44..463H} {44, 463}

\bibitem[\protect\citeauthoryear{{Hawley} \& {Krolik}}{{Hawley} \&
  {Krolik}}{2006}]{HawleyKrolik:2006}
{Hawley} J.~F.,  {Krolik} J.~H.,  2006, \mn@doi [\apj] {10.1086/500385}, \href
  {http://adsabs.harvard.edu/abs/2006ApJ...641..103H} {641, 103}

\bibitem[\protect\citeauthoryear{{Koide}, {Shibata}, {Kudoh}  \&
  {Meier}}{{Koide} et~al.}{2002}]{Koide:2002}
{Koide} S.,  {Shibata} K.,  {Kudoh} T.,   {Meier} D.~L.,  2002, \mn@doi
  [Science] {10.1126/science.1068240}, \href
  {http://adsabs.harvard.edu/abs/2002Sci...295.1688K} {295, 1688}

\bibitem[\protect\citeauthoryear{{Komissarov}, {Vlahakis}, {K{\"o}nigl}  \&
  {Barkov}}{{Komissarov} et~al.}{2009}]{Komissarov:2009}
{Komissarov} S.~S.,  {Vlahakis} N.,  {K{\"o}nigl} A.,   {Barkov} M.~V.,  2009,
  \mn@doi [\mnras] {10.1111/j.1365-2966.2009.14410.x}, \href
  {http://adsabs.harvard.edu/abs/2009MNRAS.394.1182K} {394, 1182}

\bibitem[\protect\citeauthoryear{{Komissarov}, {Vlahakis}  \&
  {K{\"o}nigl}}{{Komissarov} et~al.}{2010}]{Komissarov:2010}
{Komissarov} S.~S.,  {Vlahakis} N.,   {K{\"o}nigl} A.,  2010, \mn@doi [\mnras]
  {10.1111/j.1365-2966.2010.16779.x}, \href
  {http://adsabs.harvard.edu/abs/2010MNRAS.407...17K} {407, 17}

\bibitem[\protect\citeauthoryear{{Laurent}, {Rodriguez}, {Wilms}, {Cadolle
  Bel}, {Pottschmidt}  \& {Grinberg}}{{Laurent} et~al.}{2011}]{Laurent:2011}
{Laurent} P.,  {Rodriguez} J.,  {Wilms} J.,  {Cadolle Bel} M.,  {Pottschmidt}
  K.,   {Grinberg} V.,  2011, \mn@doi [Science] {10.1126/science.1200848},
  \href {http://adsabs.harvard.edu/abs/2011Sci...332..438L} {332, 438}

\bibitem[\protect\citeauthoryear{{Li}, {Chiueh}  \& {Begelman}}{{Li}
  et~al.}{1992}]{Li:1992}
{Li} Z.-Y.,  {Chiueh} T.,   {Begelman} M.~C.,  1992, \mn@doi [\apj]
  {10.1086/171597}, \href {http://adsabs.harvard.edu/abs/1992ApJ...394..459L}
  {394, 459}

\bibitem[\protect\citeauthoryear{{Lovelace} \& {Contopoulos}}{{Lovelace} \&
  {Contopoulos}}{1990}]{Lovelace:1990}
{Lovelace} R.~V.~E.,  {Contopoulos} J.,  1990, in {Beck} R.,  {Wielebinski} R.,
    {Kronberg} P.~P.,  eds,  IAU Symposium Vol. 140, Galactic and Intergalactic
  Magnetic Fields. p.~337

\bibitem[\protect\citeauthoryear{{Maitra}, {Markoff}, {Brocksopp}, {Noble},
  {Nowak}  \& {Wilms}}{{Maitra} et~al.}{2009}]{Maitra:2009}
{Maitra} D.,  {Markoff} S.,  {Brocksopp} C.,  {Noble} M.,  {Nowak} M.,
  {Wilms} J.,  2009, \mn@doi [\mnras] {10.1111/j.1365-2966.2009.14896.x}, \href
  {http://adsabs.harvard.edu/abs/2009MNRAS.398.1638M} {398, 1638}

\bibitem[\protect\citeauthoryear{Malzac}{Malzac}{2014}]{Malzac:2014}
Malzac J.,  2014, \mn@doi [Mon. Not. Roy. Astron. Soc.]
  {10.1093/mnras/stu1144}, 443, 299

\bibitem[\protect\citeauthoryear{{Markoff}}{{Markoff}}{2010}]{Markoff:2010}
{Markoff} S.,  2010, in {Belloni} T.,  ed.,  Lecture Notes in Physics, Berlin
  Springer Verlag Vol. 794, Lecture Notes in Physics, Berlin Springer Verlag.
  p.~143 (\mn@eprint {arXiv} {0909.2574}), \mn@doi{10.1007/978-3-540-76937-8_6}

\bibitem[\protect\citeauthoryear{{Markoff}, {Falcke}  \& {Fender}}{{Markoff}
  et~al.}{2001}]{Markoff:2001}
{Markoff} S.,  {Falcke} H.,   {Fender} R.,  2001, \mn@doi [\aap]
  {10.1051/0004-6361:20010420}, \href
  {http://adsabs.harvard.edu/abs/2001A%26A...372L..25M} {372, L25}

\bibitem[\protect\citeauthoryear{{Markoff}, {Nowak}  \& {Wilms}}{{Markoff}
  et~al.}{2005}]{Markoff:2005}
{Markoff} S.,  {Nowak} M.~A.,   {Wilms} J.,  2005, \mn@doi [\apj]
  {10.1086/497628}, \href {http://adsabs.harvard.edu/abs/2005ApJ...635.1203M}
  {635, 1203}

\bibitem[\protect\citeauthoryear{{Mart{\'{\i}}-Vidal}, {Muller}, {Vlemmings},
  {Horellou}  \& {Aalto}}{{Mart{\'{\i}}-Vidal} et~al.}{2015}]{MartiVidal:2015}
{Mart{\'{\i}}-Vidal} I.,  {Muller} S.,  {Vlemmings} W.,  {Horellou} C.,
  {Aalto} S.,  2015, \mn@doi [Science] {10.1126/science.aaa1784}, \href
  {http://adsabs.harvard.edu/abs/2015Sci...348..311M} {348, 311}

\bibitem[\protect\citeauthoryear{{McKinney}}{{McKinney}}{2006}]{McKinney:2006}
{McKinney} J.~C.,  2006, \mn@doi [\mnras] {10.1111/j.1365-2966.2006.10256.x},
  \href {http://adsabs.harvard.edu/abs/2006MNRAS.368.1561M} {368, 1561}

\bibitem[\protect\citeauthoryear{{Meier}}{{Meier}}{2003}]{Meier:2003}
{Meier} D.~L.,  2003, \mn@doi [\nar] {10.1016/S1387-6473(03)00120-9}, \href
  {http://adsabs.harvard.edu/abs/2003NewAR..47..667M} {47, 667}

\bibitem[\protect\citeauthoryear{{Meier}}{{Meier}}{2012}]{Meier:2012}
{Meier} D.~L.,  2012, {Black Hole Astrophysics: The Engine Paradigm}

\bibitem[\protect\citeauthoryear{{Mendoza} \& {Longair}}{{Mendoza} \&
  {Longair}}{2001}]{MendozaLongair:2001}
{Mendoza} S.,  {Longair} M.~S.,  2001, \mn@doi [\mnras]
  {10.1046/j.1365-8711.2001.04271.x}, \href
  {http://adsabs.harvard.edu/abs/2001MNRAS.324..149M} {324, 149}

\bibitem[\protect\citeauthoryear{{Merloni}, {Heinz}  \& {di Matteo}}{{Merloni}
  et~al.}{2003}]{Merloni:2003}
{Merloni} A.,  {Heinz} S.,   {di Matteo} T.,  2003, \mn@doi [\mnras]
  {10.1046/j.1365-2966.2003.07017.x}, \href
  {http://adsabs.harvard.edu/abs/2003MNRAS.345.1057M} {345, 1057}

\bibitem[\protect\citeauthoryear{{Mertens}, {Lobanov}, {Walker}  \&
  {Hardee}}{{Mertens} et~al.}{2016}]{Mertens:2016}
{Mertens} F.,  {Lobanov} A.~P.,  {Walker} R.~C.,   {Hardee} P.~E.,  2016,
  \mn@doi [\aap] {10.1051/0004-6361/201628829}, \href
  {http://adsabs.harvard.edu/abs/2016A%26A...595A..54M} {595, A54}

\bibitem[\protect\citeauthoryear{{Meyer}, {Sparks}, {Biretta}, {Anderson},
  {Sohn}, {van der Marel}, {Norman}  \& {Nakamura}}{{Meyer}
  et~al.}{2013}]{Meyer:2013}
{Meyer} E.~T.,  {Sparks} W.~B.,  {Biretta} J.~A.,  {Anderson} J.,  {Sohn}
  S.~T.,  {van der Marel} R.~P.,  {Norman} C.,   {Nakamura} M.,  2013, \mn@doi
  [\apjl] {10.1088/2041-8205/774/2/L21}, \href
  {http://adsabs.harvard.edu/abs/2013ApJ...774L..21M} {774, L21}

\bibitem[\protect\citeauthoryear{{Michel}}{{Michel}}{1969}]{Michel:1969}
{Michel} F.~C.,  1969, \mn@doi [\apj] {10.1086/150233}, \href
  {http://adsabs.harvard.edu/abs/1969ApJ...158..727M} {158, 727}

\bibitem[\protect\citeauthoryear{{Migliari}, {Tomsick}, {Maccarone}, {Gallo},
  {Fender}, {Nelemans}  \& {Russell}}{{Migliari} et~al.}{2006}]{Migliari:2006}
{Migliari} S.,  {Tomsick} J.~A.,  {Maccarone} T.~J.,  {Gallo} E.,  {Fender}
  R.~P.,  {Nelemans} G.,   {Russell} D.~M.,  2006, \mn@doi [\apjl]
  {10.1086/505028}, \href {http://adsabs.harvard.edu/abs/2006ApJ...643L..41M}
  {643, L41}

\bibitem[\protect\citeauthoryear{{Miller-Jones}, {Fender}  \&
  {Nakar}}{{Miller-Jones} et~al.}{2006}]{JMFN:2006}
{Miller-Jones} J.~C.~A.,  {Fender} R.~P.,   {Nakar} E.,  2006, \mn@doi [\mnras]
  {10.1111/j.1365-2966.2006.10092.x}, \href
  {http://adsabs.harvard.edu/abs/2006MNRAS.367.1432M} {367, 1432}

\bibitem[\protect\citeauthoryear{{Mo{\'s}cibrodzka}, {Falcke}  \&
  {Shiokawa}}{{Mo{\'s}cibrodzka} et~al.}{2016}]{Moscibrodzka:2016}
{Mo{\'s}cibrodzka} M.,  {Falcke} H.,   {Shiokawa} H.,  2016, \mn@doi [\aap]
  {10.1051/0004-6361/201526630}, \href
  {http://adsabs.harvard.edu/abs/2016A%26A...586A..38M} {586, A38}

\bibitem[\protect\citeauthoryear{{Nakamura} \& {Asada}}{{Nakamura} \&
  {Asada}}{2013}]{Nakamura:2013}
{Nakamura} M.,  {Asada} K.,  2013, \mn@doi [\apj]
  {10.1088/0004-637X/775/2/118}, \href
  {http://adsabs.harvard.edu/abs/2013ApJ...775..118N} {775, 118}

\bibitem[\protect\citeauthoryear{{Nemmen} \& {Tchekhovskoy}}{{Nemmen} \&
  {Tchekhovskoy}}{2015}]{NemmenTchekhovskoy:2015}
{Nemmen} R.~S.,  {Tchekhovskoy} A.,  2015, \mn@doi [\mnras]
  {10.1093/mnras/stv260}, \href
  {http://adsabs.harvard.edu/abs/2015MNRAS.449..316N} {449, 316}

\bibitem[\protect\citeauthoryear{{Parker}}{{Parker}}{1958}]{Parker:1958}
{Parker} E.~N.,  1958, \mn@doi [\apj] {10.1086/146579}, \href
  {http://adsabs.harvard.edu/abs/1958ApJ...128..664P} {128, 664}

\bibitem[\protect\citeauthoryear{{Pepe}, {Vila}  \& {Romero}}{{Pepe}
  et~al.}{2015}]{Pepe:2015}
{Pepe} C.,  {Vila} G.~S.,   {Romero} G.~E.,  2015, \mn@doi [\aap]
  {10.1051/0004-6361/201527156}, \href
  {http://adsabs.harvard.edu/abs/2015A%26A...584A..95P} {584, A95}

\bibitem[\protect\citeauthoryear{{Perlman}, {Biretta}, {Zhou}, {Sparks}  \&
  {Macchetto}}{{Perlman} et~al.}{1999}]{Perlman:1999}
{Perlman} E.~S.,  {Biretta} J.~A.,  {Zhou} F.,  {Sparks} W.~B.,   {Macchetto}
  F.~D.,  1999, \mn@doi [\aj] {10.1086/300844}, \href
  {http://adsabs.harvard.edu/abs/1999AJ....117.2185P} {117, 2185}

\bibitem[\protect\citeauthoryear{{Polko}, {Meier}  \& {Markoff}}{{Polko}
  et~al.}{2010}]{Polko:2010}
{Polko} P.,  {Meier} D.~L.,   {Markoff} S.,  2010, \mn@doi [\apj]
  {10.1088/0004-637X/723/2/1343}, \href
  {http://adsabs.harvard.edu/abs/2010ApJ...723.1343P} {723, 1343}

\bibitem[\protect\citeauthoryear{{Polko}, {Meier}  \& {Markoff}}{{Polko}
  et~al.}{2013}]{Polko:2013}
{Polko} P.,  {Meier} D.~L.,   {Markoff} S.,  2013, \mn@doi [\mnras]
  {10.1093/mnras/sts052}, \href
  {http://adsabs.harvard.edu/abs/2013MNRAS.428..587P} {428, 587}

\bibitem[\protect\citeauthoryear{{Polko}, {Meier}  \& {Markoff}}{{Polko}
  et~al.}{2014}]{Polko:2014}
{Polko} P.,  {Meier} D.~L.,   {Markoff} S.,  2014, \mn@doi [\mnras]
  {10.1093/mnras/stt2155}, \href
  {http://adsabs.harvard.edu/abs/2014MNRAS.438..959P} {438, 959}

\bibitem[\protect\citeauthoryear{{Potter} \& {Cotter}}{{Potter} \&
  {Cotter}}{2012}]{PotterCotter:2012}
{Potter} W.~J.,  {Cotter} G.,  2012, \mn@doi [\mnras]
  {10.1111/j.1365-2966.2012.20918.x}, \href
  {http://adsabs.harvard.edu/abs/2012MNRAS.423..756P} {423, 756}

\bibitem[\protect\citeauthoryear{Press, Teukolsky, Vetterling  \&
  Flannery}{Press et~al.}{1993}]{NRfortran:1993}
Press W.~H.,  Teukolsky S.~A.,  Vetterling W.~T.,   Flannery B.~P.,  1993,
  Numerical Recipes in FORTRAN; The Art of Scientific Computing, 2nd edn.
Cambridge University Press, New York, NY, USA

\bibitem[\protect\citeauthoryear{{Prieto}, {Fern{\'a}ndez-Ontiveros},
  {Markoff}, {Espada}  \& {Gonz{\'a}lez-Mart{\'{\i}}n}}{{Prieto}
  et~al.}{2016}]{Prieto:2016}
{Prieto} M.~A.,  {Fern{\'a}ndez-Ontiveros} J.~A.,  {Markoff} S.,  {Espada} D.,
   {Gonz{\'a}lez-Mart{\'{\i}}n} O.,  2016, \mn@doi [\mnras]
  {10.1093/mnras/stw166}, \href
  {http://adsabs.harvard.edu/abs/2016MNRAS.457.3801P} {457, 3801}

\bibitem[\protect\citeauthoryear{{Rawlings} \& {Saunders}}{{Rawlings} \&
  {Saunders}}{1991}]{RawlingsSaunders:1991}
{Rawlings} S.,  {Saunders} R.,  1991, \mn@doi [\nat] {10.1038/349138a0}, \href
  {http://adsabs.harvard.edu/abs/1991Natur.349..138R} {349, 138}

\bibitem[\protect\citeauthoryear{{Rees} \& {Meszaros}}{{Rees} \&
  {Meszaros}}{1994}]{ReesMeszaros:1994}
{Rees} M.~J.,  {Meszaros} P.,  1994, \mn@doi [\apjl] {10.1086/187446}, \href
  {http://adsabs.harvard.edu/abs/1994ApJ...430L..93R} {430, L93}

\bibitem[\protect\citeauthoryear{{Ressler}, {Tchekhovskoy}, {Quataert},
  {Chandra}  \& {Gammie}}{{Ressler} et~al.}{2015}]{Ressler:2015}
{Ressler} S.~M.,  {Tchekhovskoy} A.,  {Quataert} E.,  {Chandra} M.,   {Gammie}
  C.~F.,  2015, \mn@doi [\mnras] {10.1093/mnras/stv2084}, \href
  {http://adsabs.harvard.edu/abs/2015MNRAS.454.1848R} {454, 1848}

\bibitem[\protect\citeauthoryear{{Romero}, {Torres}, {Kaufman Bernad{\'o}}  \&
  {Mirabel}}{{Romero} et~al.}{2003}]{Romero:2003}
{Romero} G.~E.,  {Torres} D.~F.,  {Kaufman Bernad{\'o}} M.~M.,   {Mirabel}
  I.~F.,  2003, \mn@doi [\aap] {10.1051/0004-6361:20031314-1}, \href
  {http://adsabs.harvard.edu/abs/2003A%26A...410L...1R} {410, L1}

\bibitem[\protect\citeauthoryear{{Romero}, {Boettcher}, {Markoff}  \&
  {Tavecchio}}{{Romero} et~al.}{2017}]{Romero:2017}
{Romero} G.~E.,  {Boettcher} M.,  {Markoff} S.,   {Tavecchio} F.,  2017,
  \mn@doi [\ssr] {10.1007/s11214-016-0328-2}, \href
  {http://adsabs.harvard.edu/abs/2017SSRv..tmp....1R} {}

\bibitem[\protect\citeauthoryear{{Russell} \& {Shahbaz}}{{Russell} \&
  {Shahbaz}}{2014}]{RussellShahbaz:2014}
{Russell} D.~M.,  {Shahbaz} T.,  2014, \mn@doi [\mnras]
  {10.1093/mnras/stt2330}, \href
  {http://adsabs.harvard.edu/abs/2014MNRAS.438.2083R} {438, 2083}

\bibitem[\protect\citeauthoryear{{Russell}, {Soria}, {Miller-Jones}, {Curran},
  {Markoff}, {Russell}  \& {Sivakoff}}{{Russell} et~al.}{2014}]{Russell:2014}
{Russell} T.~D.,  {Soria} R.,  {Miller-Jones} J.~C.~A.,  {Curran} P.~A.,
  {Markoff} S.,  {Russell} D.~M.,   {Sivakoff} G.~R.,  2014, \mn@doi [\mnras]
  {10.1093/mnras/stt2498}, \href
  {http://adsabs.harvard.edu/abs/2014MNRAS.439.1390R} {439, 1390}

\bibitem[\protect\citeauthoryear{{Russell}, {Fabian}, {McNamara}  \&
  {Broderick}}{{Russell} et~al.}{2015}]{Russell:2015}
{Russell} H.~R.,  {Fabian} A.~C.,  {McNamara} B.~R.,   {Broderick} A.~E.,
  2015, \mn@doi [\mnras] {10.1093/mnras/stv954}, \href
  {http://adsabs.harvard.edu/abs/2015MNRAS.451..588R} {451, 588}

\bibitem[\protect\citeauthoryear{{Sauty} \& {Tsinganos}}{{Sauty} \&
  {Tsinganos}}{1994}]{Sauty:1994}
{Sauty} C.,  {Tsinganos} K.,  1994, \aap, \href
  {http://adsabs.harvard.edu/abs/1994A%26A...287..893S} {287, 893}

\bibitem[\protect\citeauthoryear{{Sauty}, {Cayatte}, {Lima}, {Matsakos}  \&
  {Tsinganos}}{{Sauty} et~al.}{2012}]{Sauty:2012}
{Sauty} C.,  {Cayatte} V.,  {Lima} J.~J.~G.,  {Matsakos} T.,   {Tsinganos} K.,
  2012, \mn@doi [\apjl] {10.1088/2041-8205/759/1/L1}, \href
  {http://adsabs.harvard.edu/abs/2012ApJ...759L...1S} {759, L1}

\bibitem[\protect\citeauthoryear{Stoer \& Bulirsch}{Stoer \&
  Bulirsch}{2013}]{Stoer:2013}
Stoer J.,  Bulirsch R.,  2013, Introduction to numerical analysis.
~ Vol. 12, Springer Science \& Business Media

\bibitem[\protect\citeauthoryear{{Tchekhovskoy} \& {Bromberg}}{{Tchekhovskoy}
  \& {Bromberg}}{2016}]{Tchekhovskoy:2016}
{Tchekhovskoy} A.,  {Bromberg} O.,  2016, \mn@doi [\mnras]
  {10.1093/mnrasl/slw064}, \href
  {http://adsabs.harvard.edu/abs/2016MNRAS.461L..46T} {461, L46}

\bibitem[\protect\citeauthoryear{{Tchekhovskoy}, {Narayan}  \&
  {McKinney}}{{Tchekhovskoy} et~al.}{2011}]{Tchekhovskoy:2011}
{Tchekhovskoy} A.,  {Narayan} R.,   {McKinney} J.~C.,  2011, \mn@doi [\mnras]
  {10.1111/j.1745-3933.2011.01147.x}, \href
  {http://adsabs.harvard.edu/abs/2011MNRAS.418L..79T} {418, L79}

\bibitem[\protect\citeauthoryear{{Vlahakis} \& {K{\"o}nigl}}{{Vlahakis} \&
  {K{\"o}nigl}}{2003}]{VlahakisKonigl:2003}
{Vlahakis} N.,  {K{\"o}nigl} A.,  2003, \mn@doi [\apj] {10.1086/378226}, \href
  {http://adsabs.harvard.edu/abs/2003ApJ...596.1080V} {596, 1080}

\bibitem[\protect\citeauthoryear{{Vlahakis}, {Tsinganos}, {Sauty}  \&
  {Trussoni}}{{Vlahakis} et~al.}{2000}]{VTST:2000}
{Vlahakis} N.,  {Tsinganos} K.,  {Sauty} C.,   {Trussoni} E.,  2000, \mn@doi
  [\mnras] {10.1046/j.1365-8711.2000.03703.x}, \href
  {http://adsabs.harvard.edu/abs/2000MNRAS.318..417V} {318, 417}

\bibitem[\protect\citeauthoryear{{Weber} \& {Davis}}{{Weber} \&
  {Davis}}{1967}]{WeberDavis:1967}
{Weber} E.~J.,  {Davis} Jr. L.,  1967, \mn@doi [\apj] {10.1086/149138}, \href
  {http://adsabs.harvard.edu/abs/1967ApJ...148..217W} {148, 217}

\bibitem[\protect\citeauthoryear{{Weisskopf} et~al.,}{{Weisskopf}
  et~al.}{2016}]{Weisskopf:2016}
{Weisskopf} M.~C.,  et~al., 2016, in Space Telescopes and Instrumentation 2016:
  Ultraviolet to Gamma Ray. p. 990517, \mn@doi{10.1117/12.2235240}

\bibitem[\protect\citeauthoryear{{Yuan}, {Cui}  \& {Narayan}}{{Yuan}
  et~al.}{2005}]{Yuan:2005}
{Yuan} F.,  {Cui} W.,   {Narayan} R.,  2005, \mn@doi [\apj] {10.1086/427206},
  \href {http://adsabs.harvard.edu/abs/2005ApJ...620..905Y} {620, 905}

\bibitem[\protect\citeauthoryear{{Zdziarski}, {Lubi{\'n}ski}  \&
  {Sikora}}{{Zdziarski} et~al.}{2012}]{ZLS:2012}
{Zdziarski} A.~A.,  {Lubi{\'n}ski} P.,   {Sikora} M.,  2012, \mn@doi [\mnras]
  {10.1111/j.1365-2966.2012.20903.x}, \href
  {http://adsabs.harvard.edu/abs/2012MNRAS.423..663Z} {423, 663}

\bibitem[\protect\citeauthoryear{{Zdziarski}, {Stawarz}, {Pjanka}  \&
  {Sikora}}{{Zdziarski} et~al.}{2014}]{ZSPS:2014}
{Zdziarski} A.~A.,  {Stawarz} {\L}.,  {Pjanka} P.,   {Sikora} M.,  2014,
  \mn@doi [\mnras] {10.1093/mnras/stu420}, \href
  {http://adsabs.harvard.edu/abs/2014MNRAS.440.2238Z} {440, 2238}

\makeatother
\end{thebibliography}
\end{document}